\documentclass[sigconf,xcolor,dvipsnames,usenames,screen,nonacm]{acmart}

\usepackage[inline]{enumitem}
\usepackage{cleveref}
\usepackage{amsmath}
\usepackage{bm}
\usepackage[skip=1mm]{caption}
\usepackage[table]{xcolor}
\usepackage{pgf} 
\usepackage{tcolorbox}
\usepackage{tikz}
\usepackage{placeins}

\newcommand{\header}[1]{\vspace{1mm}\noindent\textbf{#1}.}

\newcommand{\headerl}[1]{\vspace{1mm}\noindent\textit{#1}.}
\newcommand{\headerlnp}[1]{\vspace{1mm}\noindent\textit{#1}}

\newcommand{\bench}{\textsc{ERASE}}

\newcommand{\releff}{\text{RelEff}}
\newcommand{\relitems}{\text{RelItems}}

\newcommand{\na}{\cellcolor{gray!10}{n/a}}
\newcommand{\diverged}{\cellcolor{red!60}{div.}}

\newcommand*\circled[1]{\protect\tikz[baseline=(char.base)]{\protect\node[shape=circle,fill=black,draw,inner sep=0.6pt] (char) {\textcolor{white}{\footnotesize \textbf{#1}}};}}

\newcommand{\relcell}[2]{%
  \begingroup
  \ifdim\dimexpr#1pt\relax > 10pt
    \cellcolor{green!45}{\phantom{-}#2}
  \else\ifdim\dimexpr#1pt\relax > 5pt
    \cellcolor{green!35}{\phantom{-}#2}
  \else\ifdim\dimexpr#1pt\relax > -1pt
    \cellcolor{green!25}{\phantom{-}#2}
  \else\ifdim\dimexpr#1pt\relax > -5pt
    \cellcolor{yellow!40}{#2}
  \else\ifdim\dimexpr#1pt\relax > -10pt
    \cellcolor{orange!35}{#2}
  \else\ifdim\dimexpr#1pt\relax > -25pt
    \cellcolor{red!20}{#2}
  \else\ifdim\dimexpr#1pt\relax > -40pt
    \cellcolor{red!30}{#2}
  \else
    \cellcolor{red!45}{#2}
  \fi\fi\fi\fi\fi\fi\fi
  \endgroup
}

\newcommand{\addcell}[2]{%
  \begingroup
  \ifdim\dimexpr#1pt\relax > 10pt
    \cellcolor{green!45}{\phantom{-}#2}
  \else\ifdim\dimexpr#1pt\relax > 5pt
    \cellcolor{green!35}{\phantom{-}#2}
  \else\ifdim\dimexpr#1pt\relax > -1pt
    \cellcolor{green!25}{\phantom{-}#2}
  \else\ifdim\dimexpr#1pt\relax > -5pt
    \cellcolor{yellow!40}{#2}
  \else\ifdim\dimexpr#1pt\relax > -10pt
    \cellcolor{orange!35}{#2}
  \else\ifdim\dimexpr#1pt\relax > -25pt
    \cellcolor{red!20}{#2}
  \else\ifdim\dimexpr#1pt\relax > -40pt
    \cellcolor{red!30}{#2}
  \else
    \cellcolor{red!45}{#2}
  \fi\fi\fi\fi\fi\fi\fi
  \endgroup
}
\AtBeginDocument{%
  }

\setcopyright{none}
\copyrightyear{2026}
\acmYear{2026}
\acmDOI{}
\acmConference[SIGIR '26]{49th International ACM SIGIR Conference on Research and Development in Information Retrieval}{July 20--24,
  2026}{Melbourne, Australia}
\acmISBN{}

\author{Pierre Lubitzsch}
\orcid{0009-0002-4973-2328}
\affiliation{%
  \institution{BIFOLD \& TU~Berlin}
  \country{}}
\email{lubitzsch@tu-berlin.de}

\author{Maarten de Rijke}
\orcid{0000-0002-1086-0202}
\affiliation{%
  \institution{University of Amsterdam}
  \country{}}
\email{m.derijke@uva.nl}

\author{Sebastian Schelter}
\orcid{0000-0003-4722-5840}
\affiliation{%
  \institution{BIFOLD \& TU~Berlin}
  \country{}}
\email{schelter@tu-berlin.de}

\renewcommand{\shortauthors}{Lubitzsch et al.}

\begin{document}

\title[\bench{} -- A Real-World Aligned Benchmark
for Unlearning in Recommender Systems]{\bench{} -- A Real-World Aligned Benchmark\\
for Unlearning in Recommender Systems}

\begin{abstract}
Machine unlearning (MU) enables the removal of selected training data from trained models, to address privacy compliance, security, and liability issues in recommender systems. Existing MU benchmarks poorly reflect real-world recommender settings: they focus primarily on collaborative filtering, assume unrealistically large deletion requests, and overlook practical constraints such as sequential unlearning and efficiency.

We present \bench{}, a large-scale benchmark for MU in recommender systems designed to align with real-world usage. \bench{} spans three core tasks---collaborative filtering, session-based recommendation, and next-basket recommendation---and includes unlearning scenarios inspired by real-world applications, such as sequentially removing sensitive interactions or spam. The benchmark covers seven unlearning algorithms, including general-purpose and recommender-specific methods, across nine public datasets and nine state-of-the-art models. We execute \bench{} to produce more than 600 GB of reusable artifacts, such as extensive experimental logs and more than a thousand model checkpoints. 

Crucially, the artifacts that we release enable systematic analysis of where current unlearning methods succeed and where they fall short. \bench{} showcases that approximate unlearning can match retraining in some settings, but robustness varies widely across datasets and architectures. Repeated unlearning exposes weaknesses in general-purpose methods, especially for attention-based and recurrent models, while recommender-specific approaches behave more reliably.
\bench{} provides the empirical foundation to help the community assess, drive, and track progress toward practical MU in recommender systems.
\end{abstract}




\maketitle

\section{Introduction}

Modern recommender systems rely heavily on user interaction data, such as clicks, ratings, and purchases, to build personalized models. 

\header{The need for machine unlearning} Personalization enhances the user experience, but at the same time makes it challenging to adhere to privacy rights such as the ``right-to-be-forgotten'' from the GDPR~\cite{gdpr_article17} regulation in Europe and similar regulations adopted outside Europe~\cite{ccpa_faq, DigiChina_AI_Rec}. These regulations mandate that responsible, ethical, and legally compliant AI systems give their users the right to request the timely deletion of their personal data from databases and models trained on it~\cite {stoyanovich2022responsible}. Being able to efficiently remove data from models is also important for security, e.g., to quickly react to spam interactions~\cite{amazonschoice, marino2025position}, unintended data leakage~\cite{carlini2019secret}, and to handle liability issues, for example when a model consumed copyrighted content~\cite{yao2024copyright}.
The outlined challenges give rise to \emph{machine unlearning}~(MU), the problem of efficiently removing the influence of selected training data points from a machine learning model post training. Modern machine learning models are known to encode information about their training data, ranging from statistical influence to explicit memorization of individual examples~\cite{models_remember_training_data, CarliniTWJHLRBS21}. After unlearning data points, there should be no information about these points present in the model; it should behave as if they had never been in the training set. Machine unlearning is an active and growing area of research~\cite{cao_2022, ginart2019makingaiforgetyou, izzo2021approximatedatadeletionmachine, schelter2021hedgecut, wang2022efficientlymaintainingbasketrecommendations, wu2020deltagradrapidretrainingmachine, Wu_2020}.
Existing approaches range from exact retraining strategies to approximate techniques that update model parameters or intermediate statistics to remove the influence of specific data points.
However, the evaluation of unlearning algorithms is especially challenging since multiple aspects, such as model utility after unlearning, the amount of information about the unlearned data still contained in the model, and efficiency, have to be considered simultaneously.

\header{Shortcomings of existing benchmarks for unlearning in recommender systems} While MU has been studied broadly across machine learning tasks~\cite{triantafillou2024makingprogressunlearningfindings}, recommender systems pose distinct challenges. Recommendation models are trained on large-scale, highly sparse user–item interaction data, are frequently updated, and often rely on sequential or graph-based architectures. Unlearning requests in recommender systems typically arrive continuously and must be handled under strict latency and cost constraints. Current unlearning benchmarks like CURE4Rec~\cite{chen2024cure4rec} do not sufficiently address the real-world demands of unlearning in recommender systems (\Cref{sec:rqs}). They focus on collaborative filtering (CF)~\cite{bpr, ibcf, lightgcn, simrec, dccf}, overlooking tasks that are common in e-commerce and entertainment platforms, such as session-based recommendation (SBR)~\cite{hidasi2016sessionbasedrecommendationsrecurrentneural, sknn, gru4rec, srgnn, sasrec, narm} and next-basket recommendation (NBR)~\cite{rendle2010nbr, tifuknn, upcf, clea, dnntsp, sets2sets}. Existing benchmarks simulate unlearning by deleting large chunks of data—up to 5\% of all interactions—whereas actual systems face continuous, small-scale deletion requests, in-between regular retraining cycles. The choice of data to unlearn is also misrepresented: rather than being random, real scenarios often evolve around specific types of content, like interactions with sensitive items~\cite{schelter2023forget} or interactions from spammers~\cite{amazonschoice}. Promising general-purpose unlearning algorithms, e.g., from the NeurIPS 2023 unlearning competition~\cite{triantafillou2024makingprogressunlearningfindings}, remain underexplored for recommendation-specific applications. Operational efficiency is another overlooked factor: unlearning must be fast, inexpensive, and ideally executable directly on deployed models~\cite{schelter2021hedgecut}. The unlearning procedures in CURE4Rec~\cite{chen2024cure4rec} are only an order of magnitude faster than retraining, which is not attractive for real-world deployment. These limitations align with feedback that we gathered through interviews with senior team leaders at European companies operating recommender systems with millions of users. They report that data deletion requests are currently only reflected after weekly or monthly retraining cycles (and retraining itself can take up to a day), and expressed concerns about whether existing unlearning methods would be efficient enough for deployment.

\header{\bench{} -- A real-world aligned benchmark for unlearning in recommender systems} We present \bench{}, a real-world aligned benchmark for unlearning in recommender systems, building on our prior work from FAccTRec'25~\cite{lubitzsch2025towards}. We formulate four guiding questions to address shortcomings of existing benchmarks and describe our design decisions in \Cref{sec:design}. We cover NBR and SBR in addition to CF. For each task, we include multiple models, unlearning scenarios, and datasets. We consider approximate unlearning algorithms for neural recommendation models~\cite{li2023selectivecollaborativeinfluencefunction}, graph neural networks (GNNs)~\cite{Wu_2023, kun_wu_2023, dong_2024}, and methods adapted from the NeurIPS'23 unlearning competition~\cite{triantafillou2024makingprogressunlearningfindings}. Unlike existing benchmarks, \bench{} issues small unlearning requests sequentially instead of unlearning a single forget set at once, reflecting domain-specific and time-sensitive requests. The unlearned models are evaluated based on retained utility, unlearning effectiveness, and computational efficiency.

\header{Contributions} Our contributions are as follows:
\begin{itemize}[leftmargin=*,nosep]
  \item We identify shortcomings of existing unlearning benchmarks and design \bench{}, a real-world aligned benchmark for unlearning in recommender systems. \bench{} covers three recommendation tasks using two unlearning scenarios with seven unlearning algorithms across nine datasets and nine state-of-the-art models.
  
  \item We run \bench{} to produce an extensive number of reusable artifacts, enabling researchers to evaluate new unlearning algorithms without additional training of recommenders. We illustrate the value of these artifacts for assessing the current state of knowledge regarding MU for recommender systems.
  
  \item We provide our benchmark code, reusable artifacts, and usage guides under an open license at \url{https://github.com/deem-data/erase-bench}.
\end{itemize}


\section{Background on Machine Unlearning}
\label{sec:background}
Machine unlearning removes the influence of specific data from a trained model. Formally, let $D$ be the training dataset, $\mathcal{A}$ be a training algorithm, $U$ be an unlearning algorithm, $D_f \subseteq D$ be the forget set, and $D_r = D \setminus D_f$ be the retain set. $U$ is an exact unlearning algorithm iff for all $T \subseteq \mathcal{H}$:
$
    \mathbb{P}(\mathcal{A}(D_r) \in T) = \mathbb{P}(U(\mathcal{A}(D), D_r, D_f) \in T)
$
where $\mathcal{H}$ is the hypothesis space. The unlearned model should match the distribution of a model retrained from scratch on $D_r$.
For approximate unlearning, $U$ is a $(\varepsilon, \delta)$-unlearning algorithm iff for all $T \subseteq \mathcal{H}$: $\mathbb{P}(\mathcal{A}(D_r) \in T) \leq \exp(\varepsilon) \, \mathbb{P}(U(\mathcal{A}(D), D_r, D_f) \in T) + \delta$ and $\mathbb{P}(U(\mathcal{A}(D), D_r, D_f) \in T) \leq \exp(\varepsilon) \, \mathbb{P}(\mathcal{A}(D_r) \in T) + \delta$ with $\varepsilon \in [0, \infty), \delta \in [0, 1)$~\cite{triantafillou2024makingprogressunlearningfindings}. Lower $\varepsilon$ and $\delta$ values indicate stronger unlearning guarantees with $(0, 0)$-unlearning being exact unlearning.
\section{Motivation for a New Benchmark}
\label{sec:rqs}

We argue that existing unlearning benchmarks for recommender systems, such as CURE4Rec~\cite{chen2024cure4rec}, fail to reflect real-world recommendation scenarios.

\headerl{Lack of coverage of diverse recommendation tasks} In CURE4Rec~\cite{chen2024cure4rec}, the only recommendation task considered is collaborative filtering. But the field of recommendation is much broader. Many tasks are sequential and leverage temporal information to model how user preferences evolve. Important real-world applications include e-commerce and retail, which commonly rely on SBR~\cite{hidasi2016sessionbasedrecommendationsrecurrentneural, sknn, gru4rec, srgnn, sasrec, narm} and NBR~\cite{rendle2010nbr, tifuknn, upcf, clea, dnntsp, sets2sets}.
A comprehensive unlearning benchmark should also cover these tasks.

\begin{figure*}[t!]
\centering
\includegraphics[width=\textwidth]{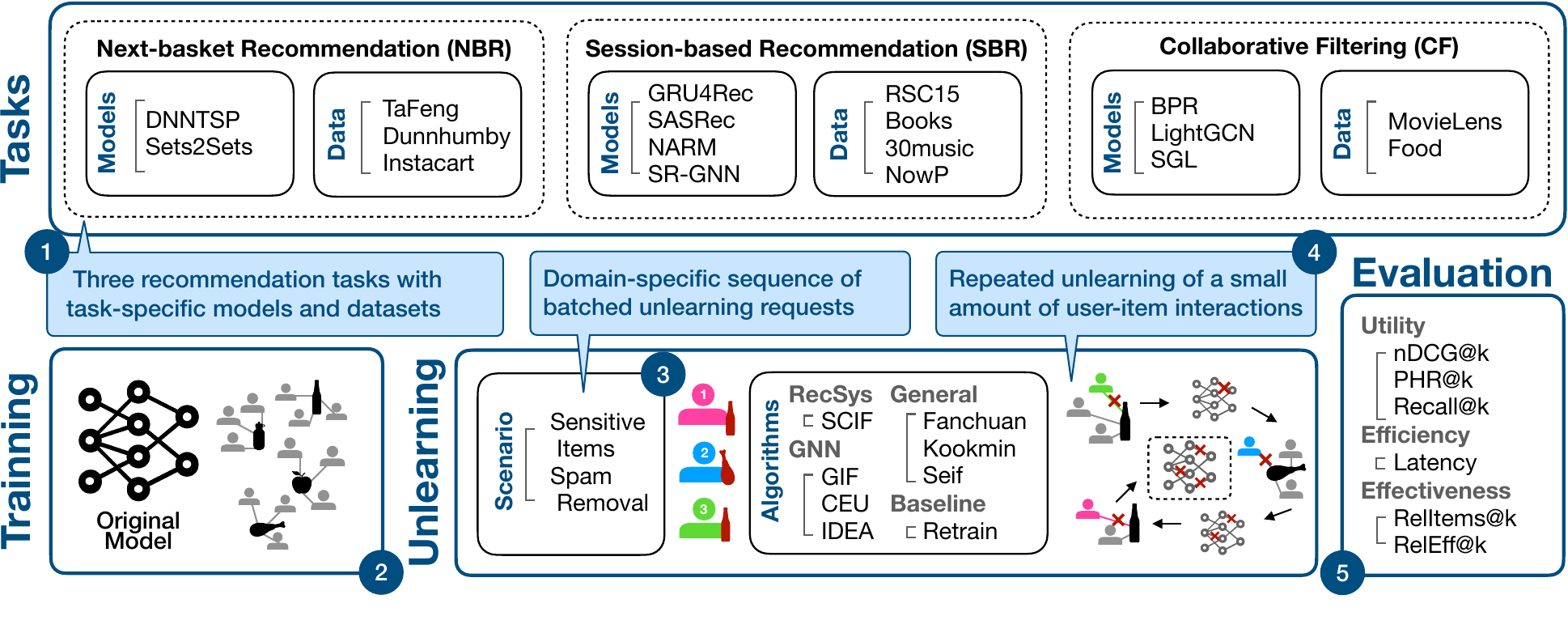}
\vspace{-8mm}
\caption{Overview of \bench{}. \circled{1} Each experiment first trains a task-specific model on a chosen dataset~\circled{2}. A sequence of unlearning requests is defined \circled{3} (based on a scenario), and an unlearning algorithm is applied repeatedly \circled{4}. Finally, the unlearned model is evaluated for utility, efficiency, and effectiveness \circled{5}, in comparison to a model retrained on the retain set.}
\label{fig:overview}
\end{figure*}

\headerl{Lack of domain-specific patterns in interactions to unlearn} In some domains, the choice of interactions to unlearn for a user follows a pattern. An example is the removal of interactions with sensitive items, for instance, alcoholic items for a user suffering from addiction~\cite{schelter2023forget,schelter2024snapcase}. Another area where the interactions follow a pattern is unlearning the interactions from malicious actors such as spammers, who try to manipulate the recommender system's behavior~\cite{amazonschoice}. Current benchmarks do not evaluate domain-specific choices of the interactions to unlearn. 

\headerl{Unrealistic forget set sizes} In real-world settings, unlearning requests typically arrive as many small updates over time, for example when users withdraw consent for personalization~\cite{schelter2021hedgecut} or when security teams remove spam users. Such requests should be handled with low latency, ideally within seconds. However, existing benchmarks focus on a single, unrealistically large unlearning request. For example, the NeurIPS unlearning competition~\cite{triantafillou2024makingprogressunlearningfindings} and CURE4Rec~\cite{chen2024cure4rec} benchmarks remove up to $5\%$ of the data at once. In practice, since models are retrained periodically anyway (e.g., weekly~\cite{aws2025personalize} or monthly~\cite{tencentrecommendation2025}), unlearning is most useful for efficiently removing small batches of data between retraining cycles.

\headerl{Insufficient focus on efficiency} Unlearning needs to be executed fast and with no or negligible utility loss. In CURE4Rec~\cite{chen2024cure4rec}, the runtime of unlearning is only roughly one order of magnitude faster than retraining from scratch, which is not efficient enough for real-world use cases. Models are often trained for a day, according to our interviews, so reducing the time needed for carrying out the unlearning request to seconds or minutes requires a decrease of at least three orders of magnitude.

\headerl{Insufficient coverage of available unlearning algorithms} The only approximate unlearning method in CURE4Rec is SCIF~\cite{li2023selectivecollaborativeinfluencefunction}, which conducts a single second-order parameter update. However, the 2023 NeurIPS unlearning competition~\cite{triantafillou2024makingprogressunlearningfindings} contains several high-scoring first-order iterative approaches to unlearning, which also need to be evaluated for recommendation tasks.

\section{The \bench{} Benchmark}
\label{sec:design}

\textbf{Design criteria.} To address the shortcomings of existing benchmarks and arrive at a wishlist for MU benchmarks for recommender systems, we pose the following guiding questions: 

\headerlnp{What is the unlearning effectiveness of approximate unlearning methods?} Due to a lack of work on evaluating approximate unlearning across different recommendation tasks and scenarios, an overview of the performance of these methods is still missing in literature. 

\headerlnp{Does approximate unlearning approach achieve robust, consistent performance across CF, SBR, and NBR recommendation tasks?} Our preliminary work~\cite{lubitzsch2025towards} indicates that unlearning algorithms are highly sensitive to hyperparameter choices. Different unlearning algorithms also have different ways to remove data from the model. An approach with robust performance across multiple tasks and scenarios is desirable. It is an open question whether any single method achieves consistently strong performance across tasks, or whether effective unlearning requires careful hyperparameter and algorithm selection per task.

\headerlnp{How does model architecture influence approximate unlearning effectiveness in recommendation systems?} Model architecture plays a central role in determining a model's learning capacity and the representations it acquires. Hence, it is crucial to understand how unlearning effectiveness differs across architecture families and whether architecture-specific unlearning methods outperform general approaches.

\headerlnp{Do current approximate unlearning algorithms achieve sufficient efficiency for real-world deployment?} Efficiency is critical for MU: an approximate method with runtime comparable to full retraining offers little practical value. 

\vspace{1mm} \noindent These questions guide the design of a real-world aligned benchmark for unlearning in recommender systems: (i)~task settings (models and datasets), (ii)~training setups, (iii)~unlearning scenarios, and (iv)~evaluation facilities. \Cref{fig:overview} provides an overview of the resulting \bench{} benchmark.

\begin{table*}[ht]
\centering

\setlength\tabcolsep{8.35pt}
\caption{Choice of datasets across recommendation tasks for \bench{}. All datasets are publicly available.}
\renewcommand{\arraystretch}{0.8}
\begin{tabular}{l l l r r r l}
\toprule
\textbf{Dataset} & \textbf{Domain} & \textbf{Task} & \textbf{\#Users/Sessions} & \textbf{\#Items} & \textbf{\#Interactions} & \textbf{Available at} \\
\midrule
TaFeng~\cite{tafeng} & Grocery shopping & NBR & 32,266 & 23,812 & 817,741 & \href{https://www.kaggle.com/datasets/chiranjivdas09/ta-feng-grocery-dataset}{Kaggle}\\
Dunnhumby~\cite{dunnhumby} & Retail shopping & NBR & 2,500 & 92,339 & 2,595,732 & \href{https://www.dunnhumby.com/source-files/}{dunnhumby.com}\\
Instacart~\cite{instacart} & Grocery shopping & NBR & 19,435 & 13,897 & 3,346,083 & \href{https://github.com/khanhnamle1994/instacart-orders/}{GitHub}\\
\midrule
RSC15~\cite{hidasi2016sessionbasedrecommendationsrecurrentneural} & E-commerce interactions & SBR & 9,249,729 & 52,739 & 33,003,944 & \href{https://www.kaggle.com/datasets/chadgostopp/recsys-challenge-2015/data}{Kaggle}\\
Books~\cite{amazonreviews} & Amazon book reviews & SBR & 10,300,000 & 4,400,000 & 29,500,000 & \href{https://www.kaggle.com/datasets/chadgostopp/recsys-challenge-2015/data}{GitHub}\\
30music~\cite{30music} & Music streams & SBR & 2,764,474 & 5,600,000 & 31,351,954 & \href{https://recsys.deib.polimi.it/datasets/}{Remap Lab}\\
NowP~\cite{nowp} & Music on social media & SBR & 738,200 & 3,734,862 & 1,291,251,677 & \href{https://zenodo.org/records/2594483}{Zenodo}\\
\midrule
MovieLens~\cite{movielens20m} & Movie ratings & CF & 138,493 & 26,744 & 20,000,263 & \href{https://www.kaggle.com/datasets/grouplens/movielens-20m-dataset}{Kaggle}\\
Food~\cite{amazonreviews} & Amazon product reviews & CF & 7,000,000 & 603,200 & 14,300,000 & \href{https://amazon-reviews-2023.github.io/}{GitHub}\\
\bottomrule
\end{tabular}
\label{tab:datasets}
\end{table*}

\header{Recommendation tasks and models} We consider a diverse set of recommendation tasks and representative models. For collaborative filtering (CF), we include classical matrix factorization~\cite{bpr} as well as two graph-based approaches, \texttt{LightGCN}~\cite{lightgcn} and \texttt{SGL}~\cite{sgl}. To cover session-based recommendation (SBR), a key task in applications such as next-item prediction in e-commerce and music streaming, we evaluate two recurrent models, \texttt{GRU4Rec}~\cite{gru4rec} and \texttt{NARM}~\cite{narm}, the attention-based model \texttt{SASRec}~\cite{sasrec}, and the graph-based model \texttt{SR-GNN}~\cite{srgnn}. For next-basket recommendation, which focuses on predicting a user’s subsequent shopping basket, we include the graph-based model \texttt{DNNTSP}~\cite{dnntsp} and the attention-based model \texttt{Sets2Sets}~\cite{sets2sets}.

\header{Datasets}  We use widely adopted, publicly available datasets for our benchmark (\Cref{tab:datasets}). The three NBR datasets relate to grocery or retail shopping. For SBR, two of the four datasets capture e-commerce interactions, while the other two are music datasets derived from social media posts about songs (NowP) or collected from internet radio stations (30music). For CF, we select two datasets containing user ratings for movies and food products.

\header{Unlearning algorithms} We include recommendation-specific as well as general unlearning algorithms in the benchmark.

\headerl{Custom unlearning algorithms for recommender systems and graph neural networks}
We do not include general exact unlearning algorithms like RecEraser~\cite{chen2022recommendation} or UltraRE~\cite{li2023ultrare}, because their running time depends on the training set size, which is by design not efficient enough to handle a lot of queries with small forget sets. We focus on approximate unlearning algorithms with a time complexity linear in the forget set size. We include the following algorithms: 

\begin{itemize}[nosep,leftmargin=*]
    \item \texttt{SCIF}~\cite{li2023selectivecollaborativeinfluencefunction} directly removes the influence of data in the forget set using influence functions. The influence of the data in the forget set is calculated similarly to a Newton step, which additionally includes retain data to prevent catastrophic forgetting via
$
        \tilde{\theta}_{z \to \bar{z}} = \tilde{\theta} - H_{\tilde{\theta}}^{-1} \nabla_{\tilde{\theta}} (\frac{1}{2 + bs}(-\ell(z, \theta) + \ell(\bar{z}, \theta) + \sum_{i = 1}^{bs} \ell(z_i, \theta)))$
    where $z_i$ are retain samples, $z$ is an unlearning sample, $\bar{z}$ is the modified unlearning sample, $H$ is the Hessian, $bs$ is the number of retain samples used for the update, and $\tilde{\theta}$ is a subset of the parameters $\theta$ of the model.
    \item \texttt{GIF}~\cite{Wu_2023} works similarly to \texttt{SCIF}, but the parameters considered for the parameter update are chosen differently: The embedding of node $u$ in the graph is contained in the parameter update if there exists a node $v$ such that $v$ is in the forget set and the distance from $u$ to $v$ is smaller than a hyperparameter $d$.
    \item \texttt{CEU}~\cite{kun_wu_2023} unlearns edges from a GNN. To get theoretical guarantees for linear models, \texttt{CEU} first adds a linear noise term $b^T \theta$ with $b \sim \mathcal{N}(0, \sigma^2 I)$ to the training loss and fine-tunes aiming to hide the real gradient residual. After that, \texttt{CEU} unlearns using an influence function.
    \item \texttt{IDEA}~\cite{dong_2024} analyses the original loss over the whole training set and the pruned loss (which excludes nodes and edges we want to unlearn). It then conducts a second-order update step based on the difference between the original and pruned objective loss.
\end{itemize}

\headerl{General unlearning algorithms for gradient-based machine learning} We additionally include the following unlearning algorithms from the NeurIPS'23 unlearning challenge~\cite{triantafillou2024makingprogressunlearningfindings}, which have not specifically been designed for recommendation models. We select these particular approaches, as they scored the highest in the final ranking of the competition:

\begin{itemize}[nosep,leftmargin=*]
    \item \texttt{Fanchuan} first minimizes the KL-diverg\-ence between the output of the model on the unlearning samples and a uniform pseudo-label.
    Next, the method varies between two modes: in the first mode, a contrastive loss maximizes the distance between representations of samples in the forget set and samples in the retain set. In the second mode, it fine-tunes the model on a subset of the retain set.
    \item \texttt{Kookmin} resets a subset of the model parameters to their states before training. This is followed by a finetuning stage on a subset of the retain set, where non-reset parameters have a smaller learning rate. The subset of parameters to reset is chosen as the $p \, |\theta|$ parameters having the most similar gradients on the forget set and a subset of the retain set where $p \in (0, 1)$ is a hyperparameter.
    \item \texttt{Seif} adds Gaussian noise to the model parameters, followed by fine-tuning on a subset of the retain set.
\end{itemize}

\header{Evaluation metrics} We discuss the dimensions and metrics to evaluate the performance of unlearning algorithms.

\headerl{Recommendation utility} We report common ranking metrics such as Recall, Personalized Hit Ratio, and Normalized Discounted Cumulative Gain.


\begin{table*}[t!]
\centering
\caption{Unlearning effectiveness for the best model per scenario-task-dataset combination. Cells with ``n/a'' indicate that an unlearning algorithm is not applicable to a particular model while ``div.'' indicates that the unlearning diverged. The effectiveness varies substantially, but for most combinations at least one algorithm matches the retrained model.}
\setlength\tabcolsep{4pt}
\setlength{\aboverulesep}{1.15pt}
\setlength{\belowrulesep}{0.85pt}
\begin{tabular}{l l cccccc  cccc }
\toprule
&& \multicolumn{6}{c}{\emph{Unlearning Interactions with Sensitive Items} (\textbf{$\relitems{}@$}$\mathbf{20\uparrow})$} & \multicolumn{4}{c}{\emph{Unlearning Spam Interactions} (\textbf{$\releff{}@$}$\mathbf{20\uparrow})$}\\[-0.5mm]
\cmidrule(r){3-8}
\cmidrule(r){9-12}
&& \multicolumn{2}{c}{\textbf{NBR}} & \multicolumn{2}{c}{\textbf{SBR}} & \multicolumn{2}{c}{\textbf{CF}} & \multicolumn{2}{c}{\textbf{NBR}} & \multicolumn{2}{c}{\textbf{SBR}} \\[-0.5mm]
\cmidrule(r){3-4}
\cmidrule(r){5-6}
\cmidrule(r){7-8}
\cmidrule(r){9-10}
\cmidrule(r){11-12}
& & \multicolumn{1}{c}{Dunnhumby} & \multicolumn{1}{c}{Instacart} & \multicolumn{1}{c}{30music} & \multicolumn{1}{c}{Books} & \multicolumn{1}{c}{MovieLens} & \multicolumn{1}{c}{Food} & \multicolumn{1}{c}{Dunnhumby} & \multicolumn{1}{c}{TaFeng} & \multicolumn{1}{c}{RSC15} & \multicolumn{1}{c}{NowP} \\[-0.5mm]
\textbf{Algorithm} & \textbf{Type} & \multicolumn{1}{c}{\texttt{Sets2Sets}} & \multicolumn{1}{c}{\texttt{DNNTSP}} & \multicolumn{1}{c}{\texttt{SR-GNN}} & \multicolumn{1}{c}{\texttt{SR-GNN}} & \multicolumn{1}{c}{\texttt{BPR}} & \multicolumn{1}{c}{\texttt{SGL}} & \multicolumn{1}{c}{\texttt{Sets2Sets}} & \multicolumn{1}{c}{\texttt{DNNTSP}} & \multicolumn{1}{c}{\texttt{SR-GNN}} & \multicolumn{1}{c}{\texttt{NARM}} \\[-0.5mm]
\midrule
\texttt{SCIF}     & RecSys  & \addcell{+0.00}{+0.00}   & \addcell{-45.01}{-45.01} & \addcell{-9.05}{-9.05}   & \addcell{-5.19}{-5.19}   & \addcell{-0.92}{-0.92}   & \addcell{+4.29}{+4.29}   & \relcell{+5.14}{+5.14}   & \relcell{-1.16}{-1.16}   & \relcell{+0.02}{+0.02}   & \relcell{-4.63}{-4.63}   \\[-0.5mm]
\texttt{GIF}      & GNN     & \na{}                     & \diverged{}               & \diverged{}               & \diverged{}               & \na{}                     & \addcell{+18.33}{+18.33} & \na{}                     & \relcell{-3.59}{-3.59}   & \diverged{}               & \na{}                     \\[-0.5mm]
\texttt{CEU}      & GNN     & \na{}                     & \addcell{-38.86}{-38.86} & \addcell{-2.26}{-2.26}   & \addcell{-21.84}{-21.84} & \na{}                     & \addcell{+3.70}{+3.70}   & \na{}                     & \relcell{-0.66}{-0.66}   & \relcell{-50.78}{-50.78} & \na{}                     \\[-0.5mm]
\texttt{IDEA}     & GNN     & \na{}                     & \addcell{-52.10}{-52.10} & \diverged{}               & \addcell{+14.17}{+14.17} & \na{}                     & \diverged{}               & \na{}                     & \relcell{-0.69}{-0.69}   & \diverged{}               & \na{}                     \\[-0.5mm]
\texttt{Fanchuan} & General & \addcell{-0.25}{-0.25}   & \addcell{-37.02}{-37.02} & \addcell{-3.54}{-3.54}   & \addcell{-19.30}{-19.30} & \addcell{-1.43}{-1.43}   & \addcell{+0.32}{+0.32}   & \relcell{+9.85}{+9.85}   & \relcell{-7.07}{-7.07}   & \relcell{-22.82}{-22.82} & \relcell{-16.99}{-16.99} \\[-0.5mm]
\texttt{Kookmin}  & General & \addcell{-0.18}{-0.18}   & \addcell{-37.40}{-37.40} & \addcell{-13.42}{-13.42} & \addcell{+5.17}{+5.17}   & \addcell{+0.06}{+0.06}   & \addcell{+2.06}{+2.06}   & \relcell{+4.90}{+4.90}   & \relcell{-2.06}{-2.06}   & \relcell{-20.54}{-20.54} & \relcell{-34.72}{-34.72} \\[-0.5mm]
\texttt{Seif}     & General & \addcell{-6.32}{-6.32}   & \addcell{-22.71}{-22.71} & \addcell{-9.56}{-9.56}   & \addcell{-2.27}{-2.27}   & \addcell{+0.07}{+0.07}   & \addcell{-2.64}{-2.64}   & \relcell{+2.74}{+2.74}   & \relcell{-11.61}{-11.61} & \relcell{-99.91}{-99.91} & \relcell{-23.34}{-23.34} \\[-0.5mm]
\bottomrule
\end{tabular}
\label{tab:effectiveness}
\end{table*}

\headerl{Unlearning effectiveness} In addition to post-unlearning utility, it is important to assess to which degree a model actually forgets the forget set in approximate unlearning. This is challenging because most approximate unlearning algorithms lack theoretical guarantees (\Cref{sec:background}). 
Common evaluation approaches compare unlearned and retrained models in weight space or output distribution. While a small distance suggests successful unlearning, a large distance is inconclusive, as model multiplicity and training randomness can yield alternative retrained models that are equally valid yet far apart by these metrics. Consequently, such comparisons cannot reliably determine unlearning effectiveness.
Another common evaluation approach uses membership inference attacks to assess unlearning effectiveness. While successful attacks indicate failed unlearning, the absence of successful attacks is inconclusive, as results depend heavily on attacker strength and cannot account for all possible adversaries.
%
%
Instead of using these evaluation methods, we empirically evaluate unlearning effectiveness with scenario-specific metrics that are easier to interpret and more meaningful for real-world usage. We include the following two scenarios: 
\begin{itemize}[nosep,leftmargin=*]
    \item \textit{Sensitive item unlearning}~\cite{schelter2023forget}: In this scenario, items from sensitive categories are removed for specific users, e.g., meat for users adopting a vegetarian diet or alcohol for users with addiction issues. Effectiveness is measured by counting how many users still receive sensitive recommendations after the unlearning request, compared to a retrained model. To evaluate recommendation utility, sensitive items are removed from the affected users’ test data.
    Let $U_s$ be the set of users that submitted a forget request for sensitive items, $R_u^k$ be the top $k$ recommended items for user $u$, and $I_s$ be the set of sensitive items. We define
    \begin{equation*}
        \text{Sensitive}@k := \frac{1}{|U_s|} \sum_{u \in U_s} [R_u^k \cap I_s \neq \emptyset] \in [0, 1]
    \end{equation*}
    as the percentage of users who submitted forget requests but still have sensitive items in their top $k$ predictions. To compare a retrained model $\theta_r$ and an unlearned model $\theta_u$ with respect to this metric, we define
    \begin{equation*}
        \relitems{}@k := \text{Sensitive}_r@k - \text{Sensitive}_u@k \in [-1, 1]
    \end{equation*}
    as the delta between the sensitive item percentage of the retrained model ($\text{Sensitive}_r@k$) and the unlearned model (Sen\-si\-tive$_u@k$). $\relitems{}@k = p \geq 0$ implies that the unlearned model successfully removes sensitive items from the top $k$ recommendations for $p \, |U_s|$ more users compared to the retrained model. Analogously, $\relitems{}@k < 0$ indicates that the unlearned model performs worse than the retrained model.
    \item \textit{Removal of poisonous data}: This scenario evolves around maliciously injected interactions in session-based or next-basket recommendation. An actor injects clicks on random items and items they want to promote, biasing the model toward these items. Successful unlearning restores model performance by removing the influence of the fake interactions, measured using standard performance metrics.
    To compare the performance of a retrained model $\theta_r$ and an unlearned model $\theta_u$, we define
    \begin{equation*}
        \releff{}@k := \frac{\text{nDCG}_u@k}{\text{nDCG}_r@k} - 1 \in [-1, \infty).
    \end{equation*}
    $\releff{}@k = p\%$ implies that the unlearned model's $\text{nDCG}@k$ is $p\%$ better than the retrained models $\text{nDCG}@k$.
\end{itemize}

\headerl{Unlearning efficiency} We measure unlearning latency, i.e., the time it takes to execute unlearning requests, both in absolute terms and in comparison to full retraining.

\header{Experimentation protocol for \bench{}} A single experiment using \bench{} combines a recommendation task, a dataset $D$, a training algorithm, an unlearning algorithm $U$, and a scenario (e.g., removal of poisonous data or unlearning sensitive item interactions). In each experiment, we sequentially unlearn a sequence of batches $D^{(0)}_f, \dots, D^{(n - 1)}_f$. Let $\theta^{(0)}$ denote the model trained on the full data $D$, then we sequentially compute $\smash{\theta^{(i+1)} \gets U(\theta^{(i)}, D^{(i)}_r, D^{(i)}_f)}$ where $\smash{D^{(i)}_r = D \setminus \bigcup_{j=0}^{i} D^{(j)}_f}$. The number of batches $n$ and the data contained in each batch depend on the scenario. For unlearning interactions with sensitive items, we sequentially unlearn by user, e.g., each batch consists of a single user's sensitive interactions to forget. Such small batches are natural, as there is an urgency to carry out the unlearning requests fast in this scenario due to ethical concerns. We choose the number of batches to account for $0.01\%$ of the interactions in total. For the removal of poisonous data it is natural to unlearn larger batches and an overall larger fraction of the dataset, as spammers typically use various accounts to pollute the data~\cite{poisoning_attack_used_various_users}. Spammers typically must inject a large volume of interactions to meaningfully influence a recommender system, and there are no privacy constraints that require companies to perform unlearning under strict latency or bounded removal-time guarantees. Thus, we add spam interactions with a size of $1\%$ of the training set for poisoning, and make each batch to unlearn contain the data of 256 spam users. Hyperparameters were tuned on RSC15 and Food or adopted from best-performing configurations reported in prior repositories. A complete overview of the hyperparameters is provided in Appendix~\ref{app:supplementary-material}.

\header{\bench{} implementation and resources}
We implement \bench{} on top of the RecBole library~\cite{zhao2021recboleunifiedcomprehensiveefficient} and release it under an open license at \url{https://github.com/deem-data/erase-bench/}. Following the protocol described above, \bench{} supports follow-up research with flexible experimentation across its models, datasets, and unlearning settings.
Beyond the benchmark code, we release the full results of running \bench{} at scale. We execute training, retraining, and unlearning runs over five random seeds, consuming more than 13,000 GPU hours and producing over 600 GB of reusable artifacts, including 1,069 model checkpoints. By providing pretrained and retrained checkpoints, we offload the most computationally expensive steps and enable the community to evaluate new unlearning algorithms using only additional unlearning runs. Specifically, we release (i)~training logs, (ii)~150 trained model checkpoints, (iii)~150 retrained model checkpoints, (iv)~769 unlearned model checkpoints, and (v)~corresponding efficiency, effectiveness, and utility metrics. We provide pointers to our artifacts (\url{https://github.com/deem-data/erase-bench/blob/main/artifacts.md}) with additional experimental details and results in Appendix~\ref{app:supplementary-material}.
\section{Example Use of \bench{}}
\label{sec:results}

We showcase how \bench{} can be used to gain insights and directions for the current progress in MU for recommender systems

\header{A snapshot of the current progress in MU for recommender systems} We analyze the experimental logs of \bench{} and reflect on our guiding questions from \Cref{sec:design}. We refer to Appendix~\ref{app:supplementary-material} for an extended discussion of the analysis results.

\begin{figure*}[t!]
\centering
\includegraphics[width=\textwidth]{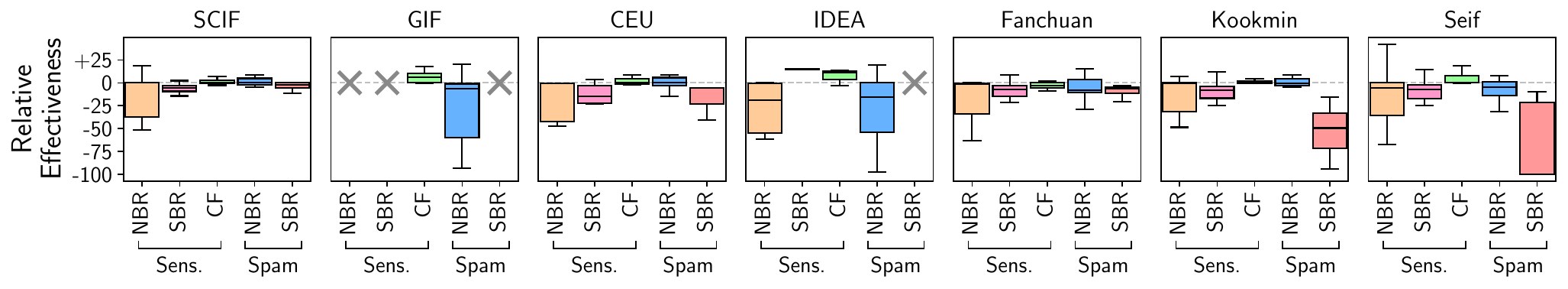}
\caption{Robustness of unlearning algorithms. We plot the distribution of relative effectiveness aggregated over tasks, datasets, models, and runs. The recommender-specific algorithm \texttt{SCIF} is the most consistent and robust approach across tasks. Algorithms with no data for a scenario-task combination are either not applicable to the respective models or diverge while unlearning.}
\label{fig:robustness}
\end{figure*}


\headerl{Unlearning effectiveness} To assess the unlearning effectiveness of the algorithms in \bench{}, we aggregate the experimental logs as follows. For each scenario-task-dataset combination, we identify the best model based on nDGC@20, and compute its mean $\releff{}@20$ in case of spam unlearning or mean $\relitems{}@20$ in case of sensitive interaction unlearning. The resulting scores in \Cref{tab:effectiveness} indicate that the effectiveness of approximate unlearning algorithms varies substantially across datasets. In most experiments, at least one approximate method matches the retrained model in both unlearning effectiveness and recommendation utility.


\headerl{Robustness} To quantify the robustness of unlearning algorithms, we collect the relative effectiveness scores ($\releff{}@20$ or $\relitems{}@20$ depending on the scenario) per algorithm for all datasets and models of a given scenario-task combination from the experimental logs of \bench{}, and plot their distributions in \Cref{fig:robustness}. Overall, no unlearning algorithm is uniformly robust across all CF, SBR, and NBR tasks; many methods show instability or performance degradation under repeated unlearning. The recommender-specific method \texttt{SCIF} stands out as the most consistent and robust approach across tasks, exhibiting low variance in unlearning effectiveness, with remaining challenges primarily in NBR sensitive item unlearning.  


\headerl{Influence of model architecture} 
Model architecture has a strong impact: the effectiveness and utility of general unlearning algorithms often degrade under repeated unlearning requests, particularly for attention-based and recurrent models. Architecture- and recommender-specific methods—most notably \texttt{SCIF}, and in some cases GNN-specific algorithms—consistently achieve higher effectiveness and utility.

\begin{figure}[h!]
\centering
\includegraphics[width=\columnwidth]{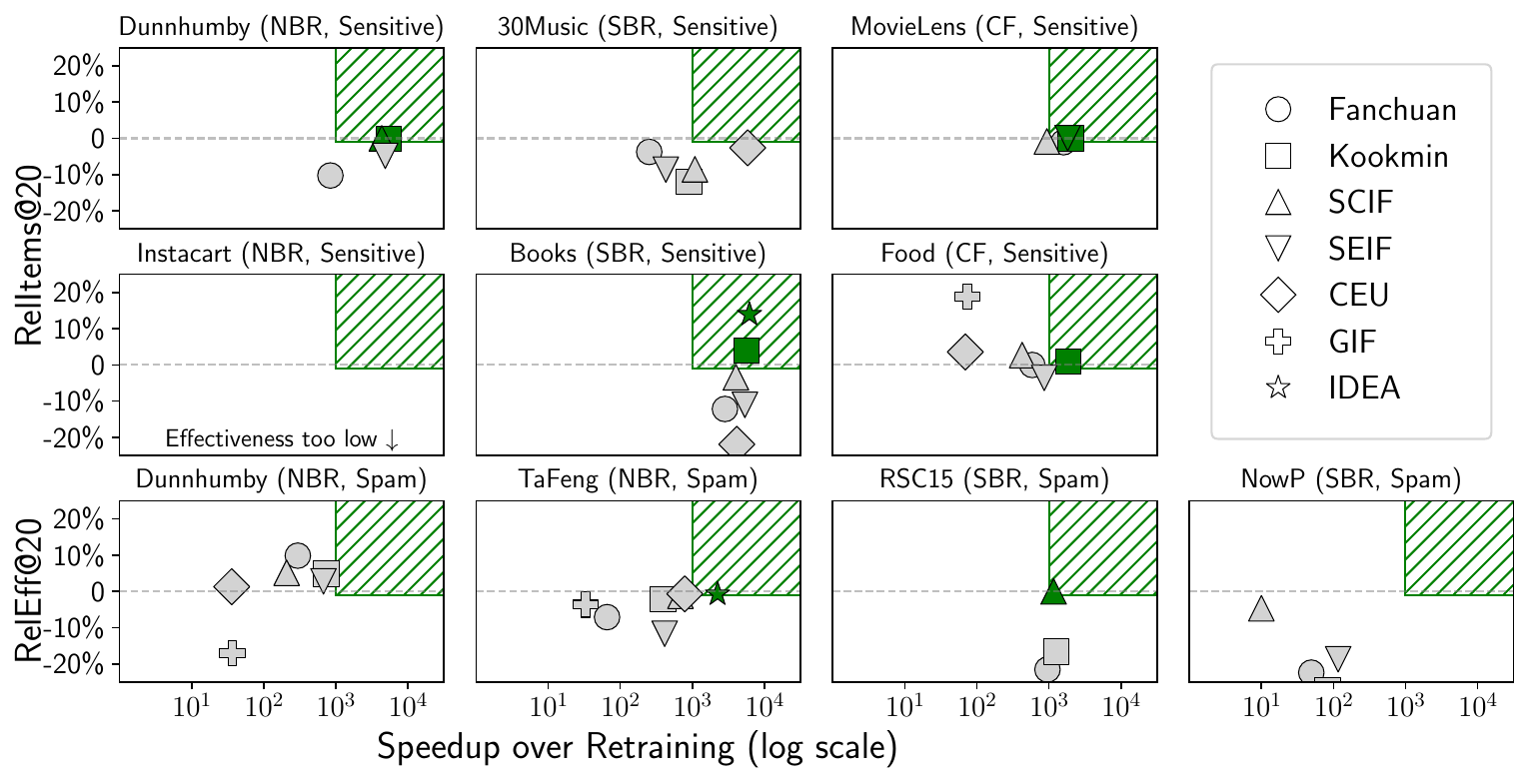}
\vspace*{-5mm}
\caption{Speedup over retraining against relative effectiveness for unlearning algorithms in \bench{}. The green area marks algorithms attractive for deployment.}
\label{fig:efficiency-first}
\vspace*{-1.5mm}
\end{figure}


\headerl{Efficiency and deployability}
For practical deployment, an unlearning algorithm must update a model substantially faster than full retraining while maintaining comparable effectiveness. To assess deployability, we extract from the \bench{} logs the mean unlearning runtime and relative effectiveness of each algorithm for the best model in each scenario–task–dataset combination. Next, we plot the resulting speedup over retraining (retraining time divided by unlearning time) against relative effectiveness in \Cref{fig:efficiency-first}.
Based on our interviews with practitioners from industry, where retraining can take up to a day, we consider a three-orders-of-magnitude runtime reduction—bringing unlearning down to minutes—sufficient for deployment, provided effectiveness remains comparable (e.g., within a 1\% margin). We indicate this target region in green as a “deployability area” in the figure and highlight methods that meet this criterion. Overall, the results show that most current unlearning algorithms fall short of real-world deployment requirements due to either excessive latency or insufficient effectiveness.


\header{Directions for follow-up research building on \bench{}} Our benchmark uncovers new research directions for MU in recommender systems. 
Since efficient, effective, and high-utility unlearning is possible in most scenarios, unlearning may be viewed as an algorithm selection problem.
This selection should account for algorithms that degrade sharply after a small number of requests. 
Stabilizing unlearning performance is another key challenge. Since unlearning is computationally cheap, ensembling differently seeded instances or various algorithms in parallel could effectively reduce variance.
Finally, as illustrated in \Cref{fig:efficiency-first}, unlearning latency must be further reduced for practical deployment, due to the lack of deployable methods for several scenarios. There has been progress on low-latency MU for neighborhood-based models~\cite{schelter2023forget,schelter2024snapcase}, but it is an open question how to accelerate approximate MU for neural network-based recommender systems. \bench{} constitutes an experimental testbed for all these research directions: newly proposed methods can directly be evaluated on our datasets, scenarios, and model checkpoints.

\section{Resource Reproducibility \& Extensibility}
\label{sec:resource}

Our benchmark can be re-run via two dedicated scripts to execute and evaluate (re)training and unlearning, see \href{https://github.com/deem-data/erase-bench/blob/main/README.md#re-running-the-benchmark}{README.md} for instructions. Both scripts offer a wide range of configuration options and automatically save log files and model checkpoints. The benchmark can also use private data; no external APIs are required. 
In the future, we will maintain \bench{} and welcome contributions. Our codebase has predefined extension points for new datasets, models, unlearning algorithms, and scenarios, as documented in \href{https://github.com/deem-data/erase-bench/blob/main/README.md#addingextending-the-benchmark}{README.md}. New models have to extend the standard RecBole model class; new unlearning methods have to extend our unlearning script and a corresponding pipeline. A new scenario requires a generator to produce forget-set artifacts for a given dataset.



\section{Conclusion}


We have introduced \bench{}, a real-world aligned benchmark for machine unlearning in recommender systems that addresses key gaps in prior benchmarks and is easy to extend. It provides a large collection of reusable artifacts, including pretrained and retrained model checkpoints, allowing researchers to avoid costly retraining and to evaluate new unlearning algorithms with only additional unlearning runs. Together, these features position \bench{} as a common testbed to systematically assess, advance, and track practical machine unlearning in recommender systems.

\begin{acks}
    This research was (partially) supported by the Dutch Research Council (NWO), under project numbers 024.004.022, NWA.1389.20.\-183, and KICH3.LTP.20.006, and the European Union under grant agreements No.\ 101070212 (FINDHR) and No.\ 101201510 (UNITE).
    
    Views and opinions expressed are those of the author(s) only and do not necessarily reflect those of their respective employers, funders and/or granting authorities.
\end{acks}

\clearpage
\bibliographystyle{ACM-Reference-Format}
\balance
\bibliography{references}

@misc{zhao2021recboleunifiedcomprehensiveefficient,
      title={RecBole: Towards a Unified, Comprehensive and Efficient Framework for Recommendation Algorithms}, 
      author={Wayne Xin Zhao and Shanlei Mu and Yupeng Hou and Zihan Lin and Yushuo Chen and Xingyu Pan and Kaiyuan Li and Yujie Lu and Hui Wang and Changxin Tian and Yingqian Min and Zhichao Feng and Xinyan Fan and Xu Chen and Pengfei Wang and Wendi Ji and Yaliang Li and Xiaoling Wang and Ji-Rong Wen},
      year={2021},
      eprint={2011.01731},
      archivePrefix={arXiv},
      primaryClass={cs.IR},
      url={https://arxiv.org/abs/2011.01731}, 
}

@article{lubitzsch2025towards,
  title={Towards a Real-World Aligned Benchmark for Unlearning in Recommender Systems},
  author={Lubitzsch, Pierre and Ovcharenko, Olga and Chen, Hao and de Rijke, Maarten and Schelter, Sebastian},
  journal={FAccTRec Workshop: Responsible Recommendation at the ACM Conference on Recommender Systems},
  year={2025}
}

@article{li2023ultrare,
  title={UltraRE: Enhancing RecEraser for Recommendation Unlearning via Error Decomposition},
  author={Li, Yuyuan and Chen, Chaochao and Zhang, Yizhao and Liu, Weiming and Lyu, Lingjuan and Zheng, Xiaolin and Meng, Dan and Wang, Jun},
  journal={Advances in Neural Information Processing Systems},
  volume={36},
  pages={12611--12625},
  year={2023}
}

@inproceedings{chen2022recommendation,
  title={Recommendation Unlearning},
  author={Chen, Chong and Sun, Fei and Zhang, Min and Ding, Bolin},
  booktitle={Proceedings of the ACM Web Conference 2022},
  pages={2768--2777},
  year={2022}
}

@article{schelter2024snapcase,
  title={Snapcase-Regain Control over Your Predictions with Low-Latency Machine Unlearning},
  author={Schelter, Sebastian and Grafberger, Stefan and de Rijke, Maarten},
  journal={Proceedings of the VLDB Endowment},
  volume={17},
  number={12},
  pages={4273--4276},
  year={2024},
  publisher={VLDB Endowment}
}

@misc{amazonschoice,
  author={Buzzfeed},
  title={“Amazon’s Choice” Does Not Necessarily Mean A Product Is Good},
  year={2019},
  url={https://www.buzzfeednews.com/article/nicolenguyen/amazons-choice-bad-products}
}

@inproceedings{carlini2019secret,
  title={The Secret Sharer: Evaluating and Testing Unintended Memorization in Neural Networks},
  author={Carlini, Nicholas and Liu, Chang and Erlingsson, {\'U}lfar and Kos, Jernej and Song, Dawn},
  booktitle={28th USENIX security symposium (USENIX security 19)},
  pages={267--284},
  year={2019}
}

@inproceedings{schelter2021hedgecut,
author = {Schelter, Sebastian and Grafberger, Stefan and Dunning, Ted},
title = {HedgeCut: Maintaining Randomised Trees for Low-Latency Machine Unlearning},
year = {2021},
isbn = {9781450383431},
publisher = {Association for Computing Machinery},
address = {New York, NY, USA},
url = {https://doi.org/10.1145/3448016.3457239},
doi = {10.1145/3448016.3457239},
booktitle = {Proceedings of the 2021 International Conference on Management of Data},
pages = {1545–1557},
numpages = {13},
location = {Virtual Event, China},
series = {SIGMOD '21}
}

@article{stoyanovich2022responsible,
  title={Responsible Data Management},
  author={Stoyanovich, Julia and Abiteboul, Serge and Howe, Bill and Jagadish, HV and Schelter, Sebastian},
  journal={Communications of the ACM},
  volume={65},
  number={6},
  pages={64--74},
  year={2022},
  publisher={ACM New York, NY, USA}
}

@inproceedings{schelter2023forget,
  title={Forget Me Now: Fast and Exact Unlearning in Neighborhood-based Recommendation},
  author={Schelter, Sebastian and Ariannezhad, Mozhdeh and de Rijke, Maarten},
  booktitle={Proceedings of the 46th International ACM SIGIR Conference on Research and Development in Information Retrieval},
  pages={2011--2015},
  year={2023}
}

@inproceedings{chen2024cure4rec,
  title={CURE4Rec: A Benchmark for Recommendation Unlearning with Deeper influence},
  author={Chen, Chaochao and Zhang, Jiaming and Zhang, Yizhao and Zhang, Li and Lyu, Lingjuan and Li, Yuyuan and Gong, Biao and Yan, Chenggang},
  booktitle={NeurIPS},
  year={2024}
}

@misc{gdpr_article17,
  author       = {{GDPR.eu}},
  title        = {Article 17: Right to be forgotten},
  howpublished = {\url{https://gdpr.eu/article-17-right-to-be-forgotten}},
  note         = {Accessed: 2026-02-12},
  year         = {2016}
}

@misc{ccpa_faq,
  author       = {{California Privacy Protection Agency}},
  title        = {California Consumer Privacy Act -- Frequently Asked Questions},
  howpublished = {\url{https://cppa.ca.gov/faq.html}},
  note         = {Accessed: 2026-02-12},
  year         = {2025}
}

@misc{DigiChina_AI_Rec,
  author = {Toner, Helen and Creemers, Rogier and Webster, Graham},
  title = {Internet Information Service Algorithmic Recommendation Management Provisions},
  howpublished = {\url{https://digichina.stanford.edu/work/translation-internet-information-servicealgorithmic-recommendation-management-provisions-opinon-seeking-draft/}},
  organization = {DigiChina, Stanford University},
  year = {2022},
  note = {Accessed: 2026-02-12}
}

@INPROCEEDINGS{cao_2022,
  author={Cao, Yinzhi and Yang, Junfeng},
  booktitle={2015 IEEE Symposium on Security and Privacy}, 
  title={Towards Making Systems Forget with Machine Unlearning}, 
  year={2015},
  volume={},
  number={},
  pages={463-480},
  doi={10.1109/SP.2015.35}
}

@misc{ginart2019makingaiforgetyou,
      title={Making AI Forget You: Data Deletion in Machine Learning}, 
      author={Antonio Ginart and Melody Y. Guan and Gregory Valiant and James Zou},
      year={2019},
      eprint={1907.05012},
      archivePrefix={arXiv},
      primaryClass={cs.LG},
      url={https://arxiv.org/abs/1907.05012}, 
}

@misc{izzo2021approximatedatadeletionmachine,
      title={Approximate Data Deletion from Machine Learning Models}, 
      author={Zachary Izzo and Mary Anne Smart and Kamalika Chaudhuri and James Zou},
      year={2021},
      eprint={2002.10077},
      archivePrefix={arXiv},
      primaryClass={cs.LG},
      url={https://arxiv.org/abs/2002.10077}, 
}

@misc{wang2022efficientlymaintainingbasketrecommendations,
      title={Efficiently Maintaining Next Basket Recommendations under Additions and Deletions of Baskets and Items}, 
      author={Benjamin Longxiang Wang and Sebastian Schelter},
      year={2022},
      eprint={2201.13313},
      archivePrefix={arXiv},
      primaryClass={cs.IR},
      url={https://arxiv.org/abs/2201.13313}, 
}

@misc{wu2020deltagradrapidretrainingmachine,
      title={DeltaGrad: Rapid retraining of machine learning models}, 
      author={Yinjun Wu and Edgar Dobriban and Susan B. Davidson},
      year={2020},
      eprint={2006.14755},
      archivePrefix={arXiv},
      primaryClass={cs.LG},
      url={https://arxiv.org/abs/2006.14755}, 
}

@inproceedings{Wu_2020, series={SIGMOD/PODS ’20},
   title={PrIU: A Provenance-Based Approach for Incrementally Updating Regression Models},
   url={http://dx.doi.org/10.1145/3318464.3380571},
   DOI={10.1145/3318464.3380571},
   booktitle={Proceedings of the 2020 ACM SIGMOD International Conference on Management of Data},
   publisher={ACM},
   author={Wu, Yinjun and Tannen, Val and Davidson, Susan B.},
   year={2020},
   month=may, pages={447–462},
   collection={SIGMOD/PODS ’20}
}

@misc{li2023selectivecollaborativeinfluencefunction,
      title={Selective and Collaborative Influence Function for Efficient Recommendation Unlearning}, 
      author={Yuyuan Li and Chaochao Chen and Xiaolin Zheng and Yizhao Zhang and Biao Gong and Jun Wang},
      year={2023},
      eprint={2304.10199},
      archivePrefix={arXiv},
      primaryClass={cs.IR},
      url={https://arxiv.org/abs/2304.10199}, 
}

@misc{triantafillou2024makingprogressunlearningfindings,
      title={Are We Making Progress in Unlearning? Findings from the First NeurIPS Unlearning Competition}, 
      author={Eleni Triantafillou and Peter Kairouz and Fabian Pedregosa and Jamie Hayes and Meghdad Kurmanji and Kairan Zhao and Vincent Dumoulin and Julio Jacques Junior and Ioannis Mitliagkas and Jun Wan and Lisheng Sun Hosoya and Sergio Escalera and Gintare Karolina Dziugaite and Peter Triantafillou and Isabelle Guyon},
      year={2024},
      eprint={2406.09073},
      archivePrefix={arXiv},
      primaryClass={cs.LG},
      url={https://arxiv.org/abs/2406.09073}, 
}

@inproceedings{dnntsp, series={KDD ’20},
   title={Predicting Temporal Sets with Deep Neural Networks},
   url={http://dx.doi.org/10.1145/3394486.3403152},
   DOI={10.1145/3394486.3403152},
   booktitle={Proceedings of the 26th ACM SIGKDD International Conference on Knowledge Discovery \& Data Mining},
   publisher={ACM},
   author={Yu, Le and Sun, Leilei and Du, Bowen and Liu, Chuanren and Xiong, Hui and Lv, Weifeng},
   year={2020},
   month=aug, pages={1083–1091},
   collection={KDD ’20} }

@inproceedings{clea,
author = {Qin, Yuqi and Wang, Pengfei and Li, Chenliang},
title = {The World is Binary: Contrastive Learning for Denoising Next Basket Recommendation},
year = {2021},
isbn = {9781450380379},
publisher = {Association for Computing Machinery},
address = {New York, NY, USA},
url = {https://doi.org/10.1145/3404835.3462836},
doi = {10.1145/3404835.3462836},
booktitle = {Proceedings of the 44th International ACM SIGIR Conference on Research and Development in Information Retrieval},
pages = {859–868},
numpages = {10},
location = {Virtual Event, Canada},
series = {SIGIR '21}
}

@inproceedings{sets2sets,
author = {Hu, Haoji and He, Xiangnan},
title = {Sets2Sets: Learning from Sequential Sets with Neural Networks},
year = {2019},
isbn = {9781450362016},
publisher = {Association for Computing Machinery},
address = {New York, NY, USA},
url = {https://doi.org/10.1145/3292500.3330979},
doi = {10.1145/3292500.3330979},
booktitle = {Proceedings of the 25th ACM SIGKDD International Conference on Knowledge Discovery \& Data Mining},
pages = {1491–1499},
numpages = {9},
location = {Anchorage, AK, USA},
series = {KDD '19}
}

@inproceedings{upcf,
author = {Faggioli, Guglielmo and Polato, Mirko and Aiolli, Fabio},
title = {Recency Aware Collaborative Filtering for Next Basket Recommendation},
year = {2020},
isbn = {9781450368612},
publisher = {Association for Computing Machinery},
address = {New York, NY, USA},
url = {https://doi.org/10.1145/3340631.3394850},
doi = {10.1145/3340631.3394850},
booktitle = {Proceedings of the 28th ACM Conference on User Modeling, Adaptation and Personalization},
pages = {80–87},
numpages = {8},
location = {Genoa, Italy},
series = {UMAP '20}
}

@inproceedings{tifuknn,
author = {Hu, Haoji and He, Xiangnan and Gao, Jinyang and Zhang, Zhi-Li},
title = {Modeling Personalized Item Frequency Information for Next-basket Recommendation},
year = {2020},
isbn = {9781450380164},
publisher = {Association for Computing Machinery},
address = {New York, NY, USA},
url = {https://doi.org/10.1145/3397271.3401066},
doi = {10.1145/3397271.3401066},
booktitle = {Proceedings of the 43rd International ACM SIGIR Conference on Research and Development in Information Retrieval},
pages = {1071–1080},
numpages = {10},
location = {Virtual Event, China},
series = {SIGIR '20}
}

@inproceedings{sknn,
author = {Jannach, Dietmar and Ludewig, Malte},
title = {When Recurrent Neural Networks meet the Neighborhood for Session-Based Recommendation},
year = {2017},
isbn = {9781450346528},
publisher = {Association for Computing Machinery},
address = {New York, NY, USA},
url = {https://doi.org/10.1145/3109859.3109872},
doi = {10.1145/3109859.3109872},
booktitle = {Proceedings of the Eleventh ACM Conference on Recommender Systems},
pages = {306–310},
numpages = {5},
location = {Como, Italy},
series = {RecSys '17}
}

@inproceedings{gru4rec,
author = {Tan, Yong Kiam and Xu, Xinxing and Liu, Yong},
title = {Improved Recurrent Neural Networks for Session-based Recommendations},
year = {2016},
isbn = {9781450347952},
publisher = {Association for Computing Machinery},
address = {New York, NY, USA},
url = {https://doi.org/10.1145/2988450.2988452},
doi = {10.1145/2988450.2988452},
booktitle = {Proceedings of the 1st Workshop on Deep Learning for Recommender Systems},
pages = {17–22},
numpages = {6},
location = {Boston, MA, USA},
series = {DLRS 2016}
}

@article{srgnn, title={Session-Based Recommendation with Graph Neural Networks}, volume={33}, url={https://ojs.aaai.org/index.php/AAAI/article/view/3804}, DOI={10.1609/aaai.v33i01.3301346}, abstractNote={&lt;p&gt;The problem of session-based recommendation aims to predict user actions based on anonymous sessions. Previous methods model a session as a sequence and estimate user representations besides item representations to make recommendations. Though achieved promising results, they are insufficient to obtain accurate user vectors in sessions and neglect complex transitions of items. To obtain accurate item embedding and take complex transitions of items into account, we propose a novel method, i.e. &lt;em&gt;Session-based Recommendation with Graph Neural Networks&lt;/em&gt;, SR-GNN for brevity. In the proposed method, session sequences are modeled as graphstructured data. Based on the session graph, GNN can capture complex transitions of items, which are difficult to be revealed by previous conventional sequential methods. Each session is then represented as the composition of the global preference and the current interest of that session using an attention network. Extensive experiments conducted on two real datasets show that SR-GNN evidently outperforms the state-of-the-art session-based recommendation methods consistently.&lt;/p&gt;}, number={01}, journal={Proceedings of the AAAI Conference on Artificial Intelligence}, author={Wu, Shu and Tang, Yuyuan and Zhu, Yanqiao and Wang, Liang and Xie, Xing and Tan, Tieniu}, year={2019}, month={Jul.}, pages={346-353} }

@INPROCEEDINGS{sasrec,
  author={Kang, Wang-Cheng and McAuley, Julian},
  booktitle={2018 IEEE International Conference on Data Mining (ICDM)}, 
  title={Self-Attentive Sequential Recommendation}, 
  year={2018},
  volume={},
  number={},
  pages={197-206},
  doi={10.1109/ICDM.2018.00035}}

@inproceedings{narm,
author = {Li, Jing and Ren, Pengjie and Chen, Zhumin and Ren, Zhaochun and Lian, Tao and Ma, Jun},
title = {Neural Attentive Session-based Recommendation},
year = {2017},
isbn = {9781450349185},
publisher = {Association for Computing Machinery},
address = {New York, NY, USA},
url = {https://doi.org/10.1145/3132847.3132926},
doi = {10.1145/3132847.3132926},
booktitle = {Proceedings of the 2017 ACM on Conference on Information and Knowledge Management},
pages = {1419–1428},
numpages = {10},
location = {Singapore, Singapore},
series = {CIKM '17}
}

@inproceedings{bpr,
author = {Rendle, Steffen and Freudenthaler, Christoph and Gantner, Zeno and Schmidt-Thieme, Lars},
title = {BPR: Bayesian Personalized Ranking from Implicit Feedback},
year = {2009},
isbn = {9780974903958},
publisher = {AUAI Press},
address = {Arlington, Virginia, USA},
booktitle = {Proceedings of the Twenty-Fifth Conference on Uncertainty in Artificial Intelligence},
pages = {452–461},
numpages = {10},
location = {Montreal, Quebec, Canada},
series = {UAI '09}
}

@inproceedings{ibcf,
author = {Sarwar, Badrul and Karypis, George and Konstan, Joseph and Riedl, John},
title = {Item-based Collaborative Filtering Recommendation Algorithms},
year = {2001},
isbn = {1581133480},
publisher = {Association for Computing Machinery},
address = {New York, NY, USA},
url = {https://doi.org/10.1145/371920.372071},
doi = {10.1145/371920.372071},
booktitle = {Proceedings of the 10th International Conference on World Wide Web},
pages = {285–295},
numpages = {11},
location = {Hong Kong, Hong Kong},
series = {WWW '01}
}

@inproceedings{lightgcn,
author = {He, Xiangnan and Deng, Kuan and Wang, Xiang and Li, Yan and Zhang, YongDong and Wang, Meng},
title = {LightGCN: Simplifying and Powering Graph Convolution Network for Recommendation},
year = {2020},
isbn = {9781450380164},
publisher = {Association for Computing Machinery},
address = {New York, NY, USA},
url = {https://doi.org/10.1145/3397271.3401063},
doi = {10.1145/3397271.3401063},
booktitle = {Proceedings of the 43rd International ACM SIGIR Conference on Research and Development in Information Retrieval},
pages = {639–648},
numpages = {10},
location = {Virtual Event, China},
series = {SIGIR '20}
}

@inproceedings{simrec, series={WWW ’23},
   title={Graph-less Collaborative Filtering},
   url={http://dx.doi.org/10.1145/3543507.3583196},
   DOI={10.1145/3543507.3583196},
   booktitle={Proceedings of the ACM Web Conference 2023},
   publisher={ACM},
   author={Xia, Lianghao and Huang, Chao and Shi, Jiao and Xu, Yong},
   year={2023},
   month=apr, pages={17–27},
   collection={WWW ’23} }

@inproceedings{dccf,
   title={Disentangled Contrastive Collaborative Filtering},
   url={http://dx.doi.org/10.1145/3539618.3591665},
   DOI={10.1145/3539618.3591665},
   booktitle={Proceedings of the 46th International ACM SIGIR Conference on Research and Development in Information Retrieval},
   publisher={ACM},
   author={Ren, Xubin and Xia, Lianghao and Zhao, Jiashu and Yin, Dawei and Huang, Chao},
   year={2023},
   month=jul, pages={1137–1146},
   collection={SIGIR ’23} }

@inproceedings{tencentrecommendation2025,
author = {Chen, Zihao and Zhang, Chenyang and Xu, Chen and Zhang, Zhao and Wang, Jiaqiang and Qian, Weining and Zhou, Aoying},
title = {Scheduling Data Processing Pipelines for Incremental Training on MLP-based Recommendation Models},
year = {2025},
isbn = {9798400715648},
publisher = {Association for Computing Machinery},
address = {New York, NY, USA},
url = {https://doi.org/10.1145/3722212.3724454},
doi = {10.1145/3722212.3724454},
booktitle = {Companion of the 2025 International Conference on Management of Data},
pages = {350–363},
numpages = {14},
location = {Berlin, Germany},
series = {SIGMOD/PODS '25}
}

@manual{aws2025personalize,
  title        = {Maintaining domain recommenders},
  author       = {{Amazon Web Services}},
  year         = {2025},
  note         = {Accessed: 2026-02-12},
  url          = {https://docs.aws.amazon.com/personalize/latest/dg/maintaining-relevance.html#maintaining-domain-recommenders}
}

@inproceedings{rendle2010nbr,
author = {Rendle, Steffen and Freudenthaler, Christoph and Schmidt-Thieme, Lars},
title = {Factorizing Personalized Markov Chains for Next-basket Recommendation},
year = {2010},
isbn = {9781605587998},
publisher = {Association for Computing Machinery},
address = {New York, NY, USA},
url = {https://doi.org/10.1145/1772690.1772773},
doi = {10.1145/1772690.1772773},
booktitle = {Proceedings of the 19th International Conference on World Wide Web},
pages = {811–820},
numpages = {10},
location = {Raleigh, North Carolina, USA},
series = {WWW '10}
}

@misc{hidasi2016sessionbasedrecommendationsrecurrentneural,
      title={Session-based Recommendations with Recurrent Neural Networks}, 
      author={Balázs Hidasi and Alexandros Karatzoglou and Linas Baltrunas and Domonkos Tikk},
      year={2016},
      eprint={1511.06939},
      archivePrefix={arXiv},
      primaryClass={cs.LG},
      url={https://arxiv.org/abs/1511.06939}, 
}

@inproceedings{Wu_2023, series={WWW ’23},
   title={GIF: A General Graph Unlearning Strategy via Influence Function},
   url={http://dx.doi.org/10.1145/3543507.3583521},
   DOI={10.1145/3543507.3583521},
   booktitle={Proceedings of the ACM Web Conference 2023},
   publisher={ACM},
   author={Wu, Jiancan and Yang, Yi and Qian, Yuchun and Sui, Yongduo and Wang, Xiang and He, Xiangnan},
   year={2023},
   month=apr, pages={651–661},
   collection={WWW ’23} }

@inproceedings{kun_wu_2023,
author = {Wu, Kun and Shen, Jie and Ning, Yue and Wang, Ting and Wang, Wendy Hui},
title = {Certified Edge Unlearning for Graph Neural Networks},
year = {2023},
isbn = {9798400701030},
publisher = {Association for Computing Machinery},
address = {New York, NY, USA},
url = {https://doi.org/10.1145/3580305.3599271},
doi = {10.1145/3580305.3599271},
booktitle = {Proceedings of the 29th ACM SIGKDD Conference on Knowledge Discovery and Data Mining},
pages = {2606–2617},
numpages = {12},
location = {Long Beach, CA, USA},
series = {KDD '23}
}

@inproceedings{dong_2024,
author = {Dong, Yushun and Zhang, Binchi and Lei, Zhenyu and Zou, Na and Li, Jundong},
title = {IDEA: A Flexible Framework of Certified Unlearning for Graph Neural Networks},
year = {2024},
isbn = {9798400704901},
publisher = {Association for Computing Machinery},
address = {New York, NY, USA},
url = {https://doi.org/10.1145/3637528.3671744},
doi = {10.1145/3637528.3671744},
booktitle = {Proceedings of the 30th ACM SIGKDD Conference on Knowledge Discovery and Data Mining},
pages = {621–630},
numpages = {10},
location = {Barcelona, Spain},
series = {KDD '24}
}

@misc{tafeng,
  author       = {Chiranjiv Das},
  title        = {Ta-Feng Grocery Dataset},
  year         = {2025},
  howpublished = {\url{https://www.kaggle.com/datasets/chiranjivdas09/ta-feng-grocery-dataset}},
  note         = {Accessed: 2026-02-12}
}

@misc{dunnhumby,
  author       = {{dunnhumby}},
  title        = {Source Files Datasets},
  year         = {2025},
  howpublished = {\url{https://www.dunnhumby.com/source-files/}},
  note         = {Accessed: 2026-02-12}
}

@misc{instacart,
  author       = {Khanh Nam Le},
  title        = {Instacart Market Basket Analysis Dataset},
  year         = {2025},
  howpublished = {\url{https://github.com/khanhnamle1994/instacart-orders/tree/master/data}},
  note         = {Accessed: 2026-02-12}
}

@inproceedings{nowp,
author = {Zangerle, Eva and Pichl, Martin and Gassler, Wolfgang and Specht, G\"{u}nther},
title = {\#nowplaying Music Dataset: Extracting Listening Behavior from Twitter},
year = {2014},
isbn = {9781450331579},
publisher = {Association for Computing Machinery},
address = {New York, NY, USA},
url = {https://doi.org/10.1145/2661714.2661719},
doi = {10.1145/2661714.2661719},
booktitle = {Proceedings of the First International Workshop on Internet-Scale Multimedia Management},
pages = {21–26},
numpages = {6},
location = {Orlando, Florida, USA},
series = {WISMM '14}
}

@article{movielens20m,
  title   = {The MovieLens Datasets: History and Context},
  author  = {F. Maxwell Harper and Joseph A. Konstan},
  journal = {ACM Transactions on Interactive Intelligent Systems (TiiS)},
  volume  = {5},
  number  = {4},
  year    = {2015},
  pages   = {19:1--19:19},
  doi     = {10.1145/2827872}
}

@article{amazonreviews,
  title={Bridging Language and Items for Retrieval and Recommendation},
  author={Hou, Yupeng and Li, Jiacheng and He, Zhankui and Yan, An and Chen, Xiusi and McAuley, Julian},
  journal={arXiv preprint arXiv:2403.03952},
  year={2024}
}

@inproceedings{sgl,
author = {Wu, Jiancan and Wang, Xiang and Feng, Fuli and He, Xiangnan and Chen, Liang and Lian, Jianxun and Xie, Xing},
title = {Self-supervised Graph Learning for Recommendation},
year = {2021},
isbn = {9781450380379},
publisher = {Association for Computing Machinery},
address = {New York, NY, USA},
url = {https://doi.org/10.1145/3404835.3462862},
doi = {10.1145/3404835.3462862},
booktitle = {Proceedings of the 44th International ACM SIGIR Conference on Research and Development in Information Retrieval},
pages = {726–735},
numpages = {10},
location = {Virtual Event, Canada},
series = {SIGIR '21}
}

@inproceedings{30music,
  author    = {Turrin, Roberto and Quadrana, Massimo and Condorelli, Andrea and Pagano, Roberto and Cremonesi, Paolo},
  title     = {30Music Listening and Playlists Dataset},
  booktitle = {Proceedings of the 9th ACM Conference on Recommender Systems (RecSys 2015), Poster Proceedings},
  year      = {2015},
  month     = {September},
  address   = {Vienna, Austria},
  series    = {CEUR Workshop Proceedings},
  volume    = {1441},
  url       = {https://ceur-ws.org/Vol-1441/recsys2015_poster13.pdf}
}

@inproceedings{mubox,
author = {Li, Xiang and Wei, Wenqi and Thuraisingham, Bhavani},
title = {MUBox: A Critical Evaluation Framework of Deep Machine Unlearning [Systematization of Knowledge Paper]},
year = {2025},
isbn = {9798400715037},
publisher = {Association for Computing Machinery},
address = {New York, NY, USA},
url = {https://doi.org/10.1145/3734436.3734454},
doi = {10.1145/3734436.3734454},
booktitle = {Proceedings of the 30th ACM Symposium on Access Control Models and Technologies},
pages = {175–188},
numpages = {14},
location = {USA},
series = {SACMAT '25}
}

@inproceedings{yao2024copyright,
 author = {Yao, Yuanshun and Xu, Xiaojun and YangLiu},
 booktitle = {Advances in Neural Information Processing Systems},
 doi = {10.52202/079017-3346},
 editor = {A. Globerson and L. Mackey and D. Belgrave and A. Fan and U. Paquet and J. Tomczak and C. Zhang},
 pages = {105425--105475},
 publisher = {Curran Associates, Inc.},
 title = {Large Language Model Unlearning},
 url = {https://proceedings.neurips.cc/paper_files/paper/2024/file/be52acf6bccf4a8c0a90fe2f5cfcead3-Paper-Conference.pdf},
 volume = {37},
 year = {2024}
}

@inproceedings{
    marino2025position,
    title={Position: Bridge the Gaps between Machine Unlearning and {AI} Regulation},
    author={Bill Marino and Meghdad Kurmanji and Nicholas D. Lane},
    booktitle={The Thirty-Ninth Annual Conference on Neural Information Processing Systems Position Paper Track},
    year={2025},
    url={https://openreview.net/forum?id=0ngi2StMwC}
}

@article{models_remember_training_data,
author = {Zhang, Chiyuan and Bengio, Samy and Hardt, Moritz and Recht, Benjamin and Vinyals, Oriol},
title = {Understanding deep learning (still) requires rethinking generalization},
year = {2021},
issue_date = {March 2021},
publisher = {Association for Computing Machinery},
address = {New York, NY, USA},
volume = {64},
number = {3},
issn = {0001-0782},
url = {https://doi.org/10.1145/3446776},
doi = {10.1145/3446776},
journal = {Commun. ACM},
month = feb,
pages = {107–115},
numpages = {9}
}

@inproceedings{CarliniTWJHLRBS21,
  title = {Extracting Training Data from Large Language Models},
  author = {Carlini, Nicholas and Tram{\`e}r, Florian and Wallace, Eric and Jagielski, Matthew and Herbert-Voss, Ariel and Lee, Katherine and Roberts, Adam and Brown, Tom B. and Song, Dawn and Erlingsson, {\`U}lfar and Oprea, Alina and Raffel, Colin},
  booktitle = {30th USENIX Security Symposium (USENIX Security 2021)},
  pages = {2633--2650},
  year = {2021},
  publisher = {USENIX Association},
  url = {https://www.usenix.org/conference/usenixsecurity21/presentation/carlini-extracting}
}

@inproceedings{poisoning_attack_used_various_users,
author = {Li, Bo and Wang, Yining and Singh, Aarti and Vorobeychik, Yevgeniy},
title = {Data poisoning attacks on factorization-based collaborative filtering},
year = {2016},
isbn = {9781510838819},
publisher = {Curran Associates Inc.},
address = {Red Hook, NY, USA},
booktitle = {Proceedings of the 30th International Conference on Neural Information Processing Systems},
pages = {1893–1901},
numpages = {9},
location = {Barcelona, Spain},
series = {NIPS'16}
}


\appendix

\section{Appendix}
\label{app:supplementary-material}

\subsection{Baseline Performmance}
Baseline performance of popularity-based, random, and neighbor\-hood based methods on the NBR, SBR, and CF tasks is reported in \Cref{tab:baseline_nbr}, \Cref{tab:baseline_sbr}, and \Cref{tab:baseline_cf}, respectively. This comparison demonstrates that the additional computational cost associated with neural models is justified by their improved utility.

\begin{table*}
\centering
\small
\caption{Recommendation utility for training baselines on the initial NBR datasets.}
\label{tab:baseline_nbr}
\setlength\tabcolsep{8.9pt}
\begin{tabular}{l l  c c  c c  c c}
\toprule
\textbf{Model} &
\textbf{Dataset} &
\multicolumn{2}{c}{\textbf{nDCG (\%)}} &
\multicolumn{2}{c}{\textbf{Recall (\%)}} &
\multicolumn{2}{c}{\textbf{Hit (\%)}} \\
\cmidrule(r){3-4}
\cmidrule(r){5-6}
\cmidrule(r){7-8}
  &   & \textbf{@$10$} & \textbf{@$20$} & \textbf{@$10$} & \textbf{@$20$} & \textbf{@$10$} & \textbf{@$20$} \\
\midrule
\textbf{GPTopFreq} & Dunnhumby & $8.76$ & $8.82$ & $8.44$ & $10.56$ & $32.98$ & $39.87$ \\
\textbf{Pop} & Dunnhumby & $6.95$ & $7.05$ & $6.93$ & $8.85$ & $29.64$ & $36.65$ \\
\textbf{Random} & Dunnhumby & $0.22$ & $0.25$ & $0.23$ & $0.39$ & $1.45$ & $2.26$ \\
\textbf{TIFUKNN} & Dunnhumby & $3.74$ & $3.85$ & $3.36$ & $4.28$ & $13.41$ & $16.55$ \\
\midrule
\textbf{GPTopFreq} & Instacart & $17.78$ & $20.00$ & $17.27$ & $25.68$ & $67.27$ & $76.99$ \\
\textbf{Pop} & Instacart & $9.86$ & $9.95$ & $7.29$ & $9.97$ & $45.89$ & $52.48$ \\
\textbf{Random} & Instacart & $0.13$ & $0.18$ & $0.12$ & $0.26$ & $1.06$ & $2.20$ \\
\textbf{TIFUKNN} & Instacart & $25.12$ & $26.91$ & $25.84$ & $33.77$ & $75.37$ & $81.13$ \\
\midrule
\textbf{GPTopFreq} & TaFeng & $7.88$ & $9.26$ & $10.71$ & $14.90$ & $33.16$ & $44.19$ \\
\textbf{Pop} & TaFeng & $9.87$ & $10.44$ & $9.94$ & $11.96$ & $29.49$ & $36.97$ \\
\textbf{Random} & TaFeng & $0.09$ & $0.13$ & $0.10$ & $0.21$ & $0.67$ & $1.42$ \\
\textbf{TIFUKNN} & TaFeng & $7.94$ & $9.76$ & $11.29$ & $17.07$ & $35.79$ & $48.54$ \\
\bottomrule
\end{tabular}
\end{table*}

\begin{table*}
\centering
\small
\caption{Recommendation utility for training baselines on the initial SBR datasets.}
\label{tab:baseline_sbr}
\setlength\tabcolsep{11.2pt}
\begin{tabular}{l l  c c  c c  c c}
\toprule
\textbf{Model} &
\textbf{Dataset} &
\multicolumn{2}{c}{\textbf{nDCG (\%)}} &
\multicolumn{2}{c}{\textbf{Recall (\%)}} &
\multicolumn{2}{c}{\textbf{Hit (\%)}} \\
\cmidrule(r){3-4}
\cmidrule(r){5-6}
\cmidrule(r){7-8}
  &   & \textbf{@$10$} & \textbf{@$20$} & \textbf{@$10$} & \textbf{@$20$} & \textbf{@$10$} & \textbf{@$20$} \\
\midrule
\textbf{Pop} & 30Music & $1.36$ & $1.24$ & $0.33$ & $0.57$ & $7.85$ & $11.32$ \\
\textbf{Random} & 30Music & $0.00$ & $0.00$ & $0.00$ & $0.00$ & $0.05$ & $0.10$ \\
\textbf{SKNN} & 30Music & $4.50$ & $5.69$ & $8.88$ & $13.56$ & $8.88$ & $13.56$ \\
\midrule
\textbf{Pop} & Books & $0.27$ & $0.36$ & $0.48$ & $0.83$ & $0.52$ & $0.90$ \\
\textbf{Random} & Books & $0.00$ & $0.00$ & $0.00$ & $0.00$ & $0.00$ & $0.00$ \\
\textbf{SKNN} & Books & $1.21$ & $1.76$ & $3.30$ & $5.46$ & $3.30$ & $5.46$ \\
\midrule
\textbf{Pop} & NowP & $1.42$ & $1.50$ & $2.00$ & $2.28$ & $4.05$ & $4.83$ \\
\textbf{Random} & NowP & $0.01$ & $0.01$ & $0.00$ & $0.00$ & $0.08$ & $0.15$ \\
\midrule
\textbf{Pop} & RSC15 & $1.38$ & $1.75$ & $2.44$ & $3.94$ & $2.45$ & $3.96$ \\
\textbf{Random} & RSC15 & $0.01$ & $0.01$ & $0.02$ & $0.04$ & $0.02$ & $0.04$ \\
\bottomrule
\end{tabular}
\end{table*}

\begin{table*}
\centering
\small
\caption{Recommendation utility for training baselines on the initial CF datasets.}
\label{tab:baseline_cf}
\setlength\tabcolsep{10pt}
\begin{tabular}{l l  c c  c c  c c}
\toprule
\textbf{Model} &
\textbf{Dataset} &
\multicolumn{2}{c}{\textbf{nDCG (\%)}} &
\multicolumn{2}{c}{\textbf{Recall (\%)}} &
\multicolumn{2}{c}{\textbf{Hit (\%)}} 
\\
\cmidrule(r){3-4}
\cmidrule(r){5-6}
\cmidrule(r){7-8}
  &   & \textbf{@$10$} & \textbf{@$20$} & \textbf{@$10$} & \textbf{@$20$} & \textbf{@$10$} & \textbf{@$20$} \\
\midrule
\textbf{ItemKNN} & Food & $0.41$ & $0.49$ & $0.70$ & $1.00$ & $0.75$ & $1.06$ \\
\textbf{Pop} & Food & $0.13$ & $0.20$ & $0.28$ & $0.56$ & $0.29$ & $0.60$ \\
\textbf{Random} & Food & $0.00$ & $0.00$ & $0.01$ & $0.01$ & $0.01$ & $0.02$ \\
\midrule
\textbf{ItemKNN} & MovieLens & $7.67$ & $9.09$ & $6.96$ & $12.11$ & $32.72$ & $45.94$ \\
\textbf{Pop} & MovieLens & $5.66$ & $6.44$ & $4.85$ & $7.96$ & $25.26$ & $35.39$ \\
\textbf{Random} & MovieLens & $0.09$ & $0.11$ & $0.05$ & $0.11$ & $0.75$ & $1.53$ \\
\bottomrule
\end{tabular}
\end{table*}

\subsection{Necessary Modifications to the Unlearning Algorithms}

Several existing unlearning algorithms require minor but important implementation adaptations to ensure numerical stability, computational efficiency, and fair comparison. In particular, many published implementations assume batch-style unlearning or full access to the retain set, which is impractical in a sequential unlearning setting. Our modifications therefore focus on making the algorithms robust under repeated small deletion requests while keeping their behavior as close as possible to the original methods.

In our codebase, unlearning is orchestrated in \path{recbole/quick_start/quick_start.py} (construction of per-step \emph{forget} / \emph{clean-forget} / \emph{retain} loaders, exclusion of previously unlearned users, and retain subsampling) and dispatched/implemented in \path{recbole/trainer/trainer.py} (method-specific update rules). A shared practical consideration across methods is that we do not use the full retain set per unlearning step. Instead, we construct a per-step retain loader by sampling complete user sessions from the remaining training data while excluding users already unlearned in earlier steps, and we cap the total number of retain interactions per step (additionally bounded by a fixed fraction of the full training interactions). This keeps runtime comparable across methods while maintaining consistent access to retain information.

\header{Stability safeguards are shared across multiple second-order and influence-function-style methods} In particular, the optional hyperparameter \texttt{max\_norm} acts as a general-purpose safeguard: it can clip the magnitude of the computed parameter update (\texttt{SCIF}, \texttt{GIF}, \texttt{CEU}, \texttt{IDEA}) or clip gradients during repair fine-tuning (\texttt{Seif}). In addition, the implementations include explicit NaN/Inf checks and early stopping of iterative solvers when divergence is detected.

\headerl{\texttt{SCIF}}
In \texttt{SCIF}, a key adaptation is selecting which parameters are updated depending on the task/model. The code updates user embeddings when explicit \texttt{user\_embedding} exists (typical CF models). For common sequential recommenders, we update a task-appropriate subset (e.g., item embeddings and final/linear layers depending on the architecture), while for models where no special-case mapping is defined (including NBR models such as Sets2Sets) the implementation defaults to updating all trainable parameters. \texttt{SCIF} uses a conjugate-gradient inverse-HVP solve with damping (\texttt{damping}) and includes NaN/Inf checks; when recurrent layers are present, we disable cuDNN for the second-order derivative path to make Hessian--vector product computation feasible. The final update can be magnitude-clipped via \texttt{max\_norm}. We carried out a hyperparameter test and found $\texttt{max\_norm} = 10$ to be the best choice.

\headerl{\texttt{GIF}}
\texttt{GIF} does not require structural changes, but it is highly sensitive to numerical instability in the Hessian-inverse approximation. Our implementation uses a Neumann-series-style approximation with damping (\texttt{gif\_damping} and \texttt{gif\_scale\_factor}), includes NaN/Inf checks throughout Hessian--vector product computation and iteration steps, and can optionally clip the final update with \texttt{max\_norm}. Despite these safeguards, \texttt{GIF} remains among the methods that diverge most frequently in practice.

\headerl{\texttt{CEU}}
Our \texttt{CEU} implementation follows the certified unlearning structure and introduces numerical stabilizers directly in the second-order computations. The update uses an $L_2$-type regularization term controlled by \texttt{ceu\_lambda} when forming the relevant Hessian(-vector) quantities, and it also adds a corresponding damping term in the Hessian--vector product computation. We apply these stabilizers during the unlearning update (not as a requirement of the initial training objective), because they substantially improve robustness of the conjugate-gradient solve and reduce divergence. Before adding certification noise with standard deviation \texttt{ceu\_sigma}, we optionally clip the influence estimate using \texttt{max\_norm}.

\headerl{\texttt{IDEA}}
\texttt{IDEA} similarly benefits from stability safeguards. Our implementation solves $(H + \lambda I)x = v$ with damping \texttt{idea\_damping}, includes NaN/Inf and non-positive-definite checks in the iterative procedure, optionally clips the update (\texttt{max\_norm}) before computing certification quantities, and applies the certified noise mechanism (with \texttt{idea\_sigma} increased if needed to satisfy the $(\epsilon,\delta)$ requirement).

\headerl{\texttt{Fanchuan}}
For \texttt{Fanchuan}, both the contrastive-learning representation and retain usage are critical. We fix the contrastive representation (via \texttt{Trainer.get\_embedding\_for\_contrastive\_learning}) as follows: user embeddings for CF (when the model exposes\linebreak \texttt{user\_embedding}), sequence/session representations for SBR (via the sequential forward pass), and basket-level representations for NBR (via \texttt{get\_basket\_embedding} when available, otherwise a for\-ward-pass fallback). To make the method feasible under sequential unlearning, we avoid full-retain repair rounds: contrastive updates draw retain batches from a shuffled index pool over the per-step retain loader, and the subsequent retain round operates on a bounded number of retain samples per unlearning call (with epochs adjusted if the requested sample budget exceeds the unique samples in the retain loader).

\headerl{\texttt{Kookmin}}
For \texttt{Kookmin}, the central hyperparameter is the per-layer reset ratio, implemented as \texttt{kookmin\_init\_rate}. During each unlearning step, we reinitialize a fraction of parameters \emph{within each tensor} based on the signed-gradient signal (using a per-layer top-$k$ thresholding rule), then perform a short retain round whose effective workload is scaled to the forget set size. Additionally, in \texttt{quick\_start.py}, we reduce the base learning rate when running \texttt{Kookmin} unlearning to prevent instability when only a subset of parameters is reset. We tune \texttt{kookmin\_init\_rate} in our experiments to avoid both collapse from overly aggressive resets and ineffectual updates from overly small resets.

\headerl{\texttt{Seif}}
\texttt{Seif} follows its erase/repair structure in our implementation (noise injection into embedding/linear-like layers followed by a bounded repair phase on retain data) and does not require additional algorithmic changes beyond these practical adaptations; optional \texttt{max\_norm} gradient clipping is supported during the repair phase.

\subsection{Extended Discussion on Preliminary Benchmark Results}
\label{sec:discussion}

\header{Why can unlearning outperform retraining on utility metrics?}
In sensitive item unlearning for CF we observe that unlearned models match or even exceed the recommendation utility of models retrained from scratch on the retained data (LightGCN, SGL). A plausible explanation is \emph{underfitting of retraining under a fixed compute budget}. In our experiments, training, retraining, and unlearning were all subject to the same maximum wall-clock time of one day. Retraining starts from random initialization (or a cold start) and may not reach the same optimum within the time limit, especially for large datasets or slower models. In contrast, unlearning methods start from the fully trained original model and apply a targeted ``forget'' update followed by a ``repair'' update on retained data. Because the repair phase continues optimization from a strong initialization, unlearned models can keep improving during repair while retrained models may still be in an earlier optimization regime. This effect is most likely when retraining is compute-limited; with sufficiently long retraining, we expect the gap to diminish.

\header{Comparative trends across methods}
Across our benchmarks, \texttt{SCIF} is the most consistently strong method, often yielding the best overall utility while maintaining good forgetting effectiveness. \texttt{GIF} behaves similarly when converging but typically incurs a modest utility degradation compared to \texttt{SCIF}; however, \texttt{GIF} tends to achieve lower sensitive-item exposure (i.e., better forgetting) than \texttt{SCIF}. \texttt{Seif} performs poorly in several multi-request regimes; we attribute this primarily to the interaction of small unlearning batch sizes with repeated noise injection, where the cumulative perturbation disrupts the model more than the repair step can compensate. Finally, \texttt{GIF} is up to an order of magnitude slower than \texttt{SCIF}, which is consistent with \texttt{GIF} updating a broader set of parameters (and/or relying on more expensive inverse-Hessian approximations) than \texttt{SCIF}'s more selective update strategy.

\header{Interpreting ``good'' and ``bad'' unlearning performance}
We evaluate unlearning along three axes: \emph{utility} (recommendation quality on retained data), \emph{effectiveness} (degree of forgetting; e.g., sensitive-item exposure), and \emph{efficiency} (time per unlearning request relative to retraining). Following prior work (e.g., CURE4REC~\cite{chen2024cure4rec} and MUBox~\cite{mubox}), we report results relative to retraining and treat retraining as the reference for exact data removal. The literature does not provide universal thresholds that map a relative difference to ``acceptable'' or ``unacceptable'' unlearning across datasets and models. Therefore, we interpret outcomes comparatively: an unlearning method is preferable when it (i) is close to retraining in utility, (ii) achieves equal or better forgetting effectiveness, and (iii) is substantially more efficient than retraining. We argue that an unlearning algorithm providing a speedup of $\geq 3$ orders of magnitude over retraining is efficient enough for practical usage. From the interviews, we learned that training time usually does not exceed one day, and the unlearning time for 1 request does not exceed a few minutes when assuming this upper bound.

\header{Why do some methods fail or degrade?}
Failure modes differ by algorithmic mechanism and by the structure of the deletion request.
For GNN-based recommenders, inverse-Hessian-style approximations can be numerically fragile; methods that rely on less stable iterative schemes may diverge when the curvature is ill-conditioned or when updates accumulate over many requests. More broadly, methods that inject noise, for certified guarantees or just for forgetting, can suffer from \emph{cumulative perturbation}: repeated noisy updates over many batches can move parameters into regions that require long repair to recover, eroding both utility and efficiency.

\header{Method-specific observations and hypotheses}

\begin{itemize}[leftmargin=*]
    \item \texttt{SCIF}: \texttt{SCIF} yields strong utility and robust forgetting across most settings. Its main limitation is runtime on large models/datasets, where inverse-Hessian-style computations (even approximated) become expensive; on the largest configurations we observe only a small efficiency advantage over retraining. This suggests that further curvature approximations or more aggressive parameter selection may be needed at scale. In spam unlearning, \texttt{SCIF}'s stability may also benefit from combining forgetting and repair into a single update rather than applying them sequentially.
    \item \texttt{GIF}: \texttt{GIF} often achieves excellent forgetting effectiveness (low sensitive exposure) but tends to reduce utility more than \texttt{SCIF} and can be slower. One explanation is that \texttt{GIF}'s parameter selection via $k$-hop neighborhoods may update a broader or noisier parameter set than \texttt{SCIF}'s recommender-tailored selection (e.g., focusing on item embeddings while freezing others). In addition, \texttt{GIF}'s Neumann-series inverse-Hessian approximation is sensitive to convergence conditions (e.g., requiring the spectral radius of $(I-\lambda H)$ to be $< 1$), which may explain occasional instability or divergence for some seeds.
    \item \texttt{CEU}: \texttt{CEU} combines noise with $L_2$-regularization in the Hessian computation to encourage positive definiteness. It succeeds for LightGCN but fails for SGL in our experiments, suggesting that the additional self-supervised components and augmentations in SGL yield a more complex optimization landscape where the approximation assumptions break down.
    \item \texttt{IDEA}: \texttt{IDEA} can exhibit similar degradation patterns to noise-based methods (e.g., \texttt{Seif}) under many requests. Its norm-depen\-dent noise, while motivated by certified guarantees in simpler settings, can still lead to cumulative perturbation and long recovery times in deep recommenders. We also observed practical issues at scale (e.g., memory usage) that limit applicability for large datasets.
    \item \texttt{Fanchuan}: For NBR sensitive-item unlearning, we observe a sharp utility drop alongside strong forgetting indicators. This suggests that the uniform pseudo-label phase can substantially distort item representations, and the subsequent repair phase may not fully recover from this catastrophic shift under the fixed time budget.
    \item \texttt{Kookmin}: \texttt{Kookmin} performs well for sensitive-item unlearning, consistent with the hypothesis that only a small parameter subset (e.g., item embeddings) needs to be reset and then repaired. In spam unlearning, performance degrades as the number of unlearning batches increases (NBR: few batches, acceptable; SBR: many batches, poor). A plausible explanation is that resets inadvertently affect parameters tied to popular items that appear in spam sessions; because these items are widely used, even small representation damage can hurt many users, and the repair phase may be insufficient to fully restore them under the compute limit.
    Also, the attack interweaves target items with popular items within sessions; consequently, harming steps that affect all items in the session can also damage representations of popular items, reducing utility broadly. A targeted variant that applies the harming stage only to identified target items could mitigate this, but it introduces an adversarial consideration: attackers could attempt to exploit the unlearning mechanism to suppress items by submitting interactions that later get removed.
    \item \texttt{Seif}: \texttt{Seif} is competitive when the number of unlearning batches is small, but degrades markedly with many sequential requests. This is consistent with fixed-noise injection accumulating over batches; the repair step cannot keep pace unless its budget is increased, at which point the benefit over retraining diminishes.
\end{itemize}

\header{Hyperparameter choice} We report the hyperparameter choices for all compared methods.

\headerl{\texttt{SCIF} and Other Influence-Function-Based Methods}
For \texttt{SCIF}, we varied the gradient clipping threshold
\[
\texttt{max\_norm} \in \{10^{-2}, 10^{-1}, 10^{0}, 10^{1}\}.
\]
For other methods relying on influence-function-based parameter updates (\texttt{GIF}, \texttt{CEU}, and \texttt{IDEA}), we used the same \texttt{max\_norm} value for consistency.
Table~\ref{tab:scif-maxnorm} reports the performance of \texttt{SCIF} under different clipping thresholds. We chose $\texttt{max\_norm} = 10$ for our experiments.

\headerl{\texttt{Fanchuan}}
The \texttt{Fanchuan} method involves several design choices related to contrastive learning. These choices can be viewed as a hyperparameter search over augmentation strategies and loss configurations. Due to space constraints, we describe the selection process and the final configuration in the main paper.

\headerl{\texttt{Kookmin}}
For \texttt{Kookmin}, we tuned the reinitialization rate:
\[
\texttt{kookmin\_init\_rate} \in \{10^{-5}, 10^{-4}, 10^{-3}, 10^{-2}, 5 \cdot 10^{-2}\}.
\]
Table~\ref{tab:kookmin-initrate} summarizes the results. We chose $\texttt{kookmin\_init\_rate} = 10^{-4}$ for our experiments.

\headerl{\texttt{Seif}}
For \texttt{Seif}, we follow the original paper and sample the noise added to the model parameters from a multivariate Gaussian distribution,
\[
\mathcal{N}(0, 0.6^2 I).
\]
We did not further tune this variance.

\headerl{Unlearning batchsize}
Unlearning interactions of $n \gg 1$ users at the same time is not realistic for sensitive item unlearning, but sensible for removing spam. Consequently, when spam interactions are identified, they are likely to originate from multiple users rather than a single account and can be merged to one unlearning request as opposed to sequential requests for single users. We test $\texttt{unlearning\_batchsize} \in \{ 2^i \mid 0 \leq i < 10 \}$ and find $2^8$ to yield the best results.

\header{Impact of model architectures}
Unlearning \emph{attention-based models} with general methods like \texttt{Kookmin} and \texttt{Seif} work well for a small number of unlearning requests, but the performance of the unlearned model drops as more unlearning requests are served. \texttt{Fanchuan} and \texttt{SCIF} generally maintain effectiveness and utility while being multiple orders of magnitude faster than training, but are sometimes too slow (only 1 order of magnitude faster than training for NowP spam).
For the \emph{recurrent model} \texttt{GRU4Rec}, only \texttt{SCIF} maintains most effectiveness and efficiency while being similarly fast as other unlearning algorithms. The general unlearning algorithms \texttt{Kookmin} and \texttt{Seif} degrade performance more than 10\% in the spam unlearning scenario, while \texttt{Fanchuan} works for spam unlearning within the 5\% error margin, but degrades performance to $<25\%$ of the retrained model in sensitive item unlearning while predicting more sensitive items than the retrained model. In NBR for \texttt{Sets2Sets}, \texttt{Kookmin} performs best, followed by \texttt{SCIF}. \texttt{Kookmin} maintains utility and predicts a similar number of sensitive items as the retrained model, likely due to the small number of unlearning requests compared to SBR or CF.
For \emph{graph-based models}, we discuss the three graph-specific unlearning algorithms. These algorithms are efficient across scenarios and tasks and have among the best effectiveness and utility in some cases (\texttt{CEU} NowP SRGNN; \texttt{GIF}, \texttt{CEU} Food LightGCN; \texttt{GIF}, \texttt{IDEA} DNNTSP). However, they are also unstable, as seen in Figure~2 in the main paper, especially \texttt{GIF} and \texttt{IDEA} diverged occasionally during the sequential execution of unlearning requests. \texttt{Seif} degraded performance too much due to noise being too big of a factor, as in every setting with multiple sequential unlearning requests. The recommender system-specific method \texttt{SCIF} maintained the highest utility and effectiveness in most experiments, but was sometimes outperformed by graph-specific algorithms. The general methods \texttt{Fanchuan} and \texttt{Kookmin} are efficient, effective, and maintain utility for CF sensitive item unlearning, but are less effective than \texttt{SCIF} in spam unlearning.

\subsection{Additional Details for ``A Snapshot of Current Progress in Machine Unlearning for Recommender Systems''}

\header{Unlearning effectiveness} For unlearning spam, we observe that \texttt{Fanchuan} is volatile with results having effectiveness close to the retrained model but also results with performance up to $22.8\%$ worse than the retrained model. \texttt{Seif} diverges likely due to noise being too oppressive. \texttt{Kookmin} is often close to the retrained model or more effective for sensitive item unlearning, but in SBR spam unlearning, degrades effectiveness by $15$--$20\%$. The graph-specific unlearning algorithms diverge often after many unlearning requests, but are close to the effectiveness of the retrained model in NBR spam unlearning and CF sensitive item unlearning. The recommender system specific method \texttt{SCIF} has an effectiveness around the retrained model consistently.

\header{Robustness} Most of the unlearning algorithms lack robustness. GNN-specific unlearning algorithms diverge in training due to unstable Hessian approximations or excessively large noise added to the parameters during the update. The general unlearning algorithms \texttt{Kookmin} and \texttt{Seif} degrade performance more for large numbers of unlearning requests than other unlearning algorithms, whereas \texttt{Fanchuan} is robust across scenarios with the exception of NBR sensitive item unlearning. The recommender-specific method \texttt{SCIF} is the most robust method with no divergences during unlearning and the lowest variance overall while being efficient, effective, and maintaining utility. An exception is NBR sensitive item unlearning for DNNTSP where all unlearning algorithms predict sensitive items for more users than the retrained model.

\header{Influence of model architecture} For attention-based models, general methods such as \texttt{Kookmin} and \texttt{Seif} perform well for few unlearning requests but degrade as requests accumulate, while \texttt{Fanchuan} and \texttt{SCIF} largely preserve effectiveness and utility. For the recurrent model \texttt{GRU4Rec}, \texttt{SCIF} is the only method that consistently maintains effectiveness, whereas general methods show notable degradation and \texttt{Fanchuan} fails under sensitive item unlearning. For graph-based models, graph-specific methods are often efficient and highly effective but can be unstable under sequential unlearning, while \texttt{SCIF} provides the most robust overall performance, with general methods remaining competitive mainly for CF sensitive item unlearning.

\subsection{Detailed Result Tables}
First, we list detailed result tables, showing recommendation performance and latency for the unlearning methods that we consider, grouped first by scenario and then by task.

\subsubsection*{Scenario: Sensitive item unlearning} \mbox{}
For the sensitive item unlearning scenario, results are reported for three tasks. For NBR, Table~\ref{tab:results_dunnhumby_alcohol} shows results on Dunnhumby and Table~\ref{tab:results_instacart_meat} on Instacart. For SBR, Table~\ref{tab:results_amazonreviewsbooks_health} reports results on Books and Table~\ref{tab:results_30music_explicit} on 30music. For CF, Table~\ref{tab:results_movielens_health} presents results on MovieLens and Table~\ref{tab:results_amazonreviewsgroceryandgourmetfood_alcohol} on Food.

\subsubsection*{Scenario: Removal of poisonous data} \mbox{}
For the removal of poisonous data scenario, we report NBR results in Table~\ref{tab:results_tafeng} (TaFeng) and Table~\ref{tab:results_dunnhumby} (Dunnhumby), and SBR results in Table~\ref{tab:results_rsc15} (RSC15) and Table~\ref{tab:results_nowp} (NowP).

Second, Table~\ref{tab:robustness} summarizes unlearning robustness per algorithm, split by scenario and task and averaged across datasets and models.

Third, we report on the influence of different hyperparameter settings. We tune the \texttt{max\_norm} parameter for \texttt{SCIF} in the sensitive item removal scenario for CF using the Food dataset with the LightGCN model. The same setting is used to tune the \texttt{kookmin\_init\_rate} parameter for \texttt{Kookmin}.
For spam removal we tune the batch size of unlearning requests.\\

\begin{itemize}
    \item Table~\ref{tab:scif-maxnorm}: The \texttt{max\_norm} parameter limits the maximum magnitude of an update step. Due to numerical instabilities, the computed update can occasionally become excessively large, which may severely degrade the model if applied without rescaling. In preliminary experiments, methods that approximate the Hessian frequently diverged when no such constraint was imposed, as the update norms grew uncontrollably.
    We initially evaluated a value of 1 for \texttt{max\_norm}, and subsequently explored the values $10^{-2}$, $10^{-1}$, and $10^{1}$ for hyperparameter tuning.
    \item Table~\ref{tab:kookmin-initrate}: The \texttt{kookmin\_init\_rate} parameter specifies the fraction of network parameters that are reset to their default initialization prior to each fine-tuning step. If this value is too large, the model’s previously acquired knowledge is removed faster than the subsequent repair phase can restore it. Conversely, if the value is too small, the model may fail to sufficiently forget the designated forget set.
    We evaluated the values $10^{-5}, 10^{-4}, 10^{-3}, 10^{-2}$, and $5 \cdot 10^{-2}$. The authors of \texttt{Kookmin} report that a value of $0.3$ performed best in their setting; however, their scenario involved a single large unlearning request followed by a long repair phase. This assumption does not hold in our setting. Under our conditions, such a high value caused catastrophic forgetting after only a few unlearning requests.
    \item Table~\ref{tab:unlearning_batchsize_results}: For sensitive item removal, the number of requests per unlearning batch is set to 1, since it is preferable to process such requests immediately rather than waiting until a larger number of requests has accumulated.
    Spam removal is different in this regard, as spam detection typically identifies multiple spam interactions at the same time. Therefore, it is reasonable to tune the hyperparameter controlling the number of unlearning requests per batch size. We consider different values of the unlearning batch size in the spam removal scenario for SBR on a subset of RSC15 with the GRU4Rec model for \texttt{SCIF}, \texttt{Fanchuan}, \texttt{Kookmin}, and \texttt{Seif}.
\end{itemize}


\begin{table*}
\centering
\small
\caption{Recommendation utility, unlearning effectiveness, and unlearning efficiency in NBR for task-specific models on the Dunnhumby dataset in the sensitive item unlearning scenario. Recommendation utility is measured by standard metrics. Unlearning efficiency is measured by $\text{Sensitive}\boldsymbol{@k}$. For efficiency, we report the average time needed to execute an unlearning request in minutes.}
\label{tab:results_dunnhumby_alcohol}
\setlength\tabcolsep{5.4pt}
\begin{tabular}{ll  c c  c c  c c  c c  c}
\toprule
 &  &
\multicolumn{2}{c}{\textbf{nDCG (\%)}$\mathbf{\uparrow}$} &
\multicolumn{2}{c}{\textbf{Recall (\%)}$\mathbf{\uparrow}$} &
\multicolumn{2}{c}{\textbf{Hit (\%)}$\mathbf{\uparrow}$} &
\multicolumn{2}{c}{\textbf{Sensitive@}$\mathbf{20\downarrow}$} &
\textbf{Time (min) $\mathbf{\downarrow}$} \\
\cmidrule(r){3-4}
\cmidrule(r){5-6}
\cmidrule(r){7-8}
\cmidrule(r){9-10}
\cmidrule(r){11-11}
\textbf{Model} &
\textbf{Algorithm} &
\textbf{@10} & \textbf{@20} &
\textbf{@10} & \textbf{@20} &
\textbf{@10} & \textbf{@20} &
\textbf{@10} & \textbf{@20} &
\textbf{Avg/Req} \\
\midrule
\textbf{DNNTSP} & \texttt{Retrain} & 7.11 \scriptsize{$\pm$ 0.31} & 7.12 \scriptsize{$\pm$ 0.37} & 7.25 \scriptsize{$\pm$ 0.24} & 8.98 \scriptsize{$\pm$ 0.42} & 30.42 \scriptsize{$\pm$ 0.78} & 36.39 \scriptsize{$\pm$ 1.11} & 0.00 \scriptsize{$\pm$ 0.00} & 0.00 \scriptsize{$\pm$ 0.00} & 95.53 \scriptsize{$\pm$ 4.08} \\
 & \texttt{Fanchuan} & 2.59 \scriptsize{$\pm$ 1.13} & 3.00 \scriptsize{$\pm$ 1.12} & 2.68 \scriptsize{$\pm$ 1.02} & 4.25 \scriptsize{$\pm$ 1.27} & 14.47 \scriptsize{$\pm$ 7.21} & 21.43 \scriptsize{$\pm$ 8.55} & 0.57 \scriptsize{$\pm$ 0.26} & 1.21 \scriptsize{$\pm$ 0.29} & 0.49 \scriptsize{$\pm$ 0.01} \\
 & \texttt{Kookmin} & \underline{5.93} \scriptsize{$\pm$ 0.52} & \underline{6.19} \scriptsize{$\pm$ 0.52} & \underline{6.94} \scriptsize{$\pm$ 0.17} & \underline{8.78} \scriptsize{$\pm$ 0.38} & \underline{29.29} \scriptsize{$\pm$ 0.59} & \underline{35.95} \scriptsize{$\pm$ 1.22} & 0.21 \scriptsize{$\pm$ 0.23} & 17.62 \scriptsize{$\pm$ 38.90} & \underline{0.02} \scriptsize{$\pm$ 0.00} \\
 & \texttt{SCIF} & \textbf{6.95} \scriptsize{$\pm$ 0.22} & \textbf{7.00} \scriptsize{$\pm$ 0.32} & \textbf{7.16} \scriptsize{$\pm$ 0.15} & \textbf{8.84} \scriptsize{$\pm$ 0.35} & \textbf{29.99} \scriptsize{$\pm$ 0.65} & \textbf{36.05} \scriptsize{$\pm$ 1.29} & \underline{0.11} \scriptsize{$\pm$ 0.16} & \underline{0.14} \scriptsize{$\pm$ 0.15} & \textbf{0.01} \scriptsize{$\pm$ 0.00} \\
 & \texttt{Seif} & 0.11 \scriptsize{$\pm$ 0.06} & 0.12 \scriptsize{$\pm$ 0.06} & 0.12 \scriptsize{$\pm$ 0.08} & 0.18 \scriptsize{$\pm$ 0.09} & 0.72 \scriptsize{$\pm$ 0.43} & 1.16 \scriptsize{$\pm$ 0.44} & \textbf{0.00} \scriptsize{$\pm$ 0.00} & \textbf{0.00} \scriptsize{$\pm$ 0.00} & 0.02 \scriptsize{$\pm$ 0.00} \\
\midrule
\textbf{Sets2Sets} & \texttt{Retrain} & 9.88 \scriptsize{$\pm$ 0.61} & 9.74 \scriptsize{$\pm$ 0.55} & 8.49 \scriptsize{$\pm$ 0.89} & 10.74 \scriptsize{$\pm$ 0.98} & 35.44 \scriptsize{$\pm$ 2.46} & 41.46 \scriptsize{$\pm$ 2.25} & 0.00 \scriptsize{$\pm$ 0.00} & 0.00 \scriptsize{$\pm$ 0.00} & 101.48 \scriptsize{$\pm$ 1.54} \\
 & \texttt{Fanchuan} & 2.38 \scriptsize{$\pm$ 1.22} & 2.33 \scriptsize{$\pm$ 1.14} & 2.12 \scriptsize{$\pm$ 1.08} & 2.44 \scriptsize{$\pm$ 1.10} & 7.98 \scriptsize{$\pm$ 3.93} & 9.09 \scriptsize{$\pm$ 4.15} & 0.18 \scriptsize{$\pm$ 0.13} & 0.25 \scriptsize{$\pm$ 0.20} & 0.12 \scriptsize{$\pm$ 0.00} \\
 & \texttt{Kookmin} & \textbf{9.30} \scriptsize{$\pm$ 0.21} & \textbf{9.38} \scriptsize{$\pm$ 0.19} & \textbf{8.27} \scriptsize{$\pm$ 0.53} & \textbf{11.01} \scriptsize{$\pm$ 0.40} & \textbf{35.73} \scriptsize{$\pm$ 0.43} & \textbf{42.70} \scriptsize{$\pm$ 0.52} & \underline{0.07} \scriptsize{$\pm$ 0.10} & \underline{0.18} \scriptsize{$\pm$ 0.13} & \textbf{0.02} \scriptsize{$\pm$ 0.00} \\
 & \texttt{SCIF} & 3.37 \scriptsize{$\pm$ 2.39} & 4.41 \scriptsize{$\pm$ 1.80} & 3.74 \scriptsize{$\pm$ 2.17} & \underline{7.11} \scriptsize{$\pm$ 1.71} & \underline{17.59} \scriptsize{$\pm$ 10.46} & \underline{29.87} \scriptsize{$\pm$ 7.04} & \textbf{0.00} \scriptsize{$\pm$ 0.00} & \textbf{0.00} \scriptsize{$\pm$ 0.00} & 0.02 \scriptsize{$\pm$ 0.00} \\
 & \texttt{Seif} & \underline{4.26} \scriptsize{$\pm$ 0.12} & \underline{4.83} \scriptsize{$\pm$ 0.13} & \underline{4.43} \scriptsize{$\pm$ 0.20} & 6.61 \scriptsize{$\pm$ 0.30} & 17.51 \scriptsize{$\pm$ 0.80} & 25.08 \scriptsize{$\pm$ 0.76} & 4.23 \scriptsize{$\pm$ 0.93} & 6.32 \scriptsize{$\pm$ 0.99} & \underline{0.02} \scriptsize{$\pm$ 0.00} \\
\bottomrule
\end{tabular}
\end{table*}

\begin{table*}
\centering
\small
\caption{Recommendation utility, unlearning effectiveness, and unlearning efficiency in NBR for task-specific models on the Instacart dataset in the sensitive item unlearning scenario. Recommendation utility is measured by standard metrics. Unlearning efficiency is measured by $\textbf{Sensitive}\boldsymbol{@k}$. For efficiency, we report the average time needed to execute an unlearning request in minutes.}
\label{tab:results_instacart_meat}
\setlength\tabcolsep{4.5pt}
\begin{tabular}{ll  c c  c c  c c  c c  c}
\toprule
 &  &
\multicolumn{2}{c}{\textbf{nDCG (\%)}$\mathbf{\uparrow}$} &
\multicolumn{2}{c}{\textbf{Recall (\%)}$\mathbf{\uparrow}$} &
\multicolumn{2}{c}{\textbf{Hit (\%)}$\mathbf{\uparrow}$} &
\multicolumn{2}{c}{\textbf{Sensitive@}$\mathbf{20\downarrow}$} &
\textbf{Time (min) $\mathbf{\downarrow}$} \\
\cmidrule(r){3-4}
\cmidrule(r){5-6}
\cmidrule(r){7-8}
\cmidrule(r){9-10}
\cmidrule(r){11-11}
\textbf{Model} &
\textbf{Algorithm} &
\textbf{@10} & \textbf{@20} &
\textbf{@10} & \textbf{@20} &
\textbf{@10} & \textbf{@20} &
\textbf{@10} & \textbf{@20} &
\textbf{Avg/Req} \\
\midrule
\textbf{DNNTSP} & \texttt{Retrain} & 29.17 \scriptsize{$\pm$ 0.21} & 30.38 \scriptsize{$\pm$ 0.16} & 24.89 \scriptsize{$\pm$ 0.08} & 32.64 \scriptsize{$\pm$ 0.07} & 77.23 \scriptsize{$\pm$ 0.11} & 83.36 \scriptsize{$\pm$ 0.04} & 12.24 \scriptsize{$\pm$ 1.70} & 37.16 \scriptsize{$\pm$ 4.59} & 1442.49 \scriptsize{$\pm$ 5.23} \\
 & \texttt{Fanchuan} & \underline{18.43} \scriptsize{$\pm$ 4.92} & \underline{20.27} \scriptsize{$\pm$ 5.44} & 17.56 \scriptsize{$\pm$ 4.39} & 24.29 \scriptsize{$\pm$ 6.32} & 62.80 \scriptsize{$\pm$ 12.49} & 71.22 \scriptsize{$\pm$ 12.57} & 43.37 \scriptsize{$\pm$ 22.53} & \underline{74.19} \scriptsize{$\pm$ 14.35} & 0.91 \scriptsize{$\pm$ 0.01} \\
 & \texttt{Kookmin} & 16.74 \scriptsize{$\pm$ 2.02} & 19.96 \scriptsize{$\pm$ 1.79} & \underline{19.66} \scriptsize{$\pm$ 1.51} & \underline{29.05} \scriptsize{$\pm$ 1.60} & \underline{70.94} \scriptsize{$\pm$ 2.62} & \underline{80.64} \scriptsize{$\pm$ 2.00} & \textbf{35.45} \scriptsize{$\pm$ 9.52} & 74.56 \scriptsize{$\pm$ 4.61} & \underline{0.03} \scriptsize{$\pm$ 0.00} \\
 & \texttt{SCIF} & \textbf{26.97} \scriptsize{$\pm$ 2.33} & \textbf{28.55} \scriptsize{$\pm$ 1.92} & \textbf{23.74} \scriptsize{$\pm$ 1.28} & \textbf{31.74} \scriptsize{$\pm$ 0.98} & \textbf{75.93} \scriptsize{$\pm$ 1.77} & \textbf{82.87} \scriptsize{$\pm$ 0.56} & 53.90 \scriptsize{$\pm$ 4.82} & 82.17 \scriptsize{$\pm$ 2.06} & \textbf{0.02} \scriptsize{$\pm$ 0.00} \\
 & \texttt{Seif} & 0.21 \scriptsize{$\pm$ 0.11} & 0.24 \scriptsize{$\pm$ 0.10} & 0.17 \scriptsize{$\pm$ 0.08} & 0.29 \scriptsize{$\pm$ 0.10} & 1.31 \scriptsize{$\pm$ 0.64} & 2.10 \scriptsize{$\pm$ 0.68} & \underline{40.05} \scriptsize{$\pm$ 54.55} & \textbf{59.88} \scriptsize{$\pm$ 54.51} & 0.03 \scriptsize{$\pm$ 0.00} \\
\midrule
\textbf{Sets2Sets} & \texttt{Retrain} & 16.89 \scriptsize{$\pm$ 8.72} & 15.78 \scriptsize{$\pm$ 8.11} & 11.90 \scriptsize{$\pm$ 6.01} & 13.35 \scriptsize{$\pm$ 6.73} & 53.04 \scriptsize{$\pm$ 24.30} & 56.21 \scriptsize{$\pm$ 24.72} & 20.85 \scriptsize{$\pm$ 44.28} & 62.56 \scriptsize{$\pm$ 23.03} & 311.44 \scriptsize{$\pm$ 65.71} \\
 & \texttt{Fanchuan} & 6.50 \scriptsize{$\pm$ 1.89} & 5.96 \scriptsize{$\pm$ 1.62} & 4.08 \scriptsize{$\pm$ 1.17} & 4.62 \scriptsize{$\pm$ 1.03} & 29.45 \scriptsize{$\pm$ 5.31} & 32.63 \scriptsize{$\pm$ 4.88} & 55.40 \scriptsize{$\pm$ 48.31} & 66.40 \scriptsize{$\pm$ 46.86} & 0.07 \scriptsize{$\pm$ 0.00} \\
 & \texttt{Kookmin} & \textbf{21.04} \scriptsize{$\pm$ 0.76} & \textbf{19.32} \scriptsize{$\pm$ 0.52} & \textbf{14.58} \scriptsize{$\pm$ 0.28} & \textbf{15.75} \scriptsize{$\pm$ 0.00} & \textbf{64.37} \scriptsize{$\pm$ 0.47} & \textbf{66.51} \scriptsize{$\pm$ 0.03} & \textbf{28.45} \scriptsize{$\pm$ 0.80} & \textbf{34.73} \scriptsize{$\pm$ 1.08} & \textbf{0.01} \scriptsize{$\pm$ 0.00} \\
 & \texttt{SCIF} & \underline{18.50} \scriptsize{$\pm$ 4.06} & \underline{16.95} \scriptsize{$\pm$ 3.61} & \underline{14.09} \scriptsize{$\pm$ 0.78} & \underline{15.12} \scriptsize{$\pm$ 0.74} & \underline{62.70} \scriptsize{$\pm$ 1.28} & \underline{64.76} \scriptsize{$\pm$ 1.02} & \underline{30.05} \scriptsize{$\pm$ 4.18} & \underline{58.87} \scriptsize{$\pm$ 31.43} & 0.02 \scriptsize{$\pm$ 0.00} \\
 & \texttt{Seif} & 8.02 \scriptsize{$\pm$ 9.17} & 7.54 \scriptsize{$\pm$ 8.52} & 5.66 \scriptsize{$\pm$ 6.16} & 6.49 \scriptsize{$\pm$ 6.93} & 29.45 \scriptsize{$\pm$ 27.16} & 32.69 \scriptsize{$\pm$ 28.12} & 70.75 \scriptsize{$\pm$ 40.09} & 80.04 \scriptsize{$\pm$ 29.32} & \underline{0.01} \scriptsize{$\pm$ 0.00} \\
\bottomrule
\end{tabular}
\end{table*}

\begin{table*}
\centering
\small
\caption{Recommendation utility, unlearning effectiveness, and unlearning efficiency in SBR for task-specific models on the Books dataset in the sensitive item unlearning scenario. Recommendation utility is measured by standard metrics. Unlearning efficiency is measured by $\text{Sensitive}\boldsymbol{@k}$. For efficiency, we report the average time needed to execute an unlearning request in minutes.}
\label{tab:results_amazonreviewsbooks_health}
\setlength\tabcolsep{5.5pt}
\begin{tabular}{ll  c c  c c  c c  c c  c}
\toprule
 &  &
\multicolumn{2}{c}{\textbf{nDCG (\%)}$\mathbf{\uparrow}$} &
\multicolumn{2}{c}{\textbf{Recall (\%)}$\mathbf{\uparrow}$} &
\multicolumn{2}{c}{\textbf{Hit (\%)}$\mathbf{\uparrow}$} &
\multicolumn{2}{c}{\textbf{Sensitive@}$\mathbf{20\downarrow}$} &
\textbf{Time (min) $\mathbf{\downarrow}$} \\
\cmidrule(r){3-4}
\cmidrule(r){5-6}
\cmidrule(r){7-8}
\cmidrule(r){9-10}
\cmidrule(r){11-11}
\textbf{Model} &
\textbf{Algorithm} &
\textbf{@10} & \textbf{@20} &
\textbf{@10} & \textbf{@20} &
\textbf{@10} & \textbf{@20} &
\textbf{@10} & \textbf{@20} &
\textbf{Avg/Req} \\
\midrule
\textbf{GRU4Rec} & \texttt{Retrain} & 2.53 \scriptsize{$\pm$ 0.08} & 3.07 \scriptsize{$\pm$ 0.10} & 4.77 \scriptsize{$\pm$ 0.15} & 6.90 \scriptsize{$\pm$ 0.22} & 4.77 \scriptsize{$\pm$ 0.15} & 6.90 \scriptsize{$\pm$ 0.22} & 58.60 \scriptsize{$\pm$ 2.69} & 77.62 \scriptsize{$\pm$ 1.71} & 428.05 \scriptsize{$\pm$ 55.45} \\
 & \texttt{Fanchuan} & 0.37 \scriptsize{$\pm$ 0.06} & 0.48 \scriptsize{$\pm$ 0.07} & 0.74 \scriptsize{$\pm$ 0.11} & 1.19 \scriptsize{$\pm$ 0.15} & 0.74 \scriptsize{$\pm$ 0.11} & 1.19 \scriptsize{$\pm$ 0.15} & 72.78 \scriptsize{$\pm$ 17.27} & 92.49 \scriptsize{$\pm$ 4.47} & 0.13 \scriptsize{$\pm$ 0.00} \\
 & \texttt{Kookmin} & \textbf{0.77} \scriptsize{$\pm$ 0.11} & \textbf{1.04} \scriptsize{$\pm$ 0.15} & \underline{1.58} \scriptsize{$\pm$ 0.23} & \underline{2.66} \scriptsize{$\pm$ 0.35} & \underline{1.58} \scriptsize{$\pm$ 0.23} & \underline{2.66} \scriptsize{$\pm$ 0.35} & 70.75 \scriptsize{$\pm$ 7.79} & 89.79 \scriptsize{$\pm$ 4.08} & \textbf{0.12} \scriptsize{$\pm$ 0.00} \\
 & \texttt{SCIF} & \underline{0.69} \scriptsize{$\pm$ 0.15} & \underline{0.99} \scriptsize{$\pm$ 0.20} & \textbf{1.61} \scriptsize{$\pm$ 0.30} & \textbf{2.81} \scriptsize{$\pm$ 0.49} & \textbf{1.61} \scriptsize{$\pm$ 0.30} & \textbf{2.81} \scriptsize{$\pm$ 0.49} & \underline{63.19} \scriptsize{$\pm$ 23.74} & \underline{82.90} \scriptsize{$\pm$ 12.70} & 0.15 \scriptsize{$\pm$ 0.00} \\
 & \texttt{Seif} & 0.00 \scriptsize{$\pm$ 0.00} & 0.00 \scriptsize{$\pm$ 0.00} & 0.00 \scriptsize{$\pm$ 0.00} & 0.01 \scriptsize{$\pm$ 0.00} & 0.00 \scriptsize{$\pm$ 0.00} & 0.01 \scriptsize{$\pm$ 0.00} & \textbf{59.02} \scriptsize{$\pm$ 35.49} & \textbf{79.33} \scriptsize{$\pm$ 29.60} & \underline{0.12} \scriptsize{$\pm$ 0.00} \\
\midrule
\textbf{NARM} & \texttt{Retrain} & 2.34 \scriptsize{$\pm$ 0.02} & 2.85 \scriptsize{$\pm$ 0.02} & 4.47 \scriptsize{$\pm$ 0.04} & 6.53 \scriptsize{$\pm$ 0.06} & 4.47 \scriptsize{$\pm$ 0.04} & 6.53 \scriptsize{$\pm$ 0.06} & 56.70 \scriptsize{$\pm$ 3.42} & 77.30 \scriptsize{$\pm$ 1.25} & 350.34 \scriptsize{$\pm$ 97.21} \\
 & \texttt{Fanchuan} & 0.51 \scriptsize{$\pm$ 0.03} & 0.65 \scriptsize{$\pm$ 0.04} & 1.01 \scriptsize{$\pm$ 0.06} & 1.59 \scriptsize{$\pm$ 0.09} & 1.01 \scriptsize{$\pm$ 0.06} & 1.59 \scriptsize{$\pm$ 0.09} & \underline{63.46} \scriptsize{$\pm$ 21.58} & \underline{82.89} \scriptsize{$\pm$ 14.81} & 0.13 \scriptsize{$\pm$ 0.00} \\
 & \texttt{Kookmin} & \underline{0.51} \scriptsize{$\pm$ 0.08} & \underline{0.70} \scriptsize{$\pm$ 0.10} & \underline{1.06} \scriptsize{$\pm$ 0.17} & \underline{1.83} \scriptsize{$\pm$ 0.29} & \underline{1.06} \scriptsize{$\pm$ 0.17} & \underline{1.83} \scriptsize{$\pm$ 0.29} & 79.93 \scriptsize{$\pm$ 9.23} & 94.55 \scriptsize{$\pm$ 3.37} & \underline{0.12} \scriptsize{$\pm$ 0.00} \\
 & \texttt{SCIF} & \textbf{0.55} \scriptsize{$\pm$ 0.15} & \textbf{0.85} \scriptsize{$\pm$ 0.16} & \textbf{1.32} \scriptsize{$\pm$ 0.28} & \textbf{2.51} \scriptsize{$\pm$ 0.34} & \textbf{1.32} \scriptsize{$\pm$ 0.28} & \textbf{2.51} \scriptsize{$\pm$ 0.34} & 66.64 \scriptsize{$\pm$ 18.71} & 83.02 \scriptsize{$\pm$ 9.73} & 0.15 \scriptsize{$\pm$ 0.00} \\
 & \texttt{Seif} & 0.00 \scriptsize{$\pm$ 0.00} & 0.00 \scriptsize{$\pm$ 0.00} & 0.00 \scriptsize{$\pm$ 0.00} & 0.00 \scriptsize{$\pm$ 0.01} & 0.00 \scriptsize{$\pm$ 0.00} & 0.00 \scriptsize{$\pm$ 0.01} & \textbf{37.14} \scriptsize{$\pm$ 19.02} & \textbf{68.34} \scriptsize{$\pm$ 15.77} & \textbf{0.12} \scriptsize{$\pm$ 0.00} \\
\midrule
\textbf{SASRec} & \texttt{Retrain} & 1.62 \scriptsize{$\pm$ 0.01} & 2.04 \scriptsize{$\pm$ 0.03} & 3.18 \scriptsize{$\pm$ 0.04} & 4.84 \scriptsize{$\pm$ 0.08} & 3.18 \scriptsize{$\pm$ 0.04} & 4.84 \scriptsize{$\pm$ 0.08} & 59.27 \scriptsize{$\pm$ 3.37} & 77.31 \scriptsize{$\pm$ 1.96} & 786.96 \scriptsize{$\pm$ 448.36} \\
 & \texttt{Fanchuan} & \textbf{0.28} \scriptsize{$\pm$ 0.03} & \textbf{0.36} \scriptsize{$\pm$ 0.04} & \textbf{0.55} \scriptsize{$\pm$ 0.06} & \textbf{0.88} \scriptsize{$\pm$ 0.10} & \textbf{0.55} \scriptsize{$\pm$ 0.06} & \textbf{0.88} \scriptsize{$\pm$ 0.10} & 45.77 \scriptsize{$\pm$ 22.73} & \underline{74.98} \scriptsize{$\pm$ 22.54} & 0.15 \scriptsize{$\pm$ 0.00} \\
 & \texttt{Kookmin} & 0.03 \scriptsize{$\pm$ 0.07} & 0.04 \scriptsize{$\pm$ 0.09} & 0.07 \scriptsize{$\pm$ 0.14} & 0.12 \scriptsize{$\pm$ 0.24} & 0.07 \scriptsize{$\pm$ 0.14} & 0.12 \scriptsize{$\pm$ 0.24} & \underline{41.79} \scriptsize{$\pm$ 53.26} & \textbf{60.00} \scriptsize{$\pm$ 54.77} & \underline{0.12} \scriptsize{$\pm$ 0.00} \\
 & \texttt{SCIF} & \underline{0.04} \scriptsize{$\pm$ 0.01} & \underline{0.08} \scriptsize{$\pm$ 0.03} & \underline{0.09} \scriptsize{$\pm$ 0.03} & \underline{0.28} \scriptsize{$\pm$ 0.08} & \underline{0.09} \scriptsize{$\pm$ 0.03} & \underline{0.28} \scriptsize{$\pm$ 0.08} & 56.97 \scriptsize{$\pm$ 11.98} & 85.20 \scriptsize{$\pm$ 6.24} & 0.13 \scriptsize{$\pm$ 0.00} \\
 & \texttt{Seif} & 0.00 \scriptsize{$\pm$ 0.00} & 0.00 \scriptsize{$\pm$ 0.00} & 0.00 \scriptsize{$\pm$ 0.00} & 0.01 \scriptsize{$\pm$ 0.00} & 0.00 \scriptsize{$\pm$ 0.00} & 0.01 \scriptsize{$\pm$ 0.00} & \textbf{40.00} \scriptsize{$\pm$ 54.77} & 80.00 \scriptsize{$\pm$ 44.72} & \textbf{0.12} \scriptsize{$\pm$ 0.00} \\
\midrule
\textbf{SRGNN} & \texttt{Retrain} & 3.56 \scriptsize{$\pm$ 0.02} & 4.12 \scriptsize{$\pm$ 0.02} & 6.06 \scriptsize{$\pm$ 0.02} & 8.29 \scriptsize{$\pm$ 0.04} & 6.06 \scriptsize{$\pm$ 0.02} & 8.29 \scriptsize{$\pm$ 0.04} & 56.49 \scriptsize{$\pm$ 1.87} & 78.02 \scriptsize{$\pm$ 0.93} & 763.09 \scriptsize{$\pm$ 73.15} \\
 & \texttt{CEU} & 0.01 \scriptsize{$\pm$ 0.01} & 0.01 \scriptsize{$\pm$ 0.01} & 0.01 \scriptsize{$\pm$ 0.01} & 0.03 \scriptsize{$\pm$ 0.02} & 0.01 \scriptsize{$\pm$ 0.01} & 0.03 \scriptsize{$\pm$ 0.02} & 83.50 \scriptsize{$\pm$ 36.77} & 99.87 \scriptsize{$\pm$ 0.21} & 0.18 \scriptsize{$\pm$ 0.01} \\
 & \texttt{Fanchuan} & \underline{0.12} \scriptsize{$\pm$ 0.02} & 0.17 \scriptsize{$\pm$ 0.03} & 0.26 \scriptsize{$\pm$ 0.06} & 0.45 \scriptsize{$\pm$ 0.11} & 0.26 \scriptsize{$\pm$ 0.06} & 0.45 \scriptsize{$\pm$ 0.11} & 61.68 \scriptsize{$\pm$ 27.69} & 97.33 \scriptsize{$\pm$ 3.40} & 0.27 \scriptsize{$\pm$ 0.00} \\
 & \texttt{IDEA} & 0.00 \scriptsize{$\pm$ 0.00} & 0.00 \scriptsize{$\pm$ 0.00} & 0.00 \scriptsize{$\pm$ 0.00} & 0.01 \scriptsize{$\pm$ 0.01} & 0.00 \scriptsize{$\pm$ 0.00} & 0.01 \scriptsize{$\pm$ 0.01} & \textbf{42.30} \scriptsize{$\pm$ 3.13} & \textbf{63.96} \scriptsize{$\pm$ 0.61} & \textbf{0.12} \scriptsize{$\pm$ 0.00} \\
 & \texttt{Kookmin} & \textbf{1.04} \scriptsize{$\pm$ 0.13} & \textbf{1.25} \scriptsize{$\pm$ 0.15} & \textbf{1.90} \scriptsize{$\pm$ 0.22} & \textbf{2.76} \scriptsize{$\pm$ 0.33} & \textbf{1.90} \scriptsize{$\pm$ 0.22} & \textbf{2.76} \scriptsize{$\pm$ 0.33} & \underline{50.42} \scriptsize{$\pm$ 6.42} & \underline{72.86} \scriptsize{$\pm$ 6.03} & \underline{0.13} \scriptsize{$\pm$ 0.01} \\
 & \texttt{SCIF} & 0.11 \scriptsize{$\pm$ 0.04} & \underline{0.22} \scriptsize{$\pm$ 0.06} & \underline{0.29} \scriptsize{$\pm$ 0.10} & \underline{0.72} \scriptsize{$\pm$ 0.22} & \underline{0.29} \scriptsize{$\pm$ 0.10} & \underline{0.72} \scriptsize{$\pm$ 0.22} & 58.65 \scriptsize{$\pm$ 23.70} & 83.21 \scriptsize{$\pm$ 10.02} & 0.19 \scriptsize{$\pm$ 0.01} \\
 & \texttt{Seif} & 0.00 \scriptsize{$\pm$ 0.00} & 0.00 \scriptsize{$\pm$ 0.00} & 0.00 \scriptsize{$\pm$ 0.00} & 0.01 \scriptsize{$\pm$ 0.00} & 0.00 \scriptsize{$\pm$ 0.00} & 0.01 \scriptsize{$\pm$ 0.00} & 64.26 \scriptsize{$\pm$ 49.02} & 80.29 \scriptsize{$\pm$ 43.65} & 0.14 \scriptsize{$\pm$ 0.00} \\
\bottomrule
\end{tabular}
\end{table*}

\begin{table*}
\centering
\small
\caption{Recommendation utility, unlearning effectiveness, and unlearning efficiency in SBR for task-specific models on the 30music dataset in the sensitive item unlearning scenario. Recommendation utility is measured by standard metrics. Unlearning efficiency is measured by $\text{Sensitive}\boldsymbol{@k}$. For efficiency, we report the average time needed to execute an unlearning request in minutes.}
\label{tab:results_30music_explicit}
\setlength\tabcolsep{4.8pt}
\begin{tabular}{ll  c c  c c  c c  c c  c}
\toprule
 &  &
\multicolumn{2}{c}{\textbf{nDCG (\%)}$\mathbf{\uparrow}$} &
\multicolumn{2}{c}{\textbf{Recall (\%)}$\mathbf{\uparrow}$} &
\multicolumn{2}{c}{\textbf{Hit (\%)}$\mathbf{\uparrow}$} &
\multicolumn{2}{c}{\textbf{Sensitive@}$\mathbf{20\downarrow}$} &
\textbf{Time (min) $\mathbf{\downarrow}$} \\
\cmidrule(r){3-4}
\cmidrule(r){5-6}
\cmidrule(r){7-8}
\cmidrule(r){9-10}
\cmidrule(r){11-11}
\textbf{Model} &
\textbf{Algorithm} &
\textbf{@10} & \textbf{@20} &
\textbf{@10} & \textbf{@20} &
\textbf{@10} & \textbf{@20} &
\textbf{@10} & \textbf{@20} &
\textbf{Avg/Req} \\
\midrule
\textbf{GRU4Rec} & \texttt{Retrain} & 14.57 \scriptsize{$\pm$ 0.28} & 15.29 \scriptsize{$\pm$ 0.28} & 19.98 \scriptsize{$\pm$ 0.35} & 22.83 \scriptsize{$\pm$ 0.39} & 19.98 \scriptsize{$\pm$ 0.35} & 22.83 \scriptsize{$\pm$ 0.39} & 1.84 \scriptsize{$\pm$ 1.18} & 4.04 \scriptsize{$\pm$ 0.96} & 932.98 \scriptsize{$\pm$ 266.80} \\
 & \texttt{Fanchuan} & \underline{1.13} \scriptsize{$\pm$ 1.10} & \underline{1.29} \scriptsize{$\pm$ 1.22} & \underline{1.94} \scriptsize{$\pm$ 1.84} & \underline{2.59} \scriptsize{$\pm$ 2.32} & \underline{1.94} \scriptsize{$\pm$ 1.84} & \underline{2.59} \scriptsize{$\pm$ 2.32} & \textbf{5.33} \scriptsize{$\pm$ 4.09} & \textbf{8.27} \scriptsize{$\pm$ 4.71} & 1.51 \scriptsize{$\pm$ 0.20} \\
 & \texttt{Kookmin} & 0.50 \scriptsize{$\pm$ 0.12} & 0.60 \scriptsize{$\pm$ 0.13} & 0.92 \scriptsize{$\pm$ 0.21} & 1.28 \scriptsize{$\pm$ 0.27} & 0.92 \scriptsize{$\pm$ 0.21} & 1.28 \scriptsize{$\pm$ 0.27} & \underline{9.10} \scriptsize{$\pm$ 3.53} & \underline{12.40} \scriptsize{$\pm$ 3.10} & \underline{1.07} \scriptsize{$\pm$ 0.40} \\
 & \texttt{SCIF} & \textbf{14.59} \scriptsize{$\pm$ 0.39} & \textbf{15.32} \scriptsize{$\pm$ 0.40} & \textbf{20.04} \scriptsize{$\pm$ 0.48} & \textbf{22.90} \scriptsize{$\pm$ 0.54} & \textbf{20.04} \scriptsize{$\pm$ 0.48} & \textbf{22.90} \scriptsize{$\pm$ 0.54} & \underline{9.10} \scriptsize{$\pm$ 3.53} & \underline{12.40} \scriptsize{$\pm$ 3.10} & \textbf{0.72} \scriptsize{$\pm$ 0.08} \\
 & \texttt{Seif} & 0.00 \scriptsize{$\pm$ 0.00} & 0.00 \scriptsize{$\pm$ 0.00} & 0.00 \scriptsize{$\pm$ 0.00} & 0.00 \scriptsize{$\pm$ 0.00} & 0.00 \scriptsize{$\pm$ 0.00} & 0.00 \scriptsize{$\pm$ 0.00} & \underline{9.10} \scriptsize{$\pm$ 3.53} & \underline{12.40} \scriptsize{$\pm$ 3.10} & 1.14 \scriptsize{$\pm$ 0.13} \\
\midrule
\textbf{NARM} & \texttt{Retrain} & 12.81 \scriptsize{$\pm$ 0.33} & 13.62 \scriptsize{$\pm$ 0.31} & 19.05 \scriptsize{$\pm$ 0.32} & 22.21 \scriptsize{$\pm$ 0.27} & 19.05 \scriptsize{$\pm$ 0.32} & 22.21 \scriptsize{$\pm$ 0.27} & 2.66 \scriptsize{$\pm$ 1.27} & 5.28 \scriptsize{$\pm$ 1.46} & 909.20 \scriptsize{$\pm$ 214.21} \\
 & \texttt{Fanchuan} & \underline{1.09} \scriptsize{$\pm$ 1.25} & \underline{1.24} \scriptsize{$\pm$ 1.40} & \underline{1.94} \scriptsize{$\pm$ 2.25} & \underline{2.54} \scriptsize{$\pm$ 2.85} & \underline{1.94} \scriptsize{$\pm$ 2.25} & \underline{2.54} \scriptsize{$\pm$ 2.85} & \underline{7.05} \scriptsize{$\pm$ 12.43} & \textbf{8.56} \scriptsize{$\pm$ 15.04} & 1.47 \scriptsize{$\pm$ 0.15} \\
 & \texttt{Kookmin} & 0.34 \scriptsize{$\pm$ 0.11} & 0.43 \scriptsize{$\pm$ 0.12} & 0.68 \scriptsize{$\pm$ 0.21} & 1.01 \scriptsize{$\pm$ 0.28} & 0.68 \scriptsize{$\pm$ 0.21} & 1.01 \scriptsize{$\pm$ 0.28} & 7.53 \scriptsize{$\pm$ 2.22} & 9.46 \scriptsize{$\pm$ 2.19} & 1.28 \scriptsize{$\pm$ 0.02} \\
 & \texttt{SCIF} & \textbf{12.65} \scriptsize{$\pm$ 0.27} & \textbf{13.45} \scriptsize{$\pm$ 0.25} & \textbf{18.87} \scriptsize{$\pm$ 0.18} & \textbf{22.05} \scriptsize{$\pm$ 0.17} & \textbf{18.87} \scriptsize{$\pm$ 0.18} & \textbf{22.05} \scriptsize{$\pm$ 0.17} & \textbf{7.04} \scriptsize{$\pm$ 1.93} & \underline{9.11} \scriptsize{$\pm$ 2.50} & \textbf{0.70} \scriptsize{$\pm$ 0.05} \\
 & \texttt{Seif} & 0.00 \scriptsize{$\pm$ 0.00} & 0.00 \scriptsize{$\pm$ 0.00} & 0.00 \scriptsize{$\pm$ 0.00} & 0.00 \scriptsize{$\pm$ 0.00} & 0.00 \scriptsize{$\pm$ 0.00} & 0.00 \scriptsize{$\pm$ 0.00} & 7.53 \scriptsize{$\pm$ 2.22} & 9.46 \scriptsize{$\pm$ 2.19} & \underline{1.21} \scriptsize{$\pm$ 0.11} \\
\midrule
\textbf{SASRec} & \texttt{Retrain} & 8.15 \scriptsize{$\pm$ 0.09} & 9.02 \scriptsize{$\pm$ 0.09} & 13.98 \scriptsize{$\pm$ 0.12} & 17.41 \scriptsize{$\pm$ 0.13} & 13.98 \scriptsize{$\pm$ 0.12} & 17.41 \scriptsize{$\pm$ 0.13} & 1.97 \scriptsize{$\pm$ 0.73} & 4.60 \scriptsize{$\pm$ 0.59} & 810.28 \scriptsize{$\pm$ 0.26} \\
 & \texttt{Fanchuan} & \underline{2.20} \scriptsize{$\pm$ 0.32} & \underline{2.56} \scriptsize{$\pm$ 0.36} & \underline{4.18} \scriptsize{$\pm$ 0.59} & \underline{5.58} \scriptsize{$\pm$ 0.77} & \underline{4.18} \scriptsize{$\pm$ 0.59} & \underline{5.58} \scriptsize{$\pm$ 0.77} & \textbf{6.34} \scriptsize{$\pm$ 3.46} & \textbf{7.95} \scriptsize{$\pm$ 3.74} & 1.80 \scriptsize{$\pm$ 0.21} \\
 & \texttt{Kookmin} & 0.06 \scriptsize{$\pm$ 0.05} & 0.09 \scriptsize{$\pm$ 0.07} & 0.13 \scriptsize{$\pm$ 0.11} & 0.23 \scriptsize{$\pm$ 0.18} & 0.13 \scriptsize{$\pm$ 0.11} & 0.23 \scriptsize{$\pm$ 0.18} & \underline{7.43} \scriptsize{$\pm$ 0.75} & \underline{9.81} \scriptsize{$\pm$ 0.46} & \textbf{0.70} \scriptsize{$\pm$ 0.50} \\
 & \texttt{SCIF} & \textbf{7.48} \scriptsize{$\pm$ 0.61} & \textbf{8.40} \scriptsize{$\pm$ 0.55} & \textbf{13.53} \scriptsize{$\pm$ 0.37} & \textbf{17.18} \scriptsize{$\pm$ 0.16} & \textbf{13.53} \scriptsize{$\pm$ 0.37} & \textbf{17.18} \scriptsize{$\pm$ 0.16} & 7.90 \scriptsize{$\pm$ 2.97} & 10.43 \scriptsize{$\pm$ 3.62} & \underline{0.83} \scriptsize{$\pm$ 0.08} \\
 & \texttt{Seif} & 0.00 \scriptsize{$\pm$ 0.00} & 0.00 \scriptsize{$\pm$ 0.00} & 0.00 \scriptsize{$\pm$ 0.00} & 0.00 \scriptsize{$\pm$ 0.01} & 0.00 \scriptsize{$\pm$ 0.00} & 0.00 \scriptsize{$\pm$ 0.01} & 8.73 \scriptsize{$\pm$ 2.67} & 11.44 \scriptsize{$\pm$ 3.28} & 1.21 \scriptsize{$\pm$ 0.07} \\
\midrule
\textbf{SRGNN} & \texttt{Retrain} & 16.30 \scriptsize{$\pm$ 0.12} & 16.87 \scriptsize{$\pm$ 0.12} & 20.78 \scriptsize{$\pm$ 0.08} & 23.06 \scriptsize{$\pm$ 0.08} & 20.78 \scriptsize{$\pm$ 0.08} & 23.06 \scriptsize{$\pm$ 0.08} & 0.74 \scriptsize{$\pm$ 0.85} & 3.22 \scriptsize{$\pm$ 0.96} & 1088.07 \scriptsize{$\pm$ 258.96} \\
 & \texttt{CEU} & 7.63 \scriptsize{$\pm$ 5.94} & 7.96 \scriptsize{$\pm$ 6.05} & 9.88 \scriptsize{$\pm$ 7.04} & 11.21 \scriptsize{$\pm$ 7.54} & 9.88 \scriptsize{$\pm$ 7.04} & 11.21 \scriptsize{$\pm$ 7.54} & \textbf{3.91} \scriptsize{$\pm$ 3.63} & \textbf{5.47} \scriptsize{$\pm$ 5.15} & \textbf{0.19} \scriptsize{$\pm$ 0.04} \\
 & \texttt{Fanchuan} & 5.27 \scriptsize{$\pm$ 0.55} & 5.53 \scriptsize{$\pm$ 0.56} & 6.97 \scriptsize{$\pm$ 0.65} & 8.02 \scriptsize{$\pm$ 0.68} & 6.97 \scriptsize{$\pm$ 0.65} & 8.02 \scriptsize{$\pm$ 0.68} & \underline{4.60} \scriptsize{$\pm$ 5.49} & \underline{6.76} \scriptsize{$\pm$ 7.01} & 4.31 \scriptsize{$\pm$ 0.43} \\
 & \texttt{Kookmin} & \underline{11.35} \scriptsize{$\pm$ 0.64} & \underline{11.82} \scriptsize{$\pm$ 0.67} & \underline{14.60} \scriptsize{$\pm$ 0.89} & \underline{16.46} \scriptsize{$\pm$ 1.05} & \underline{14.60} \scriptsize{$\pm$ 0.89} & \underline{16.46} \scriptsize{$\pm$ 1.05} & 11.67 \scriptsize{$\pm$ 4.65} & 16.63 \scriptsize{$\pm$ 5.89} & 1.21 \scriptsize{$\pm$ 0.03} \\
 & \texttt{SCIF} & \textbf{16.00} \scriptsize{$\pm$ 0.47} & \textbf{16.60} \scriptsize{$\pm$ 0.46} & \textbf{20.69} \scriptsize{$\pm$ 0.25} & \textbf{23.01} \scriptsize{$\pm$ 0.19} & \textbf{20.69} \scriptsize{$\pm$ 0.25} & \textbf{23.01} \scriptsize{$\pm$ 0.19} & 8.55 \scriptsize{$\pm$ 3.04} & 12.27 \scriptsize{$\pm$ 3.41} & \underline{1.00} \scriptsize{$\pm$ 0.05} \\
 & \texttt{Seif} & 0.00 \scriptsize{$\pm$ 0.00} & 0.00 \scriptsize{$\pm$ 0.00} & 0.00 \scriptsize{$\pm$ 0.00} & 0.00 \scriptsize{$\pm$ 0.00} & 0.00 \scriptsize{$\pm$ 0.00} & 0.00 \scriptsize{$\pm$ 0.00} & 8.73 \scriptsize{$\pm$ 3.08} & 12.77 \scriptsize{$\pm$ 3.81} & 2.53 \scriptsize{$\pm$ 0.06} \\
\bottomrule
\end{tabular}
\end{table*}

\begin{table*}
\centering
\small
\caption{Recommendation utility, unlearning effectiveness, and unlearning efficiency in CF for task-specific models on the MovieLens dataset in the sensitive item unlearning scenario. Recommendation utility is measured by standard metrics. Unlearning efficiency is measured by $\text{Sensitive}\boldsymbol{@k}$. For efficiency, we report the average time needed to execute an unlearning request in minutes.}
\label{tab:results_movielens_health}
\setlength\tabcolsep{5.2pt}
\begin{tabular}{ll  c c  c c  c c  c c  c}
\toprule
 &  &
\multicolumn{2}{c}{\textbf{nDCG (\%)}$\mathbf{\uparrow}$} &
\multicolumn{2}{c}{\textbf{Recall (\%)}$\mathbf{\uparrow}$} &
\multicolumn{2}{c}{\textbf{Hit (\%)}$\mathbf{\uparrow}$} &
\multicolumn{2}{c}{\textbf{Sensitive@}$\mathbf{20\downarrow}$} &
\textbf{Time (min) $\mathbf{\downarrow}$} \\
\cmidrule(r){3-4}
\cmidrule(r){5-6}
\cmidrule(r){7-8}
\cmidrule(r){9-10}
\cmidrule(r){11-11}
\textbf{Model} &
\textbf{Algorithm} &
\textbf{@10} & \textbf{@20} &
\textbf{@10} & \textbf{@20} &
\textbf{@10} & \textbf{@20} &
\textbf{@10} & \textbf{@20} &
\textbf{Avg/Req} \\
\midrule
\textbf{BPR} & \texttt{Retrain} & 8.06 \scriptsize{$\pm$ 0.13} & 9.54 \scriptsize{$\pm$ 0.13} & 7.33 \scriptsize{$\pm$ 0.09} & 12.63 \scriptsize{$\pm$ 0.11} & 35.19 \scriptsize{$\pm$ 0.23} & 49.36 \scriptsize{$\pm$ 0.17} & 0.03 \scriptsize{$\pm$ 0.04} & 0.06 \scriptsize{$\pm$ 0.07} & 110.69 \scriptsize{$\pm$ 28.53} \\
 & \texttt{Fanchuan} & 6.39 \scriptsize{$\pm$ 0.10} & 7.90 \scriptsize{$\pm$ 0.15} & 6.24 \scriptsize{$\pm$ 0.15} & 11.08 \scriptsize{$\pm$ 0.22} & 31.03 \scriptsize{$\pm$ 0.63} & \textbf{45.52} \scriptsize{$\pm$ 0.97} & 0.58 \scriptsize{$\pm$ 0.28} & 1.50 \scriptsize{$\pm$ 0.74} & 0.07 \scriptsize{$\pm$ 0.00} \\
 & \texttt{Kookmin} & \textbf{7.65} \scriptsize{$\pm$ 1.62} & \textbf{8.85} \scriptsize{$\pm$ 1.91} & \textbf{6.72} \scriptsize{$\pm$ 1.50} & \underline{11.25} \scriptsize{$\pm$ 2.53} & \textbf{32.12} \scriptsize{$\pm$ 6.08} & 44.51 \scriptsize{$\pm$ 8.18} & \textbf{0.00} \scriptsize{$\pm$ 0.00} & \textbf{0.00} \scriptsize{$\pm$ 0.00} & \textbf{0.05} \scriptsize{$\pm$ 0.01} \\
 & \texttt{SCIF} & \underline{7.34} \scriptsize{$\pm$ 1.68} & \underline{8.66} \scriptsize{$\pm$ 2.00} & \underline{6.66} \scriptsize{$\pm$ 1.56} & \textbf{11.44} \scriptsize{$\pm$ 2.70} & \underline{32.03} \scriptsize{$\pm$ 7.22} & \underline{44.83} \scriptsize{$\pm$ 10.03} & \underline{0.38} \scriptsize{$\pm$ 0.41} & \underline{0.98} \scriptsize{$\pm$ 0.99} & 0.12 \scriptsize{$\pm$ 0.00} \\
 & \texttt{Seif} & 0.16 \scriptsize{$\pm$ 0.03} & 0.20 \scriptsize{$\pm$ 0.03} & 0.10 \scriptsize{$\pm$ 0.03} & 0.22 \scriptsize{$\pm$ 0.03} & 1.34 \scriptsize{$\pm$ 0.14} & 2.63 \scriptsize{$\pm$ 0.14} & \textbf{0.00} \scriptsize{$\pm$ 0.00} & \textbf{0.00} \scriptsize{$\pm$ 0.00} & \underline{0.06} \scriptsize{$\pm$ 0.00} \\
\midrule
\textbf{LightGCN} & \texttt{Retrain} & 7.78 \scriptsize{$\pm$ 0.03} & 9.15 \scriptsize{$\pm$ 0.03} & 6.78 \scriptsize{$\pm$ 0.03} & 11.94 \scriptsize{$\pm$ 0.05} & 33.78 \scriptsize{$\pm$ 0.12} & 47.68 \scriptsize{$\pm$ 0.16} & 0.01 \scriptsize{$\pm$ 0.02} & 0.03 \scriptsize{$\pm$ 0.03} & 1344.00 \scriptsize{$\pm$ 80.98} \\
 & \texttt{CEU} & \underline{8.91} \scriptsize{$\pm$ 0.05} & \underline{10.50} \scriptsize{$\pm$ 0.04} & 7.93 \scriptsize{$\pm$ 0.06} & \underline{13.76} \scriptsize{$\pm$ 0.07} & \underline{37.48} \scriptsize{$\pm$ 0.17} & \underline{52.08} \scriptsize{$\pm$ 0.20} & 0.47 \scriptsize{$\pm$ 0.28} & 1.23 \scriptsize{$\pm$ 0.47} & \textbf{0.17} \scriptsize{$\pm$ 0.03} \\
 & \texttt{Fanchuan} & 7.48 \scriptsize{$\pm$ 0.16} & 8.94 \scriptsize{$\pm$ 0.22} & 6.76 \scriptsize{$\pm$ 0.16} & 11.85 \scriptsize{$\pm$ 0.35} & 33.76 \scriptsize{$\pm$ 0.51} & 48.05 \scriptsize{$\pm$ 0.74} & 4.60 \scriptsize{$\pm$ 0.34} & 5.72 \scriptsize{$\pm$ 0.46} & 0.87 \scriptsize{$\pm$ 0.26} \\
 & \texttt{GIF} & 7.30 \scriptsize{$\pm$ 0.50} & 9.07 \scriptsize{$\pm$ 0.45} & 7.13 \scriptsize{$\pm$ 0.32} & 12.75 \scriptsize{$\pm$ 0.34} & 34.75 \scriptsize{$\pm$ 0.92} & 49.84 \scriptsize{$\pm$ 0.75} & \underline{0.08} \scriptsize{$\pm$ 0.13} & \underline{0.21} \scriptsize{$\pm$ 0.25} & 2.89 \scriptsize{$\pm$ 0.78} \\
 & \texttt{Kookmin} & 8.89 \scriptsize{$\pm$ 0.15} & 10.47 \scriptsize{$\pm$ 0.17} & \underline{7.95} \scriptsize{$\pm$ 0.17} & 13.72 \scriptsize{$\pm$ 0.25} & 37.28 \scriptsize{$\pm$ 0.34} & 51.65 \scriptsize{$\pm$ 0.37} & 0.11 \scriptsize{$\pm$ 0.07} & 0.44 \scriptsize{$\pm$ 0.25} & 0.92 \scriptsize{$\pm$ 0.30} \\
 & \texttt{SCIF} & \textbf{8.99} \scriptsize{$\pm$ 0.05} & \textbf{10.58} \scriptsize{$\pm$ 0.04} & \textbf{8.00} \scriptsize{$\pm$ 0.05} & \textbf{13.84} \scriptsize{$\pm$ 0.06} & \textbf{37.68} \scriptsize{$\pm$ 0.15} & \textbf{52.20} \scriptsize{$\pm$ 0.19} & 0.45 \scriptsize{$\pm$ 0.23} & 1.21 \scriptsize{$\pm$ 0.31} & \underline{0.18} \scriptsize{$\pm$ 0.00} \\
 & \texttt{Seif} & 0.76 \scriptsize{$\pm$ 0.39} & 0.94 \scriptsize{$\pm$ 0.48} & 0.61 \scriptsize{$\pm$ 0.28} & 1.21 \scriptsize{$\pm$ 0.55} & 4.58 \scriptsize{$\pm$ 2.20} & 8.49 \scriptsize{$\pm$ 4.00} & \textbf{0.00} \scriptsize{$\pm$ 0.00} & \textbf{0.03} \scriptsize{$\pm$ 0.07} & 5.53 \scriptsize{$\pm$ 5.75} \\
\midrule
\textbf{SGL} & \texttt{Retrain} & 6.93 \scriptsize{$\pm$ 0.04} & 8.31 \scriptsize{$\pm$ 0.03} & 6.18 \scriptsize{$\pm$ 0.04} & 11.02 \scriptsize{$\pm$ 0.02} & 31.86 \scriptsize{$\pm$ 0.13} & 45.70 \scriptsize{$\pm$ 0.12} & 0.00 \scriptsize{$\pm$ 0.00} & 0.00 \scriptsize{$\pm$ 0.00} & 1370.81 \scriptsize{$\pm$ 32.45} \\
 & \texttt{CEU} & 5.58 \scriptsize{$\pm$ 0.11} & 6.35 \scriptsize{$\pm$ 0.15} & 4.69 \scriptsize{$\pm$ 0.14} & 7.76 \scriptsize{$\pm$ 0.25} & 25.12 \scriptsize{$\pm$ 0.18} & 35.46 \scriptsize{$\pm$ 0.15} & \textbf{0.00} \scriptsize{$\pm$ 0.00} & \textbf{0.00} \scriptsize{$\pm$ 0.00} & \textbf{0.52} \scriptsize{$\pm$ 0.11} \\
 & \texttt{Kookmin} & \textbf{9.09} \scriptsize{$\pm$ 0.05} & \textbf{10.73} \scriptsize{$\pm$ 0.07} & \textbf{8.17} \scriptsize{$\pm$ 0.09} & \textbf{13.96} \scriptsize{$\pm$ 0.14} & \textbf{39.36} \scriptsize{$\pm$ 0.14} & \textbf{53.96} \scriptsize{$\pm$ 0.23} & \underline{0.34} \scriptsize{$\pm$ 0.10} & 0.70 \scriptsize{$\pm$ 0.17} & \underline{2.99} \scriptsize{$\pm$ 0.85} \\
 & \texttt{SCIF} & \underline{7.97} \scriptsize{$\pm$ 0.07} & \underline{9.56} \scriptsize{$\pm$ 0.08} & \underline{7.37} \scriptsize{$\pm$ 0.08} & \underline{12.83} \scriptsize{$\pm$ 0.11} & \underline{35.54} \scriptsize{$\pm$ 0.18} & \underline{49.79} \scriptsize{$\pm$ 0.28} & \textbf{0.00} \scriptsize{$\pm$ 0.00} & \underline{0.02} \scriptsize{$\pm$ 0.03} & 5.20 \scriptsize{$\pm$ 2.48} \\
 & \texttt{Seif} & 0.10 \scriptsize{$\pm$ 0.06} & 0.16 \scriptsize{$\pm$ 0.09} & 0.04 \scriptsize{$\pm$ 0.02} & 0.14 \scriptsize{$\pm$ 0.08} & 0.80 \scriptsize{$\pm$ 0.69} & 2.30 \scriptsize{$\pm$ 1.58} & \textbf{0.00} \scriptsize{$\pm$ 0.00} & \textbf{0.00} \scriptsize{$\pm$ 0.00} & 5.04 \scriptsize{$\pm$ 1.29} \\
\bottomrule
\end{tabular}
\end{table*}

\begin{table*}[t]
\centering
\small
\caption{Recommendation utility, unlearning effectiveness, and unlearning efficiency in CF for task-specific models on the Food dataset in the sensitive item unlearning scenario. Recommendation utility is measured by standard metrics. Unlearning efficiency is measured by $\text{Sensitive}\boldsymbol{@k}$. For efficiency, we report the average time needed to execute an unlearning request in minutes.}
\label{tab:results_amazonreviewsgroceryandgourmetfood_alcohol}
\setlength\tabcolsep{5.4pt}
\begin{tabular}{ll  c c  c c  c c  c c  c}
\toprule
 &  &
\multicolumn{2}{c}{\textbf{nDCG (\%)}$\mathbf{\uparrow}$} &
\multicolumn{2}{c}{\textbf{Recall (\%)}$\mathbf{\uparrow}$} &
\multicolumn{2}{c}{\textbf{Hit (\%)}$\mathbf{\uparrow}$} &
\multicolumn{2}{c}{\textbf{Sensitive@}$\mathbf{20\downarrow}$} &
\textbf{Time (min) $\mathbf{\downarrow}$} \\
\cmidrule(r){3-4}
\cmidrule(r){5-6}
\cmidrule(r){7-8}
\cmidrule(r){9-10}
\cmidrule(r){11-11}
\textbf{Model} &
\textbf{Algorithm} &
\textbf{@10} & \textbf{@20} &
\textbf{@10} & \textbf{@20} &
\textbf{@10} & \textbf{@20} &
\textbf{@10} & \textbf{@20} &
\textbf{Avg/Req} \\
\midrule
\textbf{BPR} & \texttt{Retrain} & 0.43 \scriptsize{$\pm$ 0.01} & 0.54 \scriptsize{$\pm$ 0.01} & 0.80 \scriptsize{$\pm$ 0.02} & 1.27 \scriptsize{$\pm$ 0.03} & 0.88 \scriptsize{$\pm$ 0.02} & 1.39 \scriptsize{$\pm$ 0.03} & 34.20 \scriptsize{$\pm$ 8.15} & 50.23 \scriptsize{$\pm$ 7.05} & 147.02 \scriptsize{$\pm$ 36.90} \\
 & \texttt{Fanchuan} & 0.39 \scriptsize{$\pm$ 0.01} & 0.50 \scriptsize{$\pm$ 0.01} & 0.73 \scriptsize{$\pm$ 0.03} & 1.16 \scriptsize{$\pm$ 0.03} & 0.80 \scriptsize{$\pm$ 0.03} & 1.26 \scriptsize{$\pm$ 0.03} & 39.96 \scriptsize{$\pm$ 5.36} & 56.57 \scriptsize{$\pm$ 4.89} & 0.03 \scriptsize{$\pm$ 0.00} \\
 & \texttt{Kookmin} & \underline{0.42} \scriptsize{$\pm$ 0.01} & \underline{0.54} \scriptsize{$\pm$ 0.01} & \underline{0.80} \scriptsize{$\pm$ 0.01} & \underline{1.26} \scriptsize{$\pm$ 0.03} & \underline{0.87} \scriptsize{$\pm$ 0.01} & \underline{1.38} \scriptsize{$\pm$ 0.03} & \underline{34.01} \scriptsize{$\pm$ 8.00} & \textbf{50.17} \scriptsize{$\pm$ 6.83} & \underline{0.02} \scriptsize{$\pm$ 0.00} \\
 & \texttt{SCIF} & \textbf{0.43} \scriptsize{$\pm$ 0.01} & \textbf{0.54} \scriptsize{$\pm$ 0.01} & \textbf{0.80} \scriptsize{$\pm$ 0.02} & \textbf{1.27} \scriptsize{$\pm$ 0.03} & \textbf{0.88} \scriptsize{$\pm$ 0.02} & \textbf{1.39} \scriptsize{$\pm$ 0.03} & 34.20 \scriptsize{$\pm$ 8.15} & \underline{50.23} \scriptsize{$\pm$ 7.05} & 0.04 \scriptsize{$\pm$ 0.00} \\
 & \texttt{Seif} & 0.00 \scriptsize{$\pm$ 0.00} & 0.01 \scriptsize{$\pm$ 0.00} & 0.01 \scriptsize{$\pm$ 0.00} & 0.02 \scriptsize{$\pm$ 0.01} & 0.01 \scriptsize{$\pm$ 0.00} & 0.02 \scriptsize{$\pm$ 0.00} & \textbf{33.46} \scriptsize{$\pm$ 6.70} & 51.83 \scriptsize{$\pm$ 8.18} & \textbf{0.02} \scriptsize{$\pm$ 0.00} \\
\midrule
\textbf{LightGCN} & \texttt{Retrain} & 0.70 \scriptsize{$\pm$ 0.01} & 0.89 \scriptsize{$\pm$ 0.01} & 1.31 \scriptsize{$\pm$ 0.01} & 2.04 \scriptsize{$\pm$ 0.02} & 1.41 \scriptsize{$\pm$ 0.02} & 2.20 \scriptsize{$\pm$ 0.02} & 37.27 \scriptsize{$\pm$ 4.50} & 55.59 \scriptsize{$\pm$ 5.89} & 346.69 \scriptsize{$\pm$ 38.29} \\
 & \texttt{CEU} & \textbf{1.16} \scriptsize{$\pm$ 0.03} & \underline{1.46} \scriptsize{$\pm$ 0.03} & \underline{2.17} \scriptsize{$\pm$ 0.04} & 3.39 \scriptsize{$\pm$ 0.05} & 2.31 \scriptsize{$\pm$ 0.05} & 3.62 \scriptsize{$\pm$ 0.06} & 33.27 \scriptsize{$\pm$ 7.01} & 51.69 \scriptsize{$\pm$ 4.75} & 0.05 \scriptsize{$\pm$ 0.01} \\
 & \texttt{Fanchuan} & 1.08 \scriptsize{$\pm$ 0.01} & 1.35 \scriptsize{$\pm$ 0.01} & 2.00 \scriptsize{$\pm$ 0.02} & 3.09 \scriptsize{$\pm$ 0.02} & 2.12 \scriptsize{$\pm$ 0.01} & 3.28 \scriptsize{$\pm$ 0.02} & 44.76 \scriptsize{$\pm$ 4.71} & 60.01 \scriptsize{$\pm$ 2.93} & 0.05 \scriptsize{$\pm$ 0.00} \\
 & \texttt{GIF} & 1.10 \scriptsize{$\pm$ 0.02} & 1.42 \scriptsize{$\pm$ 0.03} & 2.11 \scriptsize{$\pm$ 0.06} & 3.36 \scriptsize{$\pm$ 0.07} & 2.24 \scriptsize{$\pm$ 0.05} & 3.58 \scriptsize{$\pm$ 0.07} & \underline{30.49} \scriptsize{$\pm$ 2.90} & \underline{48.58} \scriptsize{$\pm$ 4.30} & 0.87 \scriptsize{$\pm$ 0.18} \\
 & \texttt{IDEA} & 0.54 \scriptsize{$\pm$ 0.01} & 0.65 \scriptsize{$\pm$ 0.01} & 0.93 \scriptsize{$\pm$ 0.01} & 1.35 \scriptsize{$\pm$ 0.01} & 0.94 \scriptsize{$\pm$ 0.01} & 1.37 \scriptsize{$\pm$ 0.02} & 31.12 \scriptsize{$\pm$ 3.60} & 50.34 \scriptsize{$\pm$ 3.76} & \underline{0.04} \scriptsize{$\pm$ 0.01} \\
 & \texttt{Kookmin} & 1.15 \scriptsize{$\pm$ 0.04} & 1.46 \scriptsize{$\pm$ 0.04} & 2.17 \scriptsize{$\pm$ 0.06} & \underline{3.40} \scriptsize{$\pm$ 0.09} & \underline{2.31} \scriptsize{$\pm$ 0.06} & \underline{3.62} \scriptsize{$\pm$ 0.08} & 32.94 \scriptsize{$\pm$ 5.18} & 49.53 \scriptsize{$\pm$ 5.93} & \textbf{0.03} \scriptsize{$\pm$ 0.00} \\
 & \texttt{SCIF} & \underline{1.15} \scriptsize{$\pm$ 0.04} & \textbf{1.46} \scriptsize{$\pm$ 0.04} & \textbf{2.17} \scriptsize{$\pm$ 0.06} & \textbf{3.40} \scriptsize{$\pm$ 0.08} & \textbf{2.31} \scriptsize{$\pm$ 0.06} & \textbf{3.62} \scriptsize{$\pm$ 0.09} & 33.47 \scriptsize{$\pm$ 5.07} & 50.06 \scriptsize{$\pm$ 6.14} & 0.22 \scriptsize{$\pm$ 0.01} \\
 & \texttt{Seif} & 0.27 \scriptsize{$\pm$ 0.03} & 0.40 \scriptsize{$\pm$ 0.03} & 0.62 \scriptsize{$\pm$ 0.07} & 1.13 \scriptsize{$\pm$ 0.10} & 0.67 \scriptsize{$\pm$ 0.07} & 1.21 \scriptsize{$\pm$ 0.11} & \textbf{20.39} \scriptsize{$\pm$ 5.36} & \textbf{38.14} \scriptsize{$\pm$ 6.38} & 0.11 \scriptsize{$\pm$ 0.01} \\
\midrule
\textbf{SGL} & \texttt{Retrain} & 0.87 \scriptsize{$\pm$ 0.01} & 1.08 \scriptsize{$\pm$ 0.01} & 1.61 \scriptsize{$\pm$ 0.01} & 2.45 \scriptsize{$\pm$ 0.02} & 1.73 \scriptsize{$\pm$ 0.01} & 2.64 \scriptsize{$\pm$ 0.01} & 32.50 \scriptsize{$\pm$ 2.99} & 49.79 \scriptsize{$\pm$ 2.82} & 241.35 \scriptsize{$\pm$ 21.70} \\
 & \texttt{CEU} & 0.10 \scriptsize{$\pm$ 0.05} & 0.14 \scriptsize{$\pm$ 0.05} & 0.21 \scriptsize{$\pm$ 0.09} & 0.37 \scriptsize{$\pm$ 0.12} & 0.23 \scriptsize{$\pm$ 0.09} & 0.40 \scriptsize{$\pm$ 0.13} & \underline{28.28} \scriptsize{$\pm$ 19.86} & \underline{46.09} \scriptsize{$\pm$ 27.57} & 3.48 \scriptsize{$\pm$ 0.87} \\
 & \texttt{Fanchuan} & \textbf{1.72} \scriptsize{$\pm$ 0.01} & \textbf{2.18} \scriptsize{$\pm$ 0.01} & \textbf{3.22} \scriptsize{$\pm$ 0.01} & \textbf{5.01} \scriptsize{$\pm$ 0.02} & \textbf{3.38} \scriptsize{$\pm$ 0.01} & \textbf{5.26} \scriptsize{$\pm$ 0.02} & 33.51 \scriptsize{$\pm$ 2.66} & 49.59 \scriptsize{$\pm$ 3.73} & 0.41 \scriptsize{$\pm$ 0.01} \\
 & \texttt{GIF} & 0.10 \scriptsize{$\pm$ 0.07} & 0.13 \scriptsize{$\pm$ 0.08} & 0.20 \scriptsize{$\pm$ 0.12} & 0.35 \scriptsize{$\pm$ 0.18} & 0.21 \scriptsize{$\pm$ 0.14} & 0.38 \scriptsize{$\pm$ 0.19} & \textbf{14.78} \scriptsize{$\pm$ 17.04} & \textbf{31.46} \scriptsize{$\pm$ 28.48} & 3.27 \scriptsize{$\pm$ 0.83} \\
 & \texttt{Kookmin} & \underline{1.46} \scriptsize{$\pm$ 0.02} & \underline{1.83} \scriptsize{$\pm$ 0.03} & \underline{2.72} \scriptsize{$\pm$ 0.03} & \underline{4.20} \scriptsize{$\pm$ 0.06} & \underline{2.88} \scriptsize{$\pm$ 0.04} & \underline{4.43} \scriptsize{$\pm$ 0.06} & 29.54 \scriptsize{$\pm$ 2.19} & 48.41 \scriptsize{$\pm$ 1.93} & \textbf{0.13} \scriptsize{$\pm$ 0.01} \\
 & \texttt{SCIF} & 1.44 \scriptsize{$\pm$ 0.03} & 1.81 \scriptsize{$\pm$ 0.03} & 2.69 \scriptsize{$\pm$ 0.04} & 4.15 \scriptsize{$\pm$ 0.06} & 2.84 \scriptsize{$\pm$ 0.04} & 4.39 \scriptsize{$\pm$ 0.06} & 29.23 \scriptsize{$\pm$ 3.12} & 46.19 \scriptsize{$\pm$ 3.57} & 0.57 \scriptsize{$\pm$ 0.12} \\
 & \texttt{Seif} & 0.26 \scriptsize{$\pm$ 0.03} & 0.36 \scriptsize{$\pm$ 0.05} & 0.57 \scriptsize{$\pm$ 0.09} & 0.95 \scriptsize{$\pm$ 0.16} & 0.61 \scriptsize{$\pm$ 0.10} & 1.02 \scriptsize{$\pm$ 0.16} & 31.96 \scriptsize{$\pm$ 11.54} & 53.11 \scriptsize{$\pm$ 13.64} & \underline{0.28} \scriptsize{$\pm$ 0.03} \\
\bottomrule
\end{tabular}
\end{table*}

\begin{table*}
\centering
\small
\caption{Recommendation utility, unlearning effectiveness, and unlearning efficiency in NBR for task-specific models on the TaFeng dataset in the spam removal scenario. Recommendation utility and unlearning effectiveness are measured by standard utility metrics. For efficiency, we report the average time needed to execute an unlearning request in minutes.}
\label{tab:results_tafeng}
\setlength\tabcolsep{9.8pt}
\begin{tabular}{ll  cc cc cc c}
\toprule
 &
 &
\multicolumn{2}{c}{\textbf{nDCG (\%)}$\mathbf{\uparrow}$} &
\multicolumn{2}{c}{\textbf{Recall (\%)}$\mathbf{\uparrow}$} &
\multicolumn{2}{c}{\textbf{Hit (\%)}$\mathbf{\uparrow}$} &
\textbf{Time (min) $\mathbf{\downarrow}$} \\
\cmidrule(r){3-4}
\cmidrule(r){5-6}
\cmidrule(r){7-8}
\cmidrule(r){9-9}
\textbf{Model} & \textbf{Algorithm} &
\textbf{@10} & \textbf{@20} &
\textbf{@10} & \textbf{@20} &
\textbf{@10} & \textbf{@20} &
\textbf{Avg/Req} \\
\midrule
\textbf{DNNTSP} & \texttt{Original} & 10.73 \scriptsize{$\pm$ 0.21} & 12.13 \scriptsize{$\pm$ 0.18} & 13.52 \scriptsize{$\pm$ 0.14} & 17.96 \scriptsize{$\pm$ 0.10} & 41.34 \scriptsize{$\pm$ 0.44} & 51.47 \scriptsize{$\pm$ 0.32} & 521.38 \scriptsize{$\pm$ 37.88} \\
 & \texttt{Retrain} & 10.79 \scriptsize{$\pm$ 0.23} & 12.22 \scriptsize{$\pm$ 0.23} & 13.54 \scriptsize{$\pm$ 0.05} & 18.09 \scriptsize{$\pm$ 0.09} & 41.49 \scriptsize{$\pm$ 0.24} & 51.78 \scriptsize{$\pm$ 0.21} & 504.67 \scriptsize{$\pm$ 4.5} \\
 & \texttt{CEU} & \textbf{10.71} \scriptsize{$\pm$ 0.22} & \textbf{12.14} \scriptsize{$\pm$ 0.21} & \textbf{13.51} \scriptsize{$\pm$ 0.14} & \textbf{18.03} \scriptsize{$\pm$ 0.18} & \textbf{41.17} \scriptsize{$\pm$ 0.44} & \underline{51.44} \scriptsize{$\pm$ 0.45} & \underline{0.64} \scriptsize{$\pm$ 0.14} \\
 & \texttt{Fanchuan} & 9.70 \scriptsize{$\pm$ 0.39} & 11.36 \scriptsize{$\pm$ 0.33} & 12.74 \scriptsize{$\pm$ 0.10} & 17.78 \scriptsize{$\pm$ 0.12} & 39.54 \scriptsize{$\pm$ 0.25} & 51.11 \scriptsize{$\pm$ 0.36} & 7.69 \scriptsize{$\pm$ 0.52} \\
 & \texttt{GIF} & 10.47 \scriptsize{$\pm$ 0.64} & 11.78 \scriptsize{$\pm$ 0.50} & 13.11 \scriptsize{$\pm$ 0.43} & 17.36 \scriptsize{$\pm$ 0.38} & 40.16 \scriptsize{$\pm$ 0.86} & 49.95 \scriptsize{$\pm$ 0.80} & 15.17 \scriptsize{$\pm$ 1.18} \\
 & \texttt{IDEA} & \underline{10.70} \scriptsize{$\pm$ 0.22} & \underline{12.14} \scriptsize{$\pm$ 0.21} & \underline{13.50} \scriptsize{$\pm$ 0.12} & \underline{18.03} \scriptsize{$\pm$ 0.13} & \underline{41.12} \scriptsize{$\pm$ 0.50} & \underline{51.44} \scriptsize{$\pm$ 0.38} & \textbf{0.22} \scriptsize{$\pm$ 0.05} \\
 & \texttt{Kookmin} & 10.58 \scriptsize{$\pm$ 0.19} & 11.97 \scriptsize{$\pm$ 0.27} & 13.25 \scriptsize{$\pm$ 0.38} & 17.67 \scriptsize{$\pm$ 0.55} & 40.57 \scriptsize{$\pm$ 1.08} & 50.69 \scriptsize{$\pm$ 1.26} & 1.29 \scriptsize{$\pm$ 0.09} \\
 & \texttt{SCIF} & 10.64 \scriptsize{$\pm$ 0.23} & 12.08 \scriptsize{$\pm$ 0.20} & 13.48 \scriptsize{$\pm$ 0.15} & 18.02 \scriptsize{$\pm$ 0.12} & 41.10 \scriptsize{$\pm$ 0.48} & \textbf{51.44} \scriptsize{$\pm$ 0.37} & 0.74 \scriptsize{$\pm$ 0.09} \\
 & \texttt{Seif} & 9.43 \scriptsize{$\pm$ 1.18} & 10.80 \scriptsize{$\pm$ 0.95} & 11.68 \scriptsize{$\pm$ 0.92} & 16.01 \scriptsize{$\pm$ 0.73} & 37.37 \scriptsize{$\pm$ 2.19} & 47.60 \scriptsize{$\pm$ 1.15} & 1.22 \scriptsize{$\pm$ 0.1} \\
\midrule
\textbf{Sets2Sets} & \texttt{Original} & 8.63 \scriptsize{$\pm$ 0.32} & 9.44 \scriptsize{$\pm$ 0.43} & 9.44 \scriptsize{$\pm$ 0.41} & 12.52 \scriptsize{$\pm$ 0.81} & 32.82 \scriptsize{$\pm$ 1.42} & 40.85 \scriptsize{$\pm$ 1.97} & 213.74 \scriptsize{$\pm$ 31.83} \\
 & \texttt{Retrain} & 8.86 \scriptsize{$\pm$ 0.27} & 9.68 \scriptsize{$\pm$ 0.38} & 9.68 \scriptsize{$\pm$ 0.36} & 12.80 \scriptsize{$\pm$ 0.75} & 33.57 \scriptsize{$\pm$ 0.89} & 41.45 \scriptsize{$\pm$ 1.69} & 227.7 \scriptsize{$\pm$ 28.41} \\
 & \texttt{Fanchuan} & 7.32 \scriptsize{$\pm$ 0.97} & 7.87 \scriptsize{$\pm$ 0.98} & 7.65 \scriptsize{$\pm$ 1.14} & 9.88 \scriptsize{$\pm$ 1.25} & 28.71 \scriptsize{$\pm$ 2.03} & 35.41 \scriptsize{$\pm$ 2.23} & 0.57 \scriptsize{$\pm$ 0.06} \\
 & \texttt{Kookmin} & \underline{8.71} \scriptsize{$\pm$ 0.36} & \underline{9.53} \scriptsize{$\pm$ 0.47} & \textbf{9.52} \scriptsize{$\pm$ 0.45} & \underline{12.63} \scriptsize{$\pm$ 0.88} & \underline{33.05} \scriptsize{$\pm$ 1.52} & \textbf{41.15} \scriptsize{$\pm$ 2.11} & \underline{0.17} \scriptsize{$\pm$ 0.02} \\
 & \texttt{SCIF} & 8.69 \scriptsize{$\pm$ 0.35} & 9.51 \scriptsize{$\pm$ 0.46} & \underline{9.52} \scriptsize{$\pm$ 0.44} & 12.63 \scriptsize{$\pm$ 0.87} & \textbf{33.05} \scriptsize{$\pm$ 1.53} & \underline{41.14} \scriptsize{$\pm$ 2.10} & 0.51 \scriptsize{$\pm$ 0.06} \\
 & \texttt{Seif} & \textbf{8.77} \scriptsize{$\pm$ 0.42} & \textbf{9.61} \scriptsize{$\pm$ 0.57} & 9.49 \scriptsize{$\pm$ 0.51} & \textbf{12.65} \scriptsize{$\pm$ 1.03} & 32.97 \scriptsize{$\pm$ 1.62} & 41.11 \scriptsize{$\pm$ 2.38} & \textbf{0.16} \scriptsize{$\pm$ 0.02} \\
\bottomrule
\end{tabular}
\end{table*}

\begin{table*}
\centering
\small
\caption{Recommendation utility, unlearning effectiveness, and unlearning efficiency in NBR for task-specific models on the Dunnhumby dataset in the spam removal scenario. Recommendation utility and unlearning effectiveness are measured by standard utility metrics. For efficiency, we report the average time needed to execute an unlearning request in minutes.}
\label{tab:results_dunnhumby}
\setlength\tabcolsep{10pt}
\begin{tabular}{ll  cc  cc  cc  c}
\toprule
 &  &
\multicolumn{2}{c}{\textbf{nDCG (\%)}$\mathbf{\uparrow}$} &
\multicolumn{2}{c}{\textbf{Recall (\%)}$\mathbf{\uparrow}$} &
\multicolumn{2}{c}{\textbf{Hit (\%)}$\mathbf{\uparrow}$} &
\textbf{Time (min) $\mathbf{\downarrow}$} \\
\cmidrule(r){3-4}
\cmidrule(r){5-6}
\cmidrule(r){7-8}
\cmidrule(r){9-9}
\textbf{Model} &
\textbf{Algorithm} &
\textbf{@10} & \textbf{@20} &
\textbf{@10} & \textbf{@20} &
\textbf{@10} & \textbf{@20} &
\textbf{Avg/Req} \\
\midrule
\textbf{DNNTSP} & \texttt{Original} & 6.79 \scriptsize{$\pm$ 0.25} & 6.79 \scriptsize{$\pm$ 0.35} & 6.90 \scriptsize{$\pm$ 0.17} & 8.52 \scriptsize{$\pm$ 0.45} & 29.09 \scriptsize{$\pm$ 0.75} & 34.75 \scriptsize{$\pm$ 1.61} & 102.18 \scriptsize{$\pm$ 5.86} \\
 & \texttt{Retrain} & 6.65 \scriptsize{$\pm$ 0.65} & 6.70 \scriptsize{$\pm$ 0.67} & 6.98 \scriptsize{$\pm$ 0.17} & 8.59 \scriptsize{$\pm$ 0.46} & 29.46 \scriptsize{$\pm$ 0.56} & 35.40 \scriptsize{$\pm$ 1.51} & 100.22 \scriptsize{$\pm$ 2.39} \\
 & \texttt{CEU} & 6.55 \scriptsize{$\pm$ 0.53} & 6.64 \scriptsize{$\pm$ 0.53} & 6.93 \scriptsize{$\pm$ 0.24} & 8.63 \scriptsize{$\pm$ 0.54} & 29.11 \scriptsize{$\pm$ 0.99} & 35.12 \scriptsize{$\pm$ 1.96} & 0.49 \scriptsize{$\pm$ 0.13} \\
 & \texttt{Fanchuan} & 6.04 \scriptsize{$\pm$ 0.22} & 6.19 \scriptsize{$\pm$ 0.18} & \textbf{7.04} \scriptsize{$\pm$ 0.20} & \textbf{8.66} \scriptsize{$\pm$ 0.46} & \textbf{29.58} \scriptsize{$\pm$ 0.67} & \textbf{35.38} \scriptsize{$\pm$ 1.57} & 1.49 \scriptsize{$\pm$ 0.07} \\
 & \texttt{GIF} & \underline{6.69} \scriptsize{$\pm$ 0.35} & \underline{6.73} \scriptsize{$\pm$ 0.42} & \underline{6.97} \scriptsize{$\pm$ 0.18} & \underline{8.64} \scriptsize{$\pm$ 0.48} & \underline{29.38} \scriptsize{$\pm$ 0.81} & \underline{35.22} \scriptsize{$\pm$ 1.65} & 9.3 \scriptsize{$\pm$ 0.5} \\
 & \texttt{IDEA} & 5.52 \scriptsize{$\pm$ 2.13} & 5.95 \scriptsize{$\pm$ 1.56} & 5.78 \scriptsize{$\pm$ 2.08} & 8.22 \scriptsize{$\pm$ 0.98} & 25.23 \scriptsize{$\pm$ 6.83} & 33.93 \scriptsize{$\pm$ 3.44} & 0.49 \scriptsize{$\pm$ 0.51} \\
 & \texttt{Kookmin} & 6.69 \scriptsize{$\pm$ 0.41} & 6.68 \scriptsize{$\pm$ 0.47} & 6.97 \scriptsize{$\pm$ 0.29} & 8.52 \scriptsize{$\pm$ 0.52} & 29.30 \scriptsize{$\pm$ 1.14} & 34.79 \scriptsize{$\pm$ 1.85} & \underline{0.28} \scriptsize{$\pm$ 0.04} \\
 & \texttt{SCIF} & \textbf{6.75} \scriptsize{$\pm$ 0.33} & \textbf{6.77} \scriptsize{$\pm$ 0.40} & 6.95 \scriptsize{$\pm$ 0.22} & 8.59 \scriptsize{$\pm$ 0.53} & 29.26 \scriptsize{$\pm$ 0.81} & 35.07 \scriptsize{$\pm$ 1.88} & \textbf{0.24} \scriptsize{$\pm$ 0.03} \\
 & \texttt{Seif} & 4.99 \scriptsize{$\pm$ 0.92} & 5.14 \scriptsize{$\pm$ 0.92} & 5.00 \scriptsize{$\pm$ 1.30} & 6.53 \scriptsize{$\pm$ 1.55} & 25.35 \scriptsize{$\pm$ 1.98} & 31.78 \scriptsize{$\pm$ 2.10} & 0.32 \scriptsize{$\pm$ 0.02} \\
\midrule
\textbf{Sets2Sets} & \texttt{Original} & 10.20 \scriptsize{$\pm$ 0.58} & 10.07 \scriptsize{$\pm$ 0.57} & 8.73 \scriptsize{$\pm$ 0.63} & 10.85 \scriptsize{$\pm$ 0.90} & 35.40 \scriptsize{$\pm$ 1.50} & 41.75 \scriptsize{$\pm$ 1.90} & 92.23 \scriptsize{$\pm$ 9.47} \\
 & \texttt{Retrain} & 9.83 \scriptsize{$\pm$ 0.52} & 9.58 \scriptsize{$\pm$ 0.33} & 8.30 \scriptsize{$\pm$ 0.60} & 10.20 \scriptsize{$\pm$ 0.60} & 34.82 \scriptsize{$\pm$ 1.48} & 40.35 \scriptsize{$\pm$ 1.95} & 78.23 \scriptsize{$\pm$ 9.78} \\
 & \texttt{Fanchuan} & \textbf{10.75} \scriptsize{$\pm$ 0.91} & \textbf{10.53} \scriptsize{$\pm$ 0.66} & \textbf{9.27} \scriptsize{$\pm$ 0.41} & \textbf{11.26} \scriptsize{$\pm$ 0.57} & \textbf{35.77} \scriptsize{$\pm$ 1.07} & 41.51 \scriptsize{$\pm$ 1.69} & 0.26 \scriptsize{$\pm$ 0.03} \\
 & \texttt{Kookmin} & 10.18 \scriptsize{$\pm$ 0.65} & 10.05 \scriptsize{$\pm$ 0.61} & 8.72 \scriptsize{$\pm$ 0.67} & 10.85 \scriptsize{$\pm$ 0.95} & 35.29 \scriptsize{$\pm$ 1.49} & \underline{41.64} \scriptsize{$\pm$ 1.90} & \textbf{0.11} \scriptsize{$\pm$ 0.03} \\
 & \texttt{SCIF} & \underline{10.20} \scriptsize{$\pm$ 0.64} & \underline{10.07} \scriptsize{$\pm$ 0.60} & \underline{8.73} \scriptsize{$\pm$ 0.65} & \underline{10.85} \scriptsize{$\pm$ 0.93} & \underline{35.30} \scriptsize{$\pm$ 1.44} & \textbf{41.67} \scriptsize{$\pm$ 1.84} & 0.37 \scriptsize{$\pm$ 0.04} \\
 & \texttt{Seif} & 9.93 \scriptsize{$\pm$ 0.69} & 9.85 \scriptsize{$\pm$ 0.65} & 8.45 \scriptsize{$\pm$ 0.86} & 10.67 \scriptsize{$\pm$ 1.02} & 34.74 \scriptsize{$\pm$ 1.80} & 41.30 \scriptsize{$\pm$ 1.95} & \underline{0.12} \scriptsize{$\pm$ 0.07} \\
\bottomrule
\end{tabular}
\end{table*}
\begin{table*}
\centering
\small
\caption{Recommendation utility, unlearning effectiveness, and unlearning efficiency in SBR for task-specific models on the RSC15 dataset in the spam removal scenario. Recommendation utility and unlearning effectiveness are measured by standard utility metrics. For efficiency, we report the average time needed to execute an unlearning request in minutes.}
\label{tab:results_rsc15}
\setlength\tabcolsep{9.4pt}
\begin{tabular}{ll  cc  cc  cc c}
\toprule
 &  &
\multicolumn{2}{c}{\textbf{nDCG (\%)}$\mathbf{\uparrow}$} &
\multicolumn{2}{c}{\textbf{Recall (\%)}$\mathbf{\uparrow}$} &
\multicolumn{2}{c}{\textbf{Hit (\%)}$\mathbf{\uparrow}$} &
\textbf{Time (min) $\mathbf{\downarrow}$} \\
\cmidrule(r){3-4}
\cmidrule(r){5-6}
\cmidrule(r){7-8}
\cmidrule(r){9-9}
\textbf{Model} &
\textbf{Algorithm} &
\textbf{@10} & \textbf{@20} &
\textbf{@10} & \textbf{@20} &
\textbf{@10} & \textbf{@20} &
\textbf{Avg/Req} \\
\midrule
\textbf{GRU4Rec} & \texttt{Original} & 37.65 \scriptsize{$\pm$ 0.03} & 39.91 \scriptsize{$\pm$ 0.03} & 58.84 \scriptsize{$\pm$ 0.04} & 67.74 \scriptsize{$\pm$ 0.05} & 58.84 \scriptsize{$\pm$ 0.04} & 67.74 \scriptsize{$\pm$ 0.05} & 413.87 \scriptsize{$\pm$ 118.81} \\
 & \texttt{Retrain} & 38.46 \scriptsize{$\pm$ 0.07} & 40.77 \scriptsize{$\pm$ 0.06} & 60.10 \scriptsize{$\pm$ 0.05} & 69.19 \scriptsize{$\pm$ 0.04} & 60.10 \scriptsize{$\pm$ 0.05} & 69.19 \scriptsize{$\pm$ 0.04} & 364 \scriptsize{$\pm$ 42.05} \\
 & \texttt{Fanchuan} & \underline{37.12} \scriptsize{$\pm$ 0.14} & \underline{39.45} \scriptsize{$\pm$ 0.15} & \underline{58.48} \scriptsize{$\pm$ 0.08} & \underline{67.67} \scriptsize{$\pm$ 0.11} & \underline{58.48} \scriptsize{$\pm$ 0.08} & \underline{67.67} \scriptsize{$\pm$ 0.11} & 1.28 \scriptsize{$\pm$ 0.09} \\
 & \texttt{Kookmin} & 14.93 \scriptsize{$\pm$ 4.30} & 15.86 \scriptsize{$\pm$ 4.52} & 22.97 \scriptsize{$\pm$ 6.50} & 26.61 \scriptsize{$\pm$ 7.35} & 22.97 \scriptsize{$\pm$ 6.50} & 26.61 \scriptsize{$\pm$ 7.35} & \underline{1.16} \scriptsize{$\pm$ 0.02} \\
 & \texttt{SCIF} & \textbf{38.44} \scriptsize{$\pm$ 0.04} & \textbf{40.74} \scriptsize{$\pm$ 0.03} & \textbf{60.07} \scriptsize{$\pm$ 0.05} & \textbf{69.15} \scriptsize{$\pm$ 0.05} & \textbf{60.07} \scriptsize{$\pm$ 0.05} & \textbf{69.15} \scriptsize{$\pm$ 0.05} & 1.82 \scriptsize{$\pm$ 0.07} \\
 & \texttt{Seif} & 0.01 \scriptsize{$\pm$ 0.01} & 0.01 \scriptsize{$\pm$ 0.02} & 0.02 \scriptsize{$\pm$ 0.02} & 0.04 \scriptsize{$\pm$ 0.04} & 0.02 \scriptsize{$\pm$ 0.02} & 0.04 \scriptsize{$\pm$ 0.04} & \textbf{1.15} \scriptsize{$\pm$ 0} \\
\midrule
\textbf{NARM} & \texttt{Original} & 37.25 \scriptsize{$\pm$ 0.02} & 39.52 \scriptsize{$\pm$ 0.02} & 58.45 \scriptsize{$\pm$ 0.02} & 67.39 \scriptsize{$\pm$ 0.02} & 58.45 \scriptsize{$\pm$ 0.02} & 67.39 \scriptsize{$\pm$ 0.02} & 539.19 \scriptsize{$\pm$ 77.34} \\
 & \texttt{Retrain} & 37.99 \scriptsize{$\pm$ 0.04} & 40.31 \scriptsize{$\pm$ 0.04} & 59.62 \scriptsize{$\pm$ 0.06} & 68.75 \scriptsize{$\pm$ 0.06} & 59.62 \scriptsize{$\pm$ 0.06} & 68.75 \scriptsize{$\pm$ 0.06} & 765.09 \scriptsize{$\pm$ 106.15} \\
 & \texttt{Fanchuan} & \underline{35.94} \scriptsize{$\pm$ 0.14} & \underline{38.30} \scriptsize{$\pm$ 0.13} & \underline{57.00} \scriptsize{$\pm$ 0.15} & \underline{66.28} \scriptsize{$\pm$ 0.10} & \underline{57.00} \scriptsize{$\pm$ 0.15} & \underline{66.28} \scriptsize{$\pm$ 0.10} & 1.14 \scriptsize{$\pm$ 0.01} \\
 & \texttt{Kookmin} & 10.21 \scriptsize{$\pm$ 0.39} & 10.85 \scriptsize{$\pm$ 0.50} & 15.87 \scriptsize{$\pm$ 0.62} & 18.38 \scriptsize{$\pm$ 1.06} & 15.87 \scriptsize{$\pm$ 0.62} & 18.38 \scriptsize{$\pm$ 1.06} & \textbf{1.05} \scriptsize{$\pm$ 0.01} \\
 & \texttt{SCIF} & \textbf{38.03} \scriptsize{$\pm$ 0.03} & \textbf{40.34} \scriptsize{$\pm$ 0.02} & \textbf{59.66} \scriptsize{$\pm$ 0.03} & \textbf{68.79} \scriptsize{$\pm$ 0.02} & \textbf{59.66} \scriptsize{$\pm$ 0.03} & \textbf{68.79} \scriptsize{$\pm$ 0.02} & 1.46 \scriptsize{$\pm$ 0} \\
 & \texttt{Seif} & 0.00 \scriptsize{$\pm$ 0.01} & 0.01 \scriptsize{$\pm$ 0.01} & 0.01 \scriptsize{$\pm$ 0.02} & 0.02 \scriptsize{$\pm$ 0.03} & 0.01 \scriptsize{$\pm$ 0.02} & 0.02 \scriptsize{$\pm$ 0.03} & \underline{1.06} \scriptsize{$\pm$ 0.01} \\
\midrule
\textbf{SASRec} & \texttt{Original} & 32.42 \scriptsize{$\pm$ 0.05} & 34.82 \scriptsize{$\pm$ 0.04} & 52.31 \scriptsize{$\pm$ 0.06} & 61.79 \scriptsize{$\pm$ 0.06} & 52.31 \scriptsize{$\pm$ 0.06} & 61.79 \scriptsize{$\pm$ 0.06} & 561.27 \scriptsize{$\pm$ 57.35} \\
 & \texttt{Retrain} & 33.12 \scriptsize{$\pm$ 0.05} & 35.59 \scriptsize{$\pm$ 0.06} & 53.41 \scriptsize{$\pm$ 0.03} & 63.11 \scriptsize{$\pm$ 0.08} & 53.41 \scriptsize{$\pm$ 0.03} & 63.11 \scriptsize{$\pm$ 0.08} & 843.67 \scriptsize{$\pm$ 195.32} \\
 & \texttt{Fanchuan} & \underline{31.25} \scriptsize{$\pm$ 0.39} & \underline{33.69} \scriptsize{$\pm$ 0.40} & \underline{50.88} \scriptsize{$\pm$ 0.58} & \underline{60.51} \scriptsize{$\pm$ 0.61} & \underline{50.88} \scriptsize{$\pm$ 0.58} & \underline{60.51} \scriptsize{$\pm$ 0.61} & 1.31 \scriptsize{$\pm$ 0.08} \\
 & \texttt{Kookmin} & 2.10 \scriptsize{$\pm$ 0.46} & 2.84 \scriptsize{$\pm$ 0.57} & 4.12 \scriptsize{$\pm$ 0.94} & 7.07 \scriptsize{$\pm$ 1.42} & 4.12 \scriptsize{$\pm$ 0.94} & 7.07 \scriptsize{$\pm$ 1.42} & \textbf{1.08} \scriptsize{$\pm$ 0.1} \\
 & \texttt{SCIF} & \textbf{33.09} \scriptsize{$\pm$ 0.05} & \textbf{35.55} \scriptsize{$\pm$ 0.05} & \textbf{53.40} \scriptsize{$\pm$ 0.06} & \textbf{63.07} \scriptsize{$\pm$ 0.06} & \textbf{53.40} \scriptsize{$\pm$ 0.06} & \textbf{63.07} \scriptsize{$\pm$ 0.06} & 1.27 \scriptsize{$\pm$ 0.1} \\
 & \texttt{Seif} & 0.01 \scriptsize{$\pm$ 0.01} & 0.01 \scriptsize{$\pm$ 0.01} & 0.02 \scriptsize{$\pm$ 0.03} & 0.03 \scriptsize{$\pm$ 0.04} & 0.02 \scriptsize{$\pm$ 0.03} & 0.03 \scriptsize{$\pm$ 0.04} & \underline{1.25} \scriptsize{$\pm$ 0.08} \\
\midrule
\textbf{SRGNN} & \texttt{Original} & 39.23 \scriptsize{$\pm$ 0.01} & 41.41 \scriptsize{$\pm$ 0.01} & 60.18 \scriptsize{$\pm$ 0.01} & 68.72 \scriptsize{$\pm$ 0.01} & 60.18 \scriptsize{$\pm$ 0.01} & 68.72 \scriptsize{$\pm$ 0.01} & 660.73 \scriptsize{$\pm$ 154.37} \\
 & \texttt{Retrain} & 40.04 \scriptsize{$\pm$ 0.04} & 42.27 \scriptsize{$\pm$ 0.04} & 61.43 \scriptsize{$\pm$ 0.03} & 70.16 \scriptsize{$\pm$ 0.02} & 61.43 \scriptsize{$\pm$ 0.03} & 70.16 \scriptsize{$\pm$ 0.02} & 1441.69 \scriptsize{$\pm$ 6.4} \\
 & \texttt{CEU} & 11.46 \scriptsize{$\pm$ 9.32} & 13.30 \scriptsize{$\pm$ 10.55} & 21.02 \scriptsize{$\pm$ 16.66} & 28.28 \scriptsize{$\pm$ 21.61} & 21.02 \scriptsize{$\pm$ 16.66} & 28.28 \scriptsize{$\pm$ 21.61} & 1.81 \scriptsize{$\pm$ 0.02} \\
 & \texttt{Fanchuan} & 30.37 \scriptsize{$\pm$ 0.21} & 32.76 \scriptsize{$\pm$ 0.20} & 49.95 \scriptsize{$\pm$ 0.26} & 59.33 \scriptsize{$\pm$ 0.23} & 49.95 \scriptsize{$\pm$ 0.26} & 59.33 \scriptsize{$\pm$ 0.23} & 1.51 \scriptsize{$\pm$ 0.08} \\
 & \texttt{Kookmin} & \underline{31.55} \scriptsize{$\pm$ 2.18} & \underline{33.94} \scriptsize{$\pm$ 2.20} & \underline{51.14} \scriptsize{$\pm$ 3.20} & \underline{60.55} \scriptsize{$\pm$ 3.30} & \underline{51.14} \scriptsize{$\pm$ 3.20} & \underline{60.55} \scriptsize{$\pm$ 3.30} & \textbf{1.13} \scriptsize{$\pm$ 0.09} \\
 & \texttt{SCIF} & \textbf{40.05} \scriptsize{$\pm$ 0.01} & \textbf{42.27} \scriptsize{$\pm$ 0.01} & \textbf{61.43} \scriptsize{$\pm$ 0.01} & \textbf{70.15} \scriptsize{$\pm$ 0.01} & \textbf{61.43} \scriptsize{$\pm$ 0.01} & \textbf{70.15} \scriptsize{$\pm$ 0.01} & 1.24 \scriptsize{$\pm$ 0.11} \\
 & \texttt{Seif} & 0.03 \scriptsize{$\pm$ 0.02} & 0.04 \scriptsize{$\pm$ 0.02} & 0.07 \scriptsize{$\pm$ 0.04} & 0.10 \scriptsize{$\pm$ 0.05} & 0.07 \scriptsize{$\pm$ 0.04} & 0.10 \scriptsize{$\pm$ 0.05} & \underline{1.2} \scriptsize{$\pm$ 0.01} \\
\bottomrule
\end{tabular}
\end{table*}

\begin{table*}
\centering
\small
\caption{Recommendation utility, unlearning effectiveness, and unlearning efficiency in SBR for task-specific models on the NowP dataset in the spam removal scenario. Recommendation utility and unlearning effectiveness are measured by standard utility metrics. For efficiency, we report the average time needed to execute an unlearning request in minutes.}
\label{tab:results_nowp}
\setlength\tabcolsep{9.6pt}
\begin{tabular}{ll  c c  c c  c c  c}
\toprule
&  &
\multicolumn{2}{c}{\textbf{nDCG (\%)}$\mathbf{\uparrow}$} &
\multicolumn{2}{c}{\textbf{Recall (\%)}$\mathbf{\uparrow}$} &
\multicolumn{2}{c}{\textbf{Hit (\%)}$\mathbf{\uparrow}$} &
\textbf{Time (min) $\mathbf{\downarrow}$} \\
\cmidrule(r){3-4}
\cmidrule(r){5-6}
\cmidrule(r){7-8}
\cmidrule(r){9-9}
\textbf{Model} &
\textbf{Algorithm} &
\textbf{@10} & \textbf{@20} &
\textbf{@10} & \textbf{@20} &
\textbf{@10} & \textbf{@20} &
\textbf{Avg/Req} \\
\midrule
\textbf{GRU4Rec} & \texttt{Original} & 76.88 \scriptsize{$\pm$ 0.16} & 77.07 \scriptsize{$\pm$ 0.16} & 79.34 \scriptsize{$\pm$ 0.13} & 80.10 \scriptsize{$\pm$ 0.13} & 79.34 \scriptsize{$\pm$ 0.13} & 80.10 \scriptsize{$\pm$ 0.13} & 626.88 \scriptsize{$\pm$ 804.7} \\
 & \texttt{Retrain} & 79.69 \scriptsize{$\pm$ 0.46} & 79.93 \scriptsize{$\pm$ 0.45} & 82.48 \scriptsize{$\pm$ 0.40} & 83.44 \scriptsize{$\pm$ 0.38} & 82.48 \scriptsize{$\pm$ 0.40} & 83.44 \scriptsize{$\pm$ 0.38} & 906.05 \scriptsize{$\pm$ 3.17} \\
 & \texttt{Fanchuan} & \underline{74.05} \scriptsize{$\pm$ 1.08} & \underline{74.20} \scriptsize{$\pm$ 1.09} & \underline{76.05} \scriptsize{$\pm$ 1.15} & \underline{76.63} \scriptsize{$\pm$ 1.17} & \underline{76.05} \scriptsize{$\pm$ 1.15} & \underline{76.63} \scriptsize{$\pm$ 1.17} & \underline{24.12} \scriptsize{$\pm$ 2.36} \\
 & \texttt{Kookmin} & 66.42 \scriptsize{$\pm$ 0.35} & 66.60 \scriptsize{$\pm$ 0.34} & 68.40 \scriptsize{$\pm$ 0.34} & 69.16 \scriptsize{$\pm$ 0.34} & 68.40 \scriptsize{$\pm$ 0.34} & 69.16 \scriptsize{$\pm$ 0.34} & 28.39 \scriptsize{$\pm$ 2.15} \\
 & \texttt{SCIF} & \textbf{77.49} \scriptsize{$\pm$ 0.13} & \textbf{77.69} \scriptsize{$\pm$ 0.12} & \textbf{79.98} \scriptsize{$\pm$ 0.09} & \textbf{80.74} \scriptsize{$\pm$ 0.09} & \textbf{79.98} \scriptsize{$\pm$ 0.09} & \textbf{80.74} \scriptsize{$\pm$ 0.09} & 214.97 \scriptsize{$\pm$ 50.66} \\
 & \texttt{Seif} & 64.04 \scriptsize{$\pm$ 1.07} & 64.24 \scriptsize{$\pm$ 1.07} & 66.14 \scriptsize{$\pm$ 1.06} & 66.92 \scriptsize{$\pm$ 1.06} & 66.14 \scriptsize{$\pm$ 1.06} & 66.92 \scriptsize{$\pm$ 1.06} & \textbf{11.14} \scriptsize{$\pm$ 0.09} \\
\midrule
\textbf{NARM} & \texttt{Original} & 76.78 \scriptsize{$\pm$ 0.54} & 76.97 \scriptsize{$\pm$ 0.54} & 79.50 \scriptsize{$\pm$ 0.33} & 80.25 \scriptsize{$\pm$ 0.30} & 79.50 \scriptsize{$\pm$ 0.33} & 80.25 \scriptsize{$\pm$ 0.30} & 408.68 \scriptsize{$\pm$ 451.88} \\
 & \texttt{Retrain} & 81.14 \scriptsize{$\pm$ 0.36} & 81.35 \scriptsize{$\pm$ 0.35} & 83.83 \scriptsize{$\pm$ 0.16} & 84.67 \scriptsize{$\pm$ 0.14} & 83.83 \scriptsize{$\pm$ 0.16} & 84.67 \scriptsize{$\pm$ 0.14} & 1287.81 \scriptsize{$\pm$ 229.73} \\
 & \texttt{Fanchuan} & \underline{67.28} \scriptsize{$\pm$ 8.43} & \underline{67.53} \scriptsize{$\pm$ 8.30} & \underline{70.67} \scriptsize{$\pm$ 6.59} & \underline{71.68} \scriptsize{$\pm$ 6.06} & \underline{70.67} \scriptsize{$\pm$ 6.59} & \underline{71.68} \scriptsize{$\pm$ 6.06} & 26.15 \scriptsize{$\pm$ 2.3} \\
 & \texttt{Kookmin} & 52.84 \scriptsize{$\pm$ 1.53} & 53.11 \scriptsize{$\pm$ 1.53} & 55.16 \scriptsize{$\pm$ 1.62} & 56.20 \scriptsize{$\pm$ 1.63} & 55.16 \scriptsize{$\pm$ 1.62} & 56.20 \scriptsize{$\pm$ 1.63} & \underline{15.53} \scriptsize{$\pm$ 1.69} \\
 & \texttt{SCIF} & \textbf{77.39} \scriptsize{$\pm$ 0.59} & \textbf{77.58} \scriptsize{$\pm$ 0.58} & \textbf{80.14} \scriptsize{$\pm$ 0.36} & \textbf{80.89} \scriptsize{$\pm$ 0.33} & \textbf{80.14} \scriptsize{$\pm$ 0.36} & \textbf{80.89} \scriptsize{$\pm$ 0.33} & 128.87 \scriptsize{$\pm$ 14.4} \\
 & \texttt{Seif} & 62.15 \scriptsize{$\pm$ 2.63} & 62.37 \scriptsize{$\pm$ 2.60} & 64.72 \scriptsize{$\pm$ 2.49} & 65.59 \scriptsize{$\pm$ 2.38} & 64.72 \scriptsize{$\pm$ 2.49} & 65.59 \scriptsize{$\pm$ 2.38} & \textbf{11.09} \scriptsize{$\pm$ 0.06} \\
\midrule
\textbf{SASRec} & \texttt{Original} & 49.24 \scriptsize{$\pm$ 0.51} & 49.85 \scriptsize{$\pm$ 0.51} & 64.43 \scriptsize{$\pm$ 0.18} & 66.82 \scriptsize{$\pm$ 0.16} & 64.43 \scriptsize{$\pm$ 0.18} & 66.82 \scriptsize{$\pm$ 0.16} & 292.81 \scriptsize{$\pm$ 193.97} \\
 & \texttt{Retrain} & 58.56 \scriptsize{$\pm$ 2.08} & 59.13 \scriptsize{$\pm$ 2.03} & 71.20 \scriptsize{$\pm$ 0.49} & 73.41 \scriptsize{$\pm$ 0.33} & 71.20 \scriptsize{$\pm$ 0.49} & 73.41 \scriptsize{$\pm$ 0.33} & 1224.19 \scriptsize{$\pm$ 180.37} \\
 & \texttt{Fanchuan} & \underline{43.45} \scriptsize{$\pm$ 0.45} & \underline{44.07} \scriptsize{$\pm$ 0.43} & \underline{58.38} \scriptsize{$\pm$ 0.75} & \underline{60.81} \scriptsize{$\pm$ 0.72} & \underline{58.38} \scriptsize{$\pm$ 0.75} & \underline{60.81} \scriptsize{$\pm$ 0.72} & 38.25 \scriptsize{$\pm$ 3.9} \\
 & \texttt{Kookmin} & 25.25 \scriptsize{$\pm$ 2.79} & 25.75 \scriptsize{$\pm$ 2.78} & 32.69 \scriptsize{$\pm$ 2.94} & 34.65 \scriptsize{$\pm$ 2.93} & 32.69 \scriptsize{$\pm$ 2.94} & 34.65 \scriptsize{$\pm$ 2.93} & \underline{16.96} \scriptsize{$\pm$ 1.59} \\
 & \texttt{SCIF} & \textbf{49.64} \scriptsize{$\pm$ 0.56} & \textbf{50.25} \scriptsize{$\pm$ 0.55} & \textbf{64.95} \scriptsize{$\pm$ 0.19} & \textbf{67.35} \scriptsize{$\pm$ 0.17} & \textbf{64.95} \scriptsize{$\pm$ 0.19} & \textbf{67.35} \scriptsize{$\pm$ 0.17} & 63.85 \scriptsize{$\pm$ 6.62} \\
 & \texttt{Seif} & 0.11 \scriptsize{$\pm$ 0.09} & 0.17 \scriptsize{$\pm$ 0.12} & 0.25 \scriptsize{$\pm$ 0.19} & 0.49 \scriptsize{$\pm$ 0.30} & 0.25 \scriptsize{$\pm$ 0.19} & 0.49 \scriptsize{$\pm$ 0.30} & \textbf{11.08} \scriptsize{$\pm$ 0.06} \\
\midrule
\textbf{SRGNN} & \texttt{Original} & 70.20 \scriptsize{$\pm$ 0.18} & 70.45 \scriptsize{$\pm$ 0.18} & 72.85 \scriptsize{$\pm$ 0.13} & 73.85 \scriptsize{$\pm$ 0.11} & 72.85 \scriptsize{$\pm$ 0.13} & 73.85 \scriptsize{$\pm$ 0.11} & 355.87 \scriptsize{$\pm$ 350.49} \\
 & \texttt{Retrain} & 74.75 \scriptsize{$\pm$ 0.05} & 75.04 \scriptsize{$\pm$ 0.05} & 77.52 \scriptsize{$\pm$ 0.05} & 78.63 \scriptsize{$\pm$ 0.05} & 77.52 \scriptsize{$\pm$ 0.05} & 78.63 \scriptsize{$\pm$ 0.05} & 1046.12 \scriptsize{$\pm$ 26.54} \\
 & \texttt{CEU} & \underline{70.76} \scriptsize{$\pm$ 0.16} & \textbf{71.01} \scriptsize{$\pm$ 0.16} & \textbf{73.43} \scriptsize{$\pm$ 0.10} & \underline{74.44} \scriptsize{$\pm$ 0.08} & \textbf{73.43} \scriptsize{$\pm$ 0.10} & \underline{74.44} \scriptsize{$\pm$ 0.08} & \textbf{1.96} \scriptsize{$\pm$ 0.35} \\
 & \texttt{Fanchuan} & 69.53 \scriptsize{$\pm$ 0.84} & \underline{69.72} \scriptsize{$\pm$ 0.84} & \underline{71.59} \scriptsize{$\pm$ 0.84} & 72.35 \scriptsize{$\pm$ 0.84} & \underline{71.59} \scriptsize{$\pm$ 0.84} & 72.35 \scriptsize{$\pm$ 0.84} & 105.59 \scriptsize{$\pm$ 9.3} \\
 & \texttt{Kookmin} & 48.47 \scriptsize{$\pm$ 2.14} & 48.63 \scriptsize{$\pm$ 2.13} & 49.80 \scriptsize{$\pm$ 2.20} & 50.45 \scriptsize{$\pm$ 2.30} & 49.80 \scriptsize{$\pm$ 2.20} & 50.45 \scriptsize{$\pm$ 2.30} & 22.06 \scriptsize{$\pm$ 2.86} \\
 & \texttt{SCIF} & \textbf{70.76} \scriptsize{$\pm$ 0.16} & \textbf{71.01} \scriptsize{$\pm$ 0.16} & \textbf{73.43} \scriptsize{$\pm$ 0.10} & \textbf{74.45} \scriptsize{$\pm$ 0.08} & \textbf{73.43} \scriptsize{$\pm$ 0.10} & \textbf{74.45} \scriptsize{$\pm$ 0.08} & 56.79 \scriptsize{$\pm$ 5.88} \\
 & \texttt{Seif} & 64.85 \scriptsize{$\pm$ 4.54} & 65.12 \scriptsize{$\pm$ 4.50} & 67.31 \scriptsize{$\pm$ 4.21} & 68.40 \scriptsize{$\pm$ 4.06} & 67.31 \scriptsize{$\pm$ 4.21} & 68.40 \scriptsize{$\pm$ 4.06} & \underline{11.3} \scriptsize{$\pm$ 0.44} \\
\bottomrule
\end{tabular}
\end{table*}

\begin{table*}[t!]
\centering
\caption{Unlearning robustness for unlearning algorithms is reported by averaging unlearning effectiveness metrics across models and datasets. For sensitive item unlearning we report $\textbf{RelItems}\boldsymbol{@20 \uparrow}$. For spam removal we report relative performance compared to retrained models using $\textbf{RelEff}\boldsymbol{@20 \uparrow}$.}
\label{tab:robustness}
\setlength\tabcolsep{4.5pt}
\setlength{\aboverulesep}{1.15pt}
\setlength{\belowrulesep}{0.85pt}
\begin{tabular}{l *{5}{>{\centering\arraybackslash}p{2.9cm}}}
\toprule
& \multicolumn{3}{c}{\emph{Unlearning Interactions with Sensitive Items} (\textbf{$\relitems{}@$}$\boldsymbol{20\uparrow})$} & \multicolumn{2}{c}{\emph{Unlearning Spam Interactions} (\textbf{$\releff{}@$}$\boldsymbol{20\uparrow})$} \\
\cmidrule(r){2-4}
\cmidrule(r){5-6}
\textbf{Algorithm} & \textbf{NBR} & \textbf{SBR} & \textbf{CF} & \textbf{NBR} & \textbf{SBR} \\
\midrule
\texttt{SCIF}     & \addcell{-10.36}{-10.36 \scriptsize{$\pm$ 23.16}} & \addcell{-6.02}{-6.02 \scriptsize{$\pm$ 1.27}}   & \addcell{+1.17}{+1.17 \scriptsize{$\pm$ 2.74}}   & \relcell{+1.05}{+1.05 \scriptsize{$\pm$ 6.86}}   & \relcell{-3.76}{-3.76 \scriptsize{$\pm$ 4.98}}   \\
\texttt{GIF}      & \diverged{}                                       & \diverged{}                                       & \addcell{+8.38}{+8.38 \scriptsize{$\pm$ 9.33}}   & \relcell{-27.19}{-27.19 \scriptsize{$\pm$ 39.97}} & \diverged{}                                       \\
\texttt{CEU}      & \addcell{-20.26}{-20.26 \scriptsize{$\pm$ 27.84}} & \addcell{-21.84}{-21.84 \scriptsize{$\pm$ 0.00}} & \addcell{+1.60}{+1.60 \scriptsize{$\pm$ 2.59}}   & \relcell{-0.26}{-0.26 \scriptsize{$\pm$ 8.97}}   & \relcell{-13.99}{-13.99 \scriptsize{$\pm$ 18.24}} \\
\texttt{IDEA}     & \addcell{-26.35}{-26.35 \scriptsize{$\pm$ 36.42}} & \addcell{+14.06}{+14.06 \scriptsize{$\pm$ 0.00}} & \addcell{+5.25}{+5.25 \scriptsize{$\pm$ 0.00}}   & \relcell{-28.33}{-28.33 \scriptsize{$\pm$ 37.62}} & \diverged{}                                       \\
\texttt{Fanchuan} & \addcell{-10.58}{-10.58 \scriptsize{$\pm$ 17.70}} & \addcell{-9.36}{-9.36 \scriptsize{$\pm$ 9.66}}   & \addcell{-3.54}{-3.54 \scriptsize{$\pm$ 2.81}}   & \relcell{-5.72}{-5.72 \scriptsize{$\pm$ 12.45}}  & \relcell{-11.67}{-11.67 \scriptsize{$\pm$ 10.82}} \\
\texttt{Kookmin}  & \addcell{-6.84}{-6.84 \scriptsize{$\pm$ 27.67}}  & \addcell{-1.74}{-1.74 \scriptsize{$\pm$ 15.92}}  & \addcell{+1.08}{+1.08 \scriptsize{$\pm$ 2.54}}   & \relcell{+0.49}{+0.49 \scriptsize{$\pm$ 7.39}}   & \relcell{-42.34}{-42.34 \scriptsize{$\pm$ 27.11}} \\
\texttt{Seif}     & \addcell{-11.63}{-11.63 \scriptsize{$\pm$ 10.33}} & \addcell{+0.58}{+0.58 \scriptsize{$\pm$ 5.60}}   & \addcell{+2.10}{+2.10 \scriptsize{$\pm$ 7.64}}   & \relcell{-8.06}{-8.06 \scriptsize{$\pm$ 13.36}}  & \relcell{-64.62}{-64.62 \scriptsize{$\pm$ 41.54}} \\
\bottomrule
\end{tabular}
\end{table*}

\begin{table*}[t]
\centering
\caption{Hyperparameter test of \texttt{SCIF} for \texttt{max\_norm} in the sensitive item removal scenario for CF using the Food dataset with the LightGCN model.}
\label{tab:scif-maxnorm}
\setlength\tabcolsep{12.75pt}
\begin{tabular}{lccccc}
\toprule
\textbf{\texttt{max\_norm}} &
\textbf{nDCG@20 (\%)} $\boldsymbol{\uparrow}$ & \textbf{Recall@20 (\%)} $\boldsymbol{\uparrow}$ & \textbf{Hit@20 (\%)} $\boldsymbol{\uparrow}$ &
\textbf{Sensitive@20 (\%)} $\boldsymbol{\downarrow}$ & \textbf{Time (minutes)} $\boldsymbol{\downarrow}$ \\
\midrule
$10^{-2}$  & \textbf{1.41} & \underline{3.29} & \textbf{3.51} & \textbf{43.51} & \underline{36.30} \\
$10^{-1}$ & \textbf{1.41} & \underline{3.29} & \textbf{3.51} & \textbf{43.51} & 38.80 \\
$10^{0}$  & \textbf{1.41} & \textbf{3.30} & \textbf{3.51} & \textbf{43.51} & 39.03 \\
$10^{1}$ & \textbf{1.41} & \underline{3.29} & \textbf{3.51} & \textbf{43.51} & \textbf{36.18} \\
\bottomrule
\end{tabular}
\end{table*}

\begin{table*}[t]
\centering
\caption{Hyperparameter test of \texttt{Kookmin} for \texttt{kookmin\_init\_rate} in the sensitive item removal scenario for CF using the Food dataset with the LightGCN model.}
\label{tab:kookmin-initrate}
\setlength\tabcolsep{9pt}
\begin{tabular}{lccccc}
\toprule
\textbf{\texttt{kookmin\_init\_rate}} &
\textbf{nDCG@20 (\%)} $\boldsymbol{\uparrow}$ &
\textbf{Recall@20 (\%)} $\boldsymbol{\uparrow}$ &
\textbf{Hit@20 (\%)} $\boldsymbol{\uparrow}$ &
\textbf{Sensitive@20 (\%)} $\boldsymbol{\downarrow}$ &
\textbf{Time (minutes)} $\boldsymbol{\downarrow}$ \\
\midrule
$10^{-5}$       & \textbf{1.41} & \textbf{3.29} & \textbf{3.51} & \textbf{43.18} & \underline{8.47} \\
$10^{-4}$       & \textbf{1.41} & \textbf{3.29} & \textbf{3.51} & \textbf{43.18} & 8.49 \\
$10^{-3}$       & \textbf{1.41} & \textbf{3.29} & \textbf{3.51} & 44.48         & 8.58 \\
$10^{-2}$       & \underline{1.38} & \underline{3.22} & \underline{3.44} & \underline{43.83} & 9.68 \\
$5 \cdot 10^{-2}$ & 1.33        & 3.07        & 3.28        & 44.81         & \textbf{8.37} \\
\bottomrule
\end{tabular}
\end{table*}

\begin{table*}[t]
\centering
\setlength\tabcolsep{3.5pt}
\caption{Recommendation utility, unlearning effectiveness, and unlearning efficiency metrics for different unlearning algorithms and unlearning request batch sizes on a subset of the RSC15 dataset.}
\label{tab:unlearning_batchsize_results}
\setlength\tabcolsep{12.5pt}
\vspace{0.1cm}
\setlength\extrarowheight{-1pt}
\begin{tabular}{l c  c c  c c  c c  c}
\toprule
\textbf{Algorithm} &
\textbf{Batch Size} &
\multicolumn{2}{c}{\textbf{Recall (\%)}$\mathbf{\uparrow}$} &
\multicolumn{2}{c}{\textbf{nDCG (\%)}$\mathbf{\uparrow}$} &
\multicolumn{2}{c}{\textbf{Hit (\%)}$\mathbf{\uparrow}$} &
\textbf{Time/Request (s) $\mathbf{\downarrow}$} \\
\cmidrule(r){3-4}
\cmidrule(r){5-6}
\cmidrule(r){7-8}
\cmidrule(r){9-9}
 &  &
\textbf{@10} & \textbf{@20} &
\textbf{@10} & \textbf{@20} &
\textbf{@10} & \textbf{@20} &
 \\
\midrule
\texttt{Kookmin} & 1   & 39.35 & 47.22 & 24.40 & 26.39 & 39.35 & 47.22 & 0.7787 \\
                 & 2   & 44.77 & 52.57 & 28.04 & 30.02 & 44.77 & 52.57 & 0.3844 \\
                 & 4   & 46.43 & 54.33 & 29.10 & 31.10 & 46.43 & 54.33 & 0.1974 \\
                 & 8   & 46.84 & 54.86 & 29.35 & 31.38 & 46.84 & 54.86 & 0.1046 \\
                 & 16  & 47.89 & 55.78 & 30.21 & 32.21 & 47.89 & 55.78 & 0.0586 \\
                 & 32  & 48.62 & 56.62 & 30.85 & 32.88 & 48.62 & 56.62 & 0.0358 \\
                 & 64  & 48.86 & 56.88 & 31.05 & 33.09 & 48.86 & 56.88 & 0.0229 \\
                 & 128 & \underline{49.34} & \underline{57.19} & \underline{31.39} & \underline{33.38} & \underline{49.34} & \underline{57.19} & {0.0186} \\
                 & 256 & \textbf{49.44} & \textbf{57.47} & \textbf{31.49} & \textbf{33.53} & \textbf{49.44} & \textbf{57.47} & \underline{0.0121} \\
                 & 512 & 49.39 & 57.42 & 31.47 & 33.51 & 49.39 & 57.42 & \textbf{0.0088} \\
\midrule
\texttt{SCIF}    & 1   & 2.35  & 3.63  & 1.29  & 1.61  & 2.35  & 3.63  & 13.5533 \\
                 & 2   & 1.64  & 2.87  & 0.73  & 1.04  & 1.64  & 2.87  & 13.8718 \\
                 & 4   & 46.59 & 55.46 & 28.24 & 30.49 & 46.59 & 55.46 & 13.5540 \\
                 & 8   & 49.37 & 57.67 & 31.03 & 33.14 & 49.37 & 57.67 & 13.1827 \\
                 & 16  & 49.54 & 57.72 & \underline{31.43} & \underline{33.52} & 49.54 & 57.72 & 13.1483 \\
                 & 32  & 49.54 & \underline{57.76} & \textbf{31.48} & \textbf{33.57} & 49.54 & \underline{57.76} & 8.6703  \\
                 & 64  & \textbf{49.56} & \textbf{57.77} & \textbf{31.48} & \textbf{33.57} & \textbf{49.56} & \textbf{57.77} & 4.4762  \\
                 & 128 & \underline{49.55} & 57.73 & \textbf{31.48} & \textbf{33.57} & \underline{49.55} & 57.73 & 2.3410  \\
                 & 256 & \textbf{49.56} & 57.75 & \textbf{31.48} & \textbf{33.57} & \textbf{49.56} & 57.75 & \underline{1.2597}  \\
                 & 512 & \textbf{49.56} & 57.75 & \textbf{31.48} & \underline{33.56} & \textbf{49.56} & 57.75 & \textbf{0.7366}  \\
\midrule
\texttt{Fanchuan} & 1  & 34.94 & 42.71 & 21.05 & 23.02 & 34.94 & 42.71 & 0.9013 \\
                  & 2  & 30.36 & 38.13 & 17.78 & 19.75 & 30.36 & 38.13 & 0.4530 \\
                  & 4  & 23.44 & 30.32 & 13.72 & 15.45 & 23.44 & 30.32 & 0.3424 \\
                  & 8  & 26.57 & 34.21 & 15.88 & 17.81 & 26.57 & 34.21 & 0.1572 \\
                  & 16 & 34.40 & 42.23 & 20.84 & 22.82 & 34.40 & 42.23 & 0.0762 \\
                  & 32 & 40.07 & 48.01 & 24.92 & 26.94 & 40.07 & 48.01 & 0.0501 \\
                  & 64 & 43.28 & 51.13 & 27.30 & 29.29 & 43.28 & 51.13 & 0.0369 \\
                  & 128& \underline{44.15} & \underline{52.27} & \underline{27.83} & \underline{29.89} & \underline{44.15} & \underline{52.27} & {0.0275} \\
                  & 256& \textbf{45.09} & \textbf{52.91} & \textbf{28.40} & \textbf{30.39} & \textbf{45.09} & \textbf{52.91} & \textbf{0.0209} \\
                  & 512& 42.35 & 50.01 & 26.82 & 28.77 & 42.35 & 50.01 & \underline{0.0226} \\
\midrule
\texttt{Seif}    & 1   & 0.17  & 0.30  & 0.09  & 0.12  & 0.17  & 0.30  & 1.1711 \\
                 & 2   & 0.05  & 0.16  & 0.02  & 0.05  & 0.05  & 0.16  & 0.4479 \\
                 & 4   & 0.05  & 0.16  & 0.02  & 0.05  & 0.05  & 0.16  & 0.2781 \\
                 & 8   & 0.14  & 0.22  & 0.07  & 0.08  & 0.14  & 0.22  & 0.1681 \\
                 & 16  & 0.14  & 0.24  & 0.09  & 0.11  & 0.14  & 0.24  & 0.1211 \\
                 & 32  & 0.09  & 0.15  & 0.05  & 0.07  & 0.09  & 0.15  & 0.1004 \\
                 & 64  & 0.16  & 0.33  & 0.08  & 0.12  & 0.16  & 0.33  & 0.0667 \\
                 & 128 & 10.32 & 14.56 & 5.72  & 6.79  & 10.32 & 14.56 & 0.0351 \\
                 & 256 & \underline{32.51} & \underline{39.61} & \underline{20.64} & \underline{22.44} & \underline{32.51} & \underline{39.61} & \underline{0.0199} \\
                 & 512 & \textbf{37.48} & \textbf{44.95} & \textbf{23.99} & \textbf{25.89} & \textbf{37.48} & \textbf{44.95} & \textbf{0.0112} \\
\bottomrule
\end{tabular}
\end{table*}

\end{document}